\documentclass[a4paper,onecolumn,11pt,unpublished,noarxiv,allowtoday]{quantumarticle}
\pdfoutput=1
\usepackage[utf8]{inputenc}
\usepackage[english]{babel}
\usepackage[T1]{fontenc}
\usepackage{amsmath}
\usepackage{hyperref}
\usepackage{mathtools, nccmath}
\usepackage{float}
\usepackage{setspace}
\usepackage{hyperref}
\usepackage{outlines}
\usepackage{fancyvrb}
\usepackage{mathtools}
\usepackage{physics}
\usepackage{comment}
\usepackage{bbold}
\usepackage{amssymb}
\usepackage{cancel}
\usepackage{graphicx}
\usepackage[caption=false]{subfig}
\usepackage[usenames, dvipsnames]{color}
\usepackage{soul}
\usepackage{tensor}
\usepackage{amsthm}
\usepackage[most]{tcolorbox}
\newtcolorbox{webrulebox}{enhanced, colback=gray!20,
  colframe=gray!55, boxrule=0.4pt, arc=2.5mm,
  left=3mm, right=3mm, top=2mm, bottom=2mm}
\usepackage[caption=false]{subfig}
\usepackage{algorithm}
\usepackage{algcompatible}

\DeclarePairedDelimiter{\nint}\lfloor\rceil
\usepackage{tikz}
\usepackage{lipsum}
\usepackage{tkz-graph}
\usetikzlibrary{hobby,backgrounds,calc,trees,decorations.pathreplacing}
\usepackage{xcolor}
\usepackage{soul}
\usetikzlibrary{positioning}
\usepackage{ulem}
\usepackage{yquant}
\usepackage{caption}
\usepackage{subfig}
\useyquantlanguage{groups}
\usepackage{makecell}
\usepackage{tikz,esvect,tikz-3dplot}
\usetikzlibrary{3d,calc,intersections}
\usepackage{tikz}
\usepackage{tikz-3dplot}
\usepackage{adjustbox}
\usepackage{zx-calculus}
\usetikzlibrary{zx-calculus}
\usepackage{tikz-cd}
\usepackage[official]{eurosym}

\makeatletter

\usepackage{tensor}
\usepackage{accents}
\usetikzlibrary{angles,quotes}

\definecolor{zx_red}{RGB}{232, 165, 165}
\definecolor{zx_green}{RGB}{216, 248, 216}
\definecolor{zx_blue}{RGB}{173, 205, 240}
\definecolor{zx_hadamard}{RGB}{68, 136, 255}

\usepackage[numbers,sort&compress]{natbib}
\begin{document}

\title{Pauli web of the $\ket{Y}$ state surface code injection}

\date{\today}

\author{Kwok Ho Wan}
\email{((initials))1496((at))((9.81))mail((dot))com}
\orcid{0000-0002-1762-1001}
\affiliation{Blackett Laboratory, Imperial College London, South Kensington, London SW7 2AZ, UK}
\affiliation{Mathematical Institute, University of Oxford, Andrew Wiles Building, Woodstock Road, Oxford OX2 6GG, UK}

\author{Zhenghao Zhong}
\orcid{0000-0001-5159-1013}
\affiliation{Mathematical Institute, University of Oxford, Andrew Wiles Building, Woodstock Road, Oxford OX2 6GG, UK}
\affiliation{Blackett Laboratory, Imperial College London, South Kensington, London SW7 2AZ, UK}

\author{Ainhoa Zapirain}
\affiliation{Independent Researcher}

\begin{abstract}
We employ ZX-calculus and Pauli web to understand the $\ket{Y}$ state injection on the rotated surface code.
Under circuit-level noise, we devise an optimised schedule that
improves the logical error rate, for which we provide closed forms
using parameterised ZX-diagrams.
\end{abstract}

\maketitle

\section{Introduction}
Magic states ($\ket{T} \propto \ket{0}+\text{exp}(i\pi/4)\ket{1}$) are resource states in fault-tolerant quantum computation \cite{gottesman1997stabilizercodesquantumerror,Aaronson_2004}. One main challenge in practical fault-tolerant quantum computation is to encode such state from a bare physical $\ket{T}$ state to an encoded logical $\ket{\bar{T}}$ state on the surface code \cite{Dennis_2002,Fowler_2012}. We explore the state injection procedures \cite{Li_2015,lao_criger_magic} with ZX-calculus \cite{Coecke_2011,vandewetering2020zxcalculusworkingquantumcomputer}, specifically using Pauli webs \cite{Bombin_2024,rodatz2024floquetifyingstabilisercodesdistancepreserving}. This allows a diagrammatic way to understand theses procedures. The ZX-diagrams and Pauli webs provide hints at the importance of the upper and lower triangular located initial qubits in these protocols \cite{Li_2015,lao_criger_magic}. 

We study a simpler injected state, injecting the $\ket{Y}\propto \ket{0}+\text{exp}(i\pi/2)\ket{1}$ state in place of the $\ket{T}$ state. The choice for injecting the $\ket{Y}$ state is motivated by its Clifford nature \cite{Aaronson_2004}, this allows a full end-to-end analysis using Pauli web. In addition, we follow the Pauli web colouring scheme of \cite{rodatz2024floquetifyingstabilisercodesdistancepreserving} (opposite Pauli web colour highlighting compared to \cite{Bombin_2024}). We assume knowledge of ZX-calculus, Pauli web and error correction with the surface code throughout this text. Please refer to \cite{Bombin_2024,rodatz2024floquetifyingstabilisercodesdistancepreserving,Fowler_2012} for detailed discussions. 

The main aim of this brief communication is to show that the Pauli web produced in the Li/Lao--Criger \cite{Li_2015, lao_criger_magic} $\ket{Y}$ state injection schemes does indeed recover the logical $Y$ correlator Pauli web (see detailed discussion in \cite{Bombin_2024}) of the rotated surface code, as expected.

\section{ZX-calculus notation in brief}
\label{sec:zxintro}

ZX-calculus is a diagrammatic tool for representing and
manipulating quantum
systems~\cite{Coecke_2011,vandewetering2020zxcalculusworkingquantumcomputer}.
Figure~\ref{fig:useful_zx_notations} lists the gates, initialisations
and measurements used throughout this manuscript in ZX notation. On
top of Clifford ZX-diagrams, Pauli webs are a graphical notation that
visualises and identifies the checks, stabilisers and logical
correlators of a quantum error correcting
code~\cite{rodatz2024floquetifyingstabilisercodesdistancepreserving,
Bombin_2024}, and they are extremely
useful when verifying complicated Clifford
circuits~\cite{vandewetering2020zxcalculusworkingquantumcomputer}. Time runs from bottom to top in all
ZX-diagrams of this manuscript, except the
\texttt{tsim}~\cite{tsim2026} style ZX-diagrams, which read from
left to right.

\begin{figure}[H]
    \centering
    \resizebox{0.3\linewidth}{!}{
\begin{tikzpicture}
\coordinate [] (c0) at (1,0);
\coordinate [] (c1) at (1,-1);
\coordinate [] (c2) at (-1.5,-0.5);

\node[scale = 1.2] (nyq) at (c2) {\resizebox{.19\linewidth}{!}{%
\begin{yquantgroup}
\registers{
qubit {} q[2];
}
\circuit{
cnot q[1] | q[0];
}
\equals
\end{yquantgroup}
}
};
\node[fill=zx_green,shape=circle,draw=black] (n0) at (c0) {};
\node[fill=zx_red,shape=circle,draw=black] (n1) at (c1) {};

\begin{pgfonlayer}{background}
    \path [-,line width=0.035cm,black,opacity=1] (n0) edge node {} (n1);
    \path [-,line width=0.035cm,black,opacity=1] (0.2,0) edge node {} (1.8,0);
    \path [-,line width=0.035cm,black,opacity=1] (0.2,-1) edge node {} (1.8,-1);
\end{pgfonlayer}
\end{tikzpicture}
}
\quad
\resizebox{0.3\linewidth}{!}{
\begin{tikzpicture}
\coordinate [] (c0) at (1,0);
\coordinate [] (c1) at (1,-1);
\coordinate [] (c2) at (-1.5,-0.5);
\coordinate [] (c3) at (1,-0.5);

\node[scale = 1.2] (nyq) at (c2) {\resizebox{.19\linewidth}{!}{%
\begin{yquantgroup}
\registers{
qubit {} q[2];
}
\circuit{
zz (q[1], q[0]);
}
\equals
\end{yquantgroup}
}
};
\node[fill=zx_green,shape=circle,draw=black] (n0) at (c0) {};
\node[fill=zx_green,shape=circle,draw=black] (n1) at (c1) {};
\node[fill=yellow,shape=rectangle,draw=black] (n3) at (c3) {};
\begin{pgfonlayer}{background}
    \path [-,line width=0.035cm,black,opacity=1] (n0) edge node {} (n1);
    \path [-,line width=0.035cm,black,opacity=1] (0.2,0) edge node {} (1.8,0);
    \path [-,line width=0.035cm,black,opacity=1] (0.2,-1) edge node {} (1.8,-1);
\end{pgfonlayer}
\end{tikzpicture}
}
\quad
\resizebox{0.3\linewidth}{!}{
\begin{tikzpicture}
\coordinate [] (c0) at (1,0);
\coordinate [] (c1) at (1,-1);
\coordinate [] (c2) at (-1.5,-0.5);
\coordinate [] (c3) at (1,-0.5);

\node[scale = 1.2] (nyq) at (c2) {\resizebox{.19\linewidth}{!}{%
\begin{yquantgroup}
\registers{
qubit {} q[1];
}
\circuit{
box {$H$} q[0];
}
\equals
\end{yquantgroup}
}
};
\node[fill=yellow,shape=rectangle,draw=black] (n3) at (c3) {};
\begin{pgfonlayer}{background}
    \path [-,line width=0.035cm,black,opacity=1] (0.2,-0.5) edge node {} (1.8,-0.5);
    \path [-,line width=0.035cm,black,opacity=1] (0.2,-0.5) edge node {} (1.8,-0.5);
\end{pgfonlayer}
\end{tikzpicture}
}
\bigskip
\\
\resizebox{0.3\linewidth}{!}{
\begin{tikzpicture}
\coordinate [] (c0) at (1,0);
\coordinate [] (c1) at (1,-1);
\coordinate [] (c2) at (-1,-0.5);

\node[scale = .8] (nyq) at (c2) {\resizebox{.19\linewidth}{!}{%
\begin{yquantgroup}
\registers{
qubit {} q[3];
}
\circuit{
init {} q[0];
discard q[1];
init {} q[2];
setstyle {very thick} q;
hspace {0.25cm} -;
box {$S$} q[0];
box {$S^{\dagger}$} q[2];
hspace {0.25cm} -;
}
\equals
\end{yquantgroup}
}
};
\node[fill=zx_green,shape=circle,draw=black,scale = 0.6] (n0) at (c0) {$+\frac{\pi}{2}$};
\node[fill=zx_green,shape=circle,draw=black,scale = 0.6] (n1) at (c1) {$-\frac{\pi}{2}$};

\begin{pgfonlayer}{background}
    \path [-,line width=0.035cm,black,opacity=1] (0.2,0) edge node {} (1.8,0);
    \path [-,line width=0.035cm,black,opacity=1] (0.2,-1) edge node {} (1.8,-1);
\end{pgfonlayer}
\end{tikzpicture}
}
\quad
\resizebox{0.3\linewidth}{!}{
\begin{tikzpicture}
\coordinate [] (c0) at (1,0);
\coordinate [] (c1) at (1,-1);
\coordinate [] (c2) at (-0.75,-0.5);

\node[scale = 1.15] (nyq) at (c2) {\resizebox{.19\linewidth}{!}{%
\begin{yquantgroup}
\registers{
qubit {} q[3];
}
\circuit{
init {$\sqrt{2}\ket{+}$} q[0];
discard q[1];
init {$\sqrt{2}\ket{0}$} q[2];
setstyle {very thick} q;
hspace {0.5cm} -;
}
\equals
\end{yquantgroup}
}
};
\node[fill=zx_green,shape=circle,draw=black] (n0) at (c0) {};
\node[fill=zx_red,shape=circle,draw=black] (n1) at (c1) {};

\begin{pgfonlayer}{background}
    \path [-,line width=0.035cm,black,opacity=1] (n0) edge node {} (1.8,0);
    \path [-,line width=0.035cm,black,opacity=1] (n1) edge node {} (1.8,-1);
\end{pgfonlayer}
\end{tikzpicture}
}
\quad
\resizebox{0.3\linewidth}{!}{
\begin{tikzpicture}
\coordinate [] (c0) at (1,0);
\coordinate [] (c1) at (1,-1);
\coordinate [] (c2) at (-0.75,-0.5);

\node[scale = .85] (nyq) at (c2) {\resizebox{.19\linewidth}{!}{%
\begin{yquantgroup}
\registers{
qubit {} q[3];
}
\circuit{
init {} q[0];
discard q[1];
init {} q[2];
setstyle {very thick} q;
hspace {0.25cm} -;
dmeter {$X$} q[0];
dmeter {$Z$} q[2];
}
\end{yquantgroup}
}
};
\node[draw=none,fill=none,scale=1.1] at (0.75,-0.5) {$=$};
\node[fill=zx_green,shape=circle,draw=black,scale=0.8] (n0) at (1.9,0) {\scalebox{0.8}{$m_X\pi$}};
\node[fill=zx_red,shape=circle,draw=black,scale=0.8] (n1) at (1.9,-1) {\scalebox{0.8}{$m_Z\pi$}};
\node[draw=none,fill=none,scale=0.75] at (1.9,-1.85) {$m_X, m_Z \in \{0,1\}$};

\begin{pgfonlayer}{background}
    \path [-,line width=0.035cm,black,opacity=1] (n0) edge node {} (c0);
    \path [-,line width=0.035cm,black,opacity=1] (n1) edge node {} (c1);
\end{pgfonlayer}
\end{tikzpicture}
}
    \caption{Some useful gates (CNOT, CZ, Hadamard, $S$/$S^{\dagger}$
    phase gate), initialisations (initial states: $\ket{+}$ and
    $\ket{0}$) and measurements in the $X$ or $Z$ basis in the
    ZX-calculus notation. A measurement outcome
    $m_X, m_Z \in \{0,1\}$ enters the diagram as an $m\pi$ phase
    on the `measurement' spiders.}
    \label{fig:useful_zx_notations}
\end{figure}

Every spider is a tensor~\cite{wan2026cultivation}. The green
Z-spider with phase $\alpha$, $m$ input legs and $n$ output legs, and
its red X-spider counterpart in the Hadamard-rotated basis are (equation~\eqref{eq:spiders}):
\begin{equation}\label{eq:spiders}
\resizebox{0.82\linewidth}{!}{\begin{tikzpicture}[scale=0.8,every node/.style={minimum size=0.5cm}]
\node[fill=zx_green,shape=circle,draw=black] (z) at (0,0) {$\alpha$};
\draw[black,line width=0.035cm] (z) -- (-0.8,-1.1);
\draw[black,line width=0.035cm] (z) -- (0,-1.25);
\draw[black,line width=0.035cm] (z) -- (0.8,-1.1);
\draw[black,line width=0.035cm] (z) -- (-0.55,1.15);
\draw[black,line width=0.035cm] (z) -- (0.55,1.15);
\node[draw=none,fill=none] at (0,-1.7) {$\underbrace{\hspace{1.9cm}}_{m}$};
\node[draw=none,fill=none] at (0,1.72) {$\overbrace{\hspace{1.4cm}}^{n}$};
\node[draw=none,fill=none,anchor=west,scale=0.78] at (1.35,0)
  {$= \ket{0}^{\otimes n}\!\bra{0}^{\otimes m} +
    e^{i\alpha}\ket{1}^{\otimes n}\!\bra{1}^{\otimes m}$};
\node[fill=zx_red,shape=circle,draw=black] (x) at (8.9,0) {$\alpha$};
\draw[black,line width=0.035cm] (x) -- (8.1,-1.1);
\draw[black,line width=0.035cm] (x) -- (8.9,-1.25);
\draw[black,line width=0.035cm] (x) -- (9.7,-1.1);
\draw[black,line width=0.035cm] (x) -- (8.35,1.15);
\draw[black,line width=0.035cm] (x) -- (9.45,1.15);
\node[draw=none,fill=none] at (8.9,-1.7) {$\underbrace{\hspace{1.9cm}}_{m}$};
\node[draw=none,fill=none] at (8.9,1.72) {$\overbrace{\hspace{1.4cm}}^{n}$};
\node[draw=none,fill=none,anchor=west,scale=0.78] at (10.25,0)
  {$= \ket{+}^{\otimes n}\!\bra{+}^{\otimes m} +
    e^{i\alpha}\ket{-}^{\otimes n}\!\bra{-}^{\otimes m}$};
\end{tikzpicture}} \ .
\end{equation}
Wires contract the shared indices, spiders of the same colour that
share a wire fuse into one spider whose phase is the sum of the two,
and a diagram with no open legs contracts all the way down to a
complex number. Spiders with phases $\{0, \pm\pi/2, \pi\}$ generate Clifford
circuits/ZX-diagrams, whereas spiders with phases
$\{\pi/4, 3\pi/4, 5\pi/4, 7\pi/4\}$ generate non-Clifford
circuits/ZX-diagrams.

A Pauli web~\cite{Bombin_2024,
rodatz2024floquetifyingstabilisercodesdistancepreserving,
wan2026cultivation} decorates the edges
of a Clifford ZX-diagram: a green highlight carries a $Z$-type Pauli operator,
a red highlight an $X$-type one, and an edge highlighted in both
carries $Y \propto XZ$. The decoration must be consistent at every
spider, according to rules fixed by the spider's phase, quoted in
section~\ref{sec:pauliweb_y} below. At the initial states this fixes
which Pauli webs can terminate where: the phase-$0$ green spider $\ket{+}$
admits only $I$- (no Pauli web) or $X$-decorated terminations, and
the red $\ket{0}$ only $I$ or $Z$. A Pauli web that highlights no input or
output leg is closed, and on an error-correction circuit the closed
webs are its detectors (in the \texttt{stim}
language~\cite{gidney2021stim}) that carry syndrome information. A Pauli web with highlighted open legs is open,
and carries the logical correlators, like the logical $Z$ and $X$ webs
of figure~\ref{fig:logical_cor}. Figure~\ref{fig:web_types} shows
the two kinds on a minimal example.

\begin{figure}[H]
    \centering
    \resizebox{0.42\linewidth}{!}{\begin{tikzpicture}[scale=0.9,every node/.style={minimum size=0.42cm}]
\begin{scope}
\coordinate (tl) at (-1,1);  \coordinate (bl) at (-1,-1);
\coordinate (tr) at (3.5,1); \coordinate (br) at (3.5,-1);
\node[fill=zx_green,shape=circle,draw=black] (g1) at (0,1) {};
\node[fill=zx_green,shape=circle,draw=black] (g2) at (0,-1) {};
\node[fill=zx_green,shape=circle,draw=black] (g3) at (2.5,1) {};
\node[fill=zx_green,shape=circle,draw=black] (g4) at (2.5,-1) {};
\node[fill=zx_red,shape=circle,draw=black,scale=0.9] (k1) at (0.5,0)
  {\scalebox{0.68}{$k_1\pi$}};
\node[fill=zx_red,shape=circle,draw=black,scale=0.9] (k2) at (3,0)
  {\scalebox{0.68}{$k_2\pi$}};
\node[fill=zx_red,shape=circle,draw=black,scale=0.9] (e) at (1.25,1)
  {\scalebox{0.68}{$\pi$}};
\begin{pgfonlayer}{background}
\draw[green,line width=0.22cm,opacity=0.8] (g1.center) -- (e.center)
  -- (g3.center);
\draw[green,line width=0.22cm,opacity=0.8] (g2.center) -- (g4.center);
\draw[green,line width=0.22cm,opacity=0.8] (g1.center) -- (k1.center)
  -- (g2.center);
\draw[green,line width=0.22cm,opacity=0.8] (g3.center) -- (k2.center)
  -- (g4.center);
\draw[black,line width=0.035cm] (tl) -- (g1) -- (e) -- (g3) -- (tr);
\draw[black,line width=0.035cm] (bl) -- (g2) -- (g4) -- (br);
\draw[black,line width=0.035cm] (g1) -- (k1) -- (g2);
\draw[black,line width=0.035cm] (g3) -- (k2) -- (g4);
\end{pgfonlayer}
\node[draw=none,fill=none] at (1.25,-1.9) {\footnotesize (a) closed};
\end{scope}
\begin{scope}[shift={(7.2,0)}]
\coordinate (tl) at (-1,1);  \coordinate (bl) at (-1,-1);
\coordinate (tr) at (2.6,1); \coordinate (br) at (2.6,-1);
\node[fill=zx_green,shape=circle,draw=black] (g1) at (0,1) {};
\node[fill=zx_green,shape=circle,draw=black] (g2) at (0,-1) {};
\node[fill=zx_red,shape=circle,draw=black,scale=0.9] (k1) at (0.5,0)
  {\scalebox{0.68}{$k_1\pi$}};
\node[fill=zx_red,shape=circle,draw=black,scale=0.9] (e) at (1.4,1)
  {\scalebox{0.68}{$\pi$}};
\begin{pgfonlayer}{background}
\draw[green,line width=0.22cm,opacity=0.8] (g1.center) -- (e.center)
  -- (tr);
\draw[green,line width=0.22cm,opacity=0.8] (g2.center) -- (br);
\draw[green,line width=0.22cm,opacity=0.8] (g1.center) -- (k1.center)
  -- (g2.center);
\draw[black,line width=0.035cm] (tl) -- (g1) -- (e) -- (tr);
\draw[black,line width=0.035cm] (bl) -- (g2) -- (br);
\draw[black,line width=0.035cm] (g1) -- (k1) -- (g2);
\end{pgfonlayer}
\node[draw=none,fill=none] at (0.8,-1.9) {\footnotesize (b) open};
\end{scope}
\end{tikzpicture}}
    \caption{Closed and open Pauli webs. (a) Closed: two
    $ZZ$ parity measurements (red $k_i\pi$ `measurement' spiders) on
    the same pair of qubits. The green ($Z$-type) web forms a loop
    that reaches no open leg, so the product $k_1 \oplus k_2$ is
    deterministic (a parity check), and the $\pi$ X-spider, an $X$
    error sitting on the Pauli web, is exactly what flips it. (b) Open:
    a single parity measurement. The Pauli web reaches the output legs and
    carries the stabiliser $Z \otimes Z$ of the resultant
    state.}
    \label{fig:web_types}
\end{figure}

\section{Li/Lao--Criger state injection schemes}
Li constructed a scheme to inject states onto the regular un-rotated surface code with high fidelity \cite{Li_2015}. This was further generalised to the rotated surface code by Lao and Criger \cite{lao_criger_magic}. We will work exclusively with the Lao--Criger scheme with an injected $\ket{Y}$ state physical qubit on the corner of a rotated surface code. In figure \ref{fig:inj_1}, we write the initial state proposed by \cite{lao_criger_magic} in the ZX-calculus notation. Time goes from bottom to top in all ZX-diagrams, except the \texttt{tsim}~\cite{tsim2026} style ZX-diagrams, which read from left to right. For illustration purposes, a distance $d=5$ surface code will be used here.
\subsection{Initialisation}
\begin{figure}[!h]
    \centering     
    \tdplotsetmaincoords{70}{23}
    \resizebox{0.4\linewidth}{!}{
   
    \begin{tikzpicture}[tdplot_main_coords,every node/.style={minimum size=1cm}]
    \def\hei{0}
    \def\ttl{t0}
    \begin{scope}
        
    \node[fill=zx_green,shape=circle,draw=black] (n4\ttl) at (2,18,\hei) {$\pi/2$};
    \node[fill=zx_green,shape=circle,draw=black] (n3\ttl) at (2,14,\hei) {};
    \node[fill=zx_green,shape=circle,draw=black] (n2\ttl) at (2,10,\hei) {};
    \node[fill=zx_green,shape=circle,draw=black] (n1\ttl) at (2,6,\hei) {};
    \node[fill=zx_green,shape=circle,draw=black] (n0\ttl) at (2,2,\hei) {};

    \node[fill=zx_red,shape=circle,draw=black] (n9\ttl) at (6,18,\hei) {};
    \node[fill=zx_green,shape=circle,draw=black] (n8\ttl) at (6,14,\hei) {};
    \node[fill=zx_green,shape=circle,draw=black] (n7\ttl) at (6,10,\hei) {};
    \node[fill=zx_green,shape=circle,draw=black] (n6\ttl) at (6,6,\hei) {};
    \node[fill=zx_green,shape=circle,draw=black] (n5\ttl) at (6,2,\hei) {}; 

    \node[fill=zx_red,shape=circle,draw=black] (n14\ttl) at (10,18,\hei) {};
    \node[fill=zx_red,shape=circle,draw=black] (n13\ttl) at (10,14,\hei) {};
    \node[fill=zx_green,shape=circle,draw=black] (n12\ttl) at (10,10,\hei) {};
    \node[fill=zx_green,shape=circle,draw=black] (n11\ttl) at (10,6,\hei) {};
    \node[fill=zx_green,shape=circle,draw=black] (n10\ttl) at (10,2,\hei) {};

    \node[fill=zx_red,shape=circle,draw=black] (n19\ttl) at (14,18,\hei) {};
    \node[fill=zx_red,shape=circle,draw=black] (n18\ttl) at (14,14,\hei) {};
    \node[fill=zx_red,shape=circle,draw=black] (n17\ttl) at (14,10,\hei) {};
    \node[fill=zx_green,shape=circle,draw=black] (n16\ttl) at (14,6,\hei) {};
    \node[fill=zx_green,shape=circle,draw=black] (n15\ttl) at (14,2,\hei) {}; 

    \node[fill=zx_red,shape=circle,draw=black] (n24\ttl) at (18,18,\hei) {};
    \node[fill=zx_red,shape=circle,draw=black] (n23\ttl) at (18,14,\hei) {};
    \node[fill=zx_red,shape=circle,draw=black] (n22\ttl) at (18,10,\hei) {};
    \node[fill=zx_red,shape=circle,draw=black] (n21\ttl) at (18,6,\hei) {};
    \node[fill=zx_green,shape=circle,draw=black] (n20\ttl) at (18,2,\hei) {};
    
    \end{scope}

    \end{tikzpicture}
    }
    \caption{\label{fig:inj_1}The initial state to the injection schemes \cite{Li_2015,lao_criger_magic} written as a ZX-diagram. Upper left corner qubit initialised in the $\ket{Y}\propto \ket{0}+i\ket{1}$ state ($\frac{\pi}{2}$ phase green Z-spider). The lower triangular and diagonally located qubits in the $\ket{+}$ state and the upper triangular located qubits in the $\ket{0}$ state. }
\end{figure}
For a distance $d$ surface code, a square grid of $d \times d$ physical (data) qubits need to be initialised. A bare physical $\ket{Y}$ state is initialised in the top left corner. The rest of the $d^2-1$ data qubits will be initialised in $\ket{+}$ state if they are located in the the lower triangular and main diagonal, and in the $\ket{0}$ state if they sit along the upper triangular locations.  

\subsection{Encoder circuit and logical correlators}
After the state initialisation, the surface code encoding circuit is applied to the state. This encoding circuit following the notation of \cite{Bombin_2024} is shown in figure \ref{fig:inj_encoder}. This constitutes a full round of parity measurement. Measuring all the $XXXX$/$XX$ plaquettes parity measurements in the first layer followed by the $ZZZZ$/$ZZ$ equivalents in the second layer.

\begin{figure}[!h]
    \centering     
    \tdplotsetmaincoords{70}{23}
    \resizebox{0.4\linewidth}{!}{
   
    \begin{tikzpicture}[tdplot_main_coords,every node/.style={minimum size=1cm}]
     
    \def\hei{7.5}
    \def\ttl{t0}
    \begin{scope}
        
    \node[fill=none,shape=circle,draw=none] (n4\ttl) at (2,18,\hei) {};
    \node[fill=none,shape=circle,draw=none]  (n3\ttl) at (2,14,\hei) {};
    \node[fill=none,shape=circle,draw=none]  (n2\ttl) at (2,10,\hei) {};
    \node[fill=none,shape=circle,draw=none]  (n1\ttl) at (2,6,\hei) {};
   \node[fill=none,shape=circle,draw=none]  (n0\ttl) at (2,2,\hei) {};

   \node[fill=none,shape=circle,draw=none] (n9\ttl) at (6,18,\hei) {};
    \node[fill=none,shape=circle,draw=none]  (n8\ttl) at (6,14,\hei) {};
    \node[fill=none,shape=circle,draw=none]  (n7\ttl) at (6,10,\hei) {};
    \node[fill=none,shape=circle,draw=none]  (n6\ttl) at (6,6,\hei) {};
    \node[fill=none,shape=circle,draw=none] (n5\ttl) at (6,2,\hei) {}; 

   \node[fill=none,shape=circle,draw=none]  (n14\ttl) at (10,18,\hei) {};
   \node[fill=none,shape=circle,draw=none] (n13\ttl) at (10,14,\hei) {};
   \node[fill=none,shape=circle,draw=none]  (n12\ttl) at (10,10,\hei) {};
    \node[fill=none,shape=circle,draw=none]  (n11\ttl) at (10,6,\hei) {};
    \node[fill=none,shape=circle,draw=none]  (n10\ttl) at (10,2,\hei) {};

    \node[fill=none,shape=circle,draw=none] (n19\ttl) at (14,18,\hei) {};
    \node[fill=none,shape=circle,draw=none] (n18\ttl) at (14,14,\hei) {};
    \node[fill=none,shape=circle,draw=none] (n17\ttl) at (14,10,\hei) {};
    \node[fill=none,shape=circle,draw=none]  (n16\ttl) at (14,6,\hei) {};
    \node[fill=none,shape=circle,draw=none]  (n15\ttl) at (14,2,\hei) {}; 

    \node[fill=none,shape=circle,draw=none]  (n24\ttl) at (18,18,\hei) {};
    \node[fill=none,shape=circle,draw=none]  (n23\ttl) at (18,14,\hei) {};
    \node[fill=none,shape=circle,draw=none]  (n22\ttl) at (18,10,\hei) {};
    \node[fill=none,shape=circle,draw=none] (n21\ttl) at (18,6,\hei) {};
    \node[fill=none,shape=circle,draw=none]  (n20\ttl) at (18,2,\hei) {};
    
    \end{scope}

    \def\hei{10}
    \def\ttl{t1}
    \begin{scope}
        
    \node[fill=zx_red,shape=circle,draw=black] (n4\ttl) at (2,18,\hei) {};

    \node[fill=zx_red,shape=circle,draw=black] (n3\ttl) at (2,14,\hei) {};
    \node[fill=zx_red,shape=circle,draw=black] (n2\ttl) at (2,10,\hei) {};
    \node[fill=zx_red,shape=circle,draw=black] (n1\ttl) at (2,6,\hei) {};
    \node[fill=zx_red,shape=circle,draw=black] (n0\ttl) at (2,2,\hei) {};

    \node[fill=zx_red,shape=circle,draw=black] (n9\ttl) at (6,18,\hei) {};
    \node[fill=zx_red,shape=circle,draw=black] (n8\ttl) at (6,14,\hei) {};
    \node[fill=zx_red,shape=circle,draw=black] (n7\ttl) at (6,10,\hei) {};
    \node[fill=zx_red,shape=circle,draw=black] (n6\ttl) at (6,6,\hei) {};
    \node[fill=zx_red,shape=circle,draw=black] (n5\ttl) at (6,2,\hei) {}; 

    \node[fill=zx_red,shape=circle,draw=black] (n14\ttl) at (10,18,\hei) {};
    \node[fill=zx_red,shape=circle,draw=black] (n13\ttl) at (10,14,\hei) {};
    \node[fill=zx_red,shape=circle,draw=black] (n12\ttl) at (10,10,\hei) {};
    \node[fill=zx_red,shape=circle,draw=black] (n11\ttl) at (10,6,\hei) {};
    \node[fill=zx_red,shape=circle,draw=black] (n10\ttl) at (10,2,\hei) {};

    \node[fill=zx_red,shape=circle,draw=black] (n19\ttl) at (14,18,\hei) {};
    \node[fill=zx_red,shape=circle,draw=black] (n18\ttl) at (14,14,\hei) {};
    \node[fill=zx_red,shape=circle,draw=black] (n17\ttl) at (14,10,\hei) {};
    \node[fill=zx_red,shape=circle,draw=black] (n16\ttl) at (14,6,\hei) {};
    \node[fill=zx_red,shape=circle,draw=black] (n15\ttl) at (14,2,\hei) {}; 

    \node[fill=zx_red,shape=circle,draw=black] (n24\ttl) at (18,18,\hei) {};
    \node[fill=zx_red,shape=circle,draw=black] (n23\ttl) at (18,14,\hei) {};
    \node[fill=zx_red,shape=circle,draw=black] (n22\ttl) at (18,10,\hei) {};
    \node[fill=zx_red,shape=circle,draw=black] (n21\ttl) at (18,6,\hei) {};
    \node[fill=zx_red,shape=circle,draw=black] (n20\ttl) at (18,2,\hei) {};
    
    \end{scope}
    \begin{scope}
        \node[fill=zx_green,circle,draw=black] (b0\ttl) at (4,4,\hei) {};
        \node[fill=zx_green,circle,draw=black] (b1\ttl) at (4,12,\hei) {};
        \node[fill=zx_green,circle,draw=black] (b2\ttl) at (4,20,\hei) {};

        \node[fill=zx_green,circle,draw=black] (b3\ttl) at (8,0,\hei) {};
        \node[fill=zx_green,circle,draw=black] (b4\ttl) at (8,8,\hei) {};
        \node[fill=zx_green,circle,draw=black] (b5\ttl) at (8,16,\hei) {};

        \node[fill=zx_green,circle,draw=black] (b6\ttl) at (12,4,\hei) {};
        \node[fill=zx_green,circle,draw=black] (b7\ttl) at (12,12,\hei) {};
        \node[fill=zx_green,circle,draw=black] (b8\ttl) at (12,20,\hei) {};

        \node[fill=zx_green,circle,draw=black] (b9\ttl) at (16,0,\hei) {};
        \node[fill=zx_green,circle,draw=black] (b10\ttl) at (16,8,\hei) {};
        \node[fill=zx_green,circle,draw=black] (b11\ttl) at (16,16,\hei) {};

    \end{scope}
    \begin{scope}
        
    \draw[black,fill=black,opacity=1,line width = 0.05cm] (b0\ttl) to (n0\ttl);
    \draw[black,fill=black,opacity=1,line width = 0.05cm] (b0\ttl) to (n1\ttl);
    \draw[black,fill=black,opacity=1,line width = 0.05cm] (b0\ttl) to (n5\ttl);
    \draw[black,fill=black,opacity=1,line width = 0.05cm] (b0\ttl) to (n6\ttl);

    \draw[black,fill=black,opacity=1,line width = 0.05cm] (b1\ttl) to (n2\ttl);
    \draw[black,fill=black,opacity=1,line width = 0.05cm] (b1\ttl) to (n3\ttl);
    \draw[black,fill=black,opacity=1,line width = 0.05cm] (b1\ttl) to (n7\ttl);
    \draw[black,fill=black,opacity=1,line width = 0.05cm] (b1\ttl) to (n8\ttl);
    
    \draw[black,fill=black,opacity=1,line width = 0.05cm] (b2\ttl) to (n4\ttl);
    \draw[black,fill=black,opacity=1,line width = 0.05cm] (b2\ttl) to (n9\ttl);
    
    \draw[black,fill=black,opacity=1,line width = 0.05cm] (b3\ttl) to (n5\ttl);
    \draw[black,fill=black,opacity=1,line width = 0.05cm] (b3\ttl) to (n10\ttl);

    \draw[black,fill=black,opacity=1,line width = 0.05cm] (b4\ttl) to (n6\ttl);
    \draw[black,fill=black,opacity=1,line width = 0.05cm] (b4\ttl) to (n7\ttl);
    \draw[black,fill=black,opacity=1,line width = 0.05cm] (b4\ttl) to (n11\ttl);
    \draw[black,fill=black,opacity=1,line width = 0.05cm] (b4\ttl) to (n12\ttl);

    \draw[black,fill=black,opacity=1,line width = 0.05cm] (b5\ttl) to (n8\ttl);
    \draw[black,fill=black,opacity=1,line width = 0.05cm] (b5\ttl) to (n9\ttl);
    \draw[black,fill=black,opacity=1,line width = 0.05cm] (b5\ttl) to (n13\ttl);
    \draw[black,fill=black,opacity=1,line width = 0.05cm] (b5\ttl) to (n14\ttl);

    \draw[black,fill=black,opacity=1,line width = 0.05cm] (b6\ttl) to (n10\ttl);
    \draw[black,fill=black,opacity=1,line width = 0.05cm] (b6\ttl) to (n11\ttl);
    \draw[black,fill=black,opacity=1,line width = 0.05cm] (b6\ttl) to (n15\ttl);
    \draw[black,fill=black,opacity=1,line width = 0.05cm] (b6\ttl) to (n16\ttl);

    \draw[black,fill=black,opacity=1,line width = 0.05cm] (b7\ttl) to (n12\ttl);
    \draw[black,fill=black,opacity=1,line width = 0.05cm] (b7\ttl) to (n13\ttl);
    \draw[black,fill=black,opacity=1,line width = 0.05cm] (b7\ttl) to (n17\ttl);
    \draw[black,fill=black,opacity=1,line width = 0.05cm] (b7\ttl) to (n18\ttl);
    
    \draw[black,fill=black,opacity=1,line width = 0.05cm] (b8\ttl) to (n14\ttl);
    \draw[black,fill=black,opacity=1,line width = 0.05cm] (b8\ttl) to (n19\ttl);

    \draw[black,fill=black,opacity=1,line width = 0.05cm] (b9\ttl) to (n15\ttl);
    \draw[black,fill=black,opacity=1,line width = 0.05cm] (b9\ttl) to (n20\ttl);

    \draw[black,fill=black,opacity=1,line width = 0.05cm] (b10\ttl) to (n16\ttl);
    \draw[black,fill=black,opacity=1,line width = 0.05cm] (b10\ttl) to (n17\ttl);
    \draw[black,fill=black,opacity=1,line width = 0.05cm] (b10\ttl) to (n21\ttl);
    \draw[black,fill=black,opacity=1,line width = 0.05cm] (b10\ttl) to (n22\ttl);

    \draw[black,fill=black,opacity=1,line width = 0.05cm] (b11\ttl) to (n18\ttl);
    \draw[black,fill=black,opacity=1,line width = 0.05cm] (b11\ttl) to (n19\ttl);
    \draw[black,fill=black,opacity=1,line width = 0.05cm] (b11\ttl) to (n23\ttl);
    \draw[black,fill=black,opacity=1,line width = 0.05cm] (b11\ttl) to (n24\ttl);

    \end{scope}

    \def\hei{20}
    \def\ttl{t2}
    \begin{scope}
        
    \node[fill=zx_green,shape=circle,draw=black] (n4\ttl) at (2,18,\hei) {};

    \node[fill=zx_green,shape=circle,draw=black] (n3\ttl) at (2,14,\hei) {};
    \node[fill=zx_green,shape=circle,draw=black] (n2\ttl) at (2,10,\hei) {};
    \node[fill=zx_green,shape=circle,draw=black] (n1\ttl) at (2,6,\hei) {};
    \node[fill=zx_green,shape=circle,draw=black] (n0\ttl) at (2,2,\hei) {};

    \node[fill=zx_green,shape=circle,draw=black] (n9\ttl) at (6,18,\hei) {};
    \node[fill=zx_green,shape=circle,draw=black] (n8\ttl) at (6,14,\hei) {};
    \node[fill=zx_green,shape=circle,draw=black] (n7\ttl) at (6,10,\hei) {};
    \node[fill=zx_green,shape=circle,draw=black] (n6\ttl) at (6,6,\hei) {};
    \node[fill=zx_green,shape=circle,draw=black] (n5\ttl) at (6,2,\hei) {}; 

    \node[fill=zx_green,shape=circle,draw=black] (n14\ttl) at (10,18,\hei) {};
    \node[fill=zx_green,shape=circle,draw=black] (n13\ttl) at (10,14,\hei) {};
    \node[fill=zx_green,shape=circle,draw=black] (n12\ttl) at (10,10,\hei) {};
    \node[fill=zx_green,shape=circle,draw=black] (n11\ttl) at (10,6,\hei) {};
    \node[fill=zx_green,shape=circle,draw=black] (n10\ttl) at (10,2,\hei) {};

    \node[fill=zx_green,shape=circle,draw=black] (n19\ttl) at (14,18,\hei) {};
    \node[fill=zx_green,shape=circle,draw=black] (n18\ttl) at (14,14,\hei) {};
    \node[fill=zx_green,shape=circle,draw=black] (n17\ttl) at (14,10,\hei) {};
    \node[fill=zx_green,shape=circle,draw=black] (n16\ttl) at (14,6,\hei) {};
    \node[fill=zx_green,shape=circle,draw=black] (n15\ttl) at (14,2,\hei) {}; 

    \node[fill=zx_green,shape=circle,draw=black] (n24\ttl) at (18,18,\hei) {};
    \node[fill=zx_green,shape=circle,draw=black] (n23\ttl) at (18,14,\hei) {};
    \node[fill=zx_green,shape=circle,draw=black] (n22\ttl) at (18,10,\hei) {};
    \node[fill=zx_green,shape=circle,draw=black] (n21\ttl) at (18,6,\hei) {};
    \node[fill=zx_green,shape=circle,draw=black] (n20\ttl) at (18,2,\hei) {};
    
    \end{scope}
    \begin{scope}

        \node[fill=zx_red,circle,draw=black] (a0\ttl) at (0,4,\hei) {};
        \node[fill=zx_red,circle,draw=black] (a1\ttl) at (0,12,\hei) {};

        \node[fill=zx_red,circle,draw=black] (a2\ttl) at (4,8,\hei) {};
        \node[fill=zx_red,circle,draw=black] (a3\ttl) at (4,16,\hei) {};

        \node[fill=zx_red,circle,draw=black] (a4\ttl) at (8,4,\hei) {};
        \node[fill=zx_red,circle,draw=black] (a5\ttl) at (8,12,\hei) {};

        \node[fill=zx_red,circle,draw=black] (a6\ttl) at (12,8,\hei) {};
        \node[fill=zx_red,circle,draw=black] (a7\ttl) at (12,16,\hei) {};

        \node[fill=zx_red,circle,draw=black] (a8\ttl) at (16,4,\hei) {};
        \node[fill=zx_red,circle,draw=black] (a9\ttl) at (16,12,\hei) {};

        \node[fill=zx_red,circle,draw=black] (a10\ttl) at (20,8,\hei) {};
        \node[fill=zx_red,circle,draw=black] (a11\ttl) at (20,16,\hei) {};
    \end{scope}
    \begin{scope}
        
    \draw[black,fill=black,opacity=1,line width = 0.05cm] (a0\ttl) to (n0\ttl);
    \draw[black,fill=black,opacity=1,line width = 0.05cm] (a0\ttl) to (n1\ttl);

    \draw[black,fill=black,opacity=1,line width = 0.05cm] (a1\ttl) to (n2\ttl);
    \draw[black,fill=black,opacity=1,line width = 0.05cm] (a1\ttl) to (n3\ttl);

    \draw[black,fill=black,opacity=1,line width = 0.05cm] (a2\ttl) to (n1\ttl);
    \draw[black,fill=black,opacity=1,line width = 0.05cm] (a2\ttl) to (n2\ttl);
    \draw[black,fill=black,opacity=1,line width = 0.05cm] (a2\ttl) to (n6\ttl);
    \draw[black,fill=black,opacity=1,line width = 0.05cm] (a2\ttl) to (n7\ttl);

    \draw[black,fill=black,opacity=1,line width = 0.05cm] (a3\ttl) to (n3\ttl);
    \draw[black,fill=black,opacity=1,line width = 0.05cm] (a3\ttl) to (n4\ttl);
    \draw[black,fill=black,opacity=1,line width = 0.05cm] (a3\ttl) to (n8\ttl);
    \draw[black,fill=black,opacity=1,line width = 0.05cm] (a3\ttl) to (n9\ttl);

    \draw[black,fill=black,opacity=1,line width = 0.05cm] (a4\ttl) to (n5\ttl);
    \draw[black,fill=black,opacity=1,line width = 0.05cm] (a4\ttl) to (n6\ttl);
    \draw[black,fill=black,opacity=1,line width = 0.05cm] (a4\ttl) to (n10\ttl);
    \draw[black,fill=black,opacity=1,line width = 0.05cm] (a4\ttl) to (n11\ttl);
    
    \draw[black,fill=black,opacity=1,line width = 0.05cm] (a5\ttl) to (n7\ttl);
    \draw[black,fill=black,opacity=1,line width = 0.05cm] (a5\ttl) to (n8\ttl);
    \draw[black,fill=black,opacity=1,line width = 0.05cm] (a5\ttl) to (n12\ttl);
    \draw[black,fill=black,opacity=1,line width = 0.05cm] (a5\ttl) to (n13\ttl);

    \draw[black,fill=black,opacity=1,line width = 0.05cm] (a6\ttl) to (n11\ttl);
    \draw[black,fill=black,opacity=1,line width = 0.05cm] (a6\ttl) to (n12\ttl);
    \draw[black,fill=black,opacity=1,line width = 0.05cm] (a6\ttl) to (n16\ttl);
    \draw[black,fill=black,opacity=1,line width = 0.05cm] (a6\ttl) to (n17\ttl);

    \draw[black,fill=black,opacity=1,line width = 0.05cm] (a7\ttl) to (n13\ttl);
    \draw[black,fill=black,opacity=1,line width = 0.05cm] (a7\ttl) to (n14\ttl);
    \draw[black,fill=black,opacity=1,line width = 0.05cm] (a7\ttl) to (n18\ttl);
    \draw[black,fill=black,opacity=1,line width = 0.05cm] (a7\ttl) to (n19\ttl);

    \draw[black,fill=black,opacity=1,line width = 0.05cm] (a8\ttl) to (n15\ttl);
    \draw[black,fill=black,opacity=1,line width = 0.05cm] (a8\ttl) to (n16\ttl);
    \draw[black,fill=black,opacity=1,line width = 0.05cm] (a8\ttl) to (n20\ttl);
    \draw[black,fill=black,opacity=1,line width = 0.05cm] (a8\ttl) to (n21\ttl);

    \draw[black,fill=black,opacity=1,line width = 0.05cm] (a9\ttl) to (n17\ttl);
    \draw[black,fill=black,opacity=1,line width = 0.05cm] (a9\ttl) to (n18\ttl);
    \draw[black,fill=black,opacity=1,line width = 0.05cm] (a9\ttl) to (n22\ttl);
    \draw[black,fill=black,opacity=1,line width = 0.05cm] (a9\ttl) to (n23\ttl);
    
    \draw[black,fill=black,opacity=1,line width = 0.05cm] (a10\ttl) to (n21\ttl);
    \draw[black,fill=black,opacity=1,line width = 0.05cm] (a10\ttl) to (n22\ttl);

    \draw[black,fill=black,opacity=1,line width = 0.05cm] (a11\ttl) to (n23\ttl);
    \draw[black,fill=black,opacity=1,line width = 0.05cm] (a11\ttl) to (n24\ttl);

    \end{scope}

    \def\hei{22.5}
    \def\ttl{t3}
    \begin{scope}
        
    \node[fill=none,shape=circle,draw=none] (n4\ttl) at (2,18,\hei) {};

    \node[fill=none,shape=circle,draw=none] (n3\ttl) at (2,14,\hei) {};
    \node[fill=none,shape=circle,draw=none] (n2\ttl) at (2,10,\hei) {};
    \node[fill=none,shape=circle,draw=none] (n1\ttl) at (2,6,\hei) {};
    \node[fill=none,shape=circle,draw=none] (n0\ttl) at (2,2,\hei) {};

    \node[fill=none,shape=circle,draw=none] (n9\ttl) at (6,18,\hei) {};
    \node[fill=none,shape=circle,draw=none] (n8\ttl) at (6,14,\hei) {};
    \node[fill=none,shape=circle,draw=none] (n7\ttl) at (6,10,\hei) {};
    \node[fill=none,shape=circle,draw=none] (n6\ttl) at (6,6,\hei) {};
    \node[fill=none,shape=circle,draw=none] (n5\ttl) at (6,2,\hei) {}; 

    \node[fill=none,shape=circle,draw=none] (n14\ttl) at (10,18,\hei) {};
    \node[fill=none,shape=circle,draw=none] (n13\ttl) at (10,14,\hei) {};
    \node[fill=none,shape=circle,draw=none] (n12\ttl) at (10,10,\hei) {};
    \node[fill=none,shape=circle,draw=none] (n11\ttl) at (10,6,\hei) {};
    \node[fill=none,shape=circle,draw=none] (n10\ttl) at (10,2,\hei) {};

    \node[fill=none,shape=circle,draw=none] (n19\ttl) at (14,18,\hei) {};
    \node[fill=none,shape=circle,draw=none] (n18\ttl) at (14,14,\hei) {};
    \node[fill=none,shape=circle,draw=none] (n17\ttl) at (14,10,\hei) {};
    \node[fill=none,shape=circle,draw=none] (n16\ttl) at (14,6,\hei) {};
    \node[fill=none,shape=circle,draw=none] (n15\ttl) at (14,2,\hei) {}; 

    \node[fill=none,shape=circle,draw=none] (n24\ttl) at (18,18,\hei) {};
    \node[fill=none,shape=circle,draw=none] (n23\ttl) at (18,14,\hei) {};
    \node[fill=none,shape=circle,draw=none] (n22\ttl) at (18,10,\hei) {};
    \node[fill=none,shape=circle,draw=none] (n21\ttl) at (18,6,\hei) {};
    \node[fill=none,shape=circle,draw=none] (n20\ttl) at (18,2,\hei) {};
    
    \end{scope}

    \def\tlow{t0}
    \def\thig{t1}
    \begin{pgfonlayer}{background}
    \foreach \x in {0,...,24}
    {
    \draw[black,fill=black,opacity=1,line width = 0.05cm] (n\x\tlow) to (n\x\thig);
    }
    \end{pgfonlayer} 

    \def\tlow{t1}
    \def\thig{t2}
    \begin{pgfonlayer}{background}
    \foreach \x in {0,...,24}
    {
    \draw[black,fill=black,opacity=1,line width = 0.05cm] (n\x\tlow) to (n\x\thig);
    }
    \end{pgfonlayer}

    \def\tlow{t2}
    \def\thig{t3}
    \begin{pgfonlayer}{background}
    \foreach \x in {0,...,24}
    {
    \draw[black,fill=black,opacity=1,line width = 0.05cm] (n\x\tlow) to (n\x\thig);
    }
    \end{pgfonlayer}

    \def\tlow{t1}

    \def\tlow{t2}

    \end{tikzpicture}
    }
    \caption{\label{fig:inj_encoder}An error-free (distance $d=5$) rotated surface code encoder circuit. This consists of X-type followed by Z-type plaquette parity measurements (for detailed discussion, see \cite{Fowler_2012,Bombin_2024}).}
\end{figure}

\begin{figure}[!h]
    \centering     
    \tdplotsetmaincoords{70}{23}

    \subfloat[][\label{fig:logical_cor_Z}Logical $Z$ correlator.]{
    \resizebox{0.4\linewidth}{!}{
   
    \begin{tikzpicture}[tdplot_main_coords,every node/.style={minimum size=1cm}]
     
    \def\hei{7.5}
    \def\ttl{t0}
    \begin{scope}
        
    \node[fill=none,shape=circle,draw=none] (n4\ttl) at (2,18,\hei) {};
    \node[fill=none,shape=circle,draw=none]  (n3\ttl) at (2,14,\hei) {};
    \node[fill=none,shape=circle,draw=none]  (n2\ttl) at (2,10,\hei) {};
    \node[fill=none,shape=circle,draw=none]  (n1\ttl) at (2,6,\hei) {};
   \node[fill=none,shape=circle,draw=none]  (n0\ttl) at (2,2,\hei) {};

   \node[fill=none,shape=circle,draw=none] (n9\ttl) at (6,18,\hei) {};
    \node[fill=none,shape=circle,draw=none]  (n8\ttl) at (6,14,\hei) {};
    \node[fill=none,shape=circle,draw=none]  (n7\ttl) at (6,10,\hei) {};
    \node[fill=none,shape=circle,draw=none]  (n6\ttl) at (6,6,\hei) {};
    \node[fill=none,shape=circle,draw=none] (n5\ttl) at (6,2,\hei) {}; 

   \node[fill=none,shape=circle,draw=none]  (n14\ttl) at (10,18,\hei) {};
   \node[fill=none,shape=circle,draw=none] (n13\ttl) at (10,14,\hei) {};
   \node[fill=none,shape=circle,draw=none]  (n12\ttl) at (10,10,\hei) {};
    \node[fill=none,shape=circle,draw=none]  (n11\ttl) at (10,6,\hei) {};
    \node[fill=none,shape=circle,draw=none]  (n10\ttl) at (10,2,\hei) {};

    \node[fill=none,shape=circle,draw=none] (n19\ttl) at (14,18,\hei) {};
    \node[fill=none,shape=circle,draw=none] (n18\ttl) at (14,14,\hei) {};
    \node[fill=none,shape=circle,draw=none] (n17\ttl) at (14,10,\hei) {};
    \node[fill=none,shape=circle,draw=none]  (n16\ttl) at (14,6,\hei) {};
    \node[fill=none,shape=circle,draw=none]  (n15\ttl) at (14,2,\hei) {}; 

    \node[fill=none,shape=circle,draw=none]  (n24\ttl) at (18,18,\hei) {};
    \node[fill=none,shape=circle,draw=none]  (n23\ttl) at (18,14,\hei) {};
    \node[fill=none,shape=circle,draw=none]  (n22\ttl) at (18,10,\hei) {};
    \node[fill=none,shape=circle,draw=none] (n21\ttl) at (18,6,\hei) {};
    \node[fill=none,shape=circle,draw=none]  (n20\ttl) at (18,2,\hei) {};
    
    \end{scope}

    \def\hei{10}
    \def\ttl{t1}
    \begin{scope}
        
    \node[fill=zx_red,shape=circle,draw=black] (n4\ttl) at (2,18,\hei) {};

    \node[fill=zx_red,shape=circle,draw=black] (n3\ttl) at (2,14,\hei) {};
    \node[fill=zx_red,shape=circle,draw=black] (n2\ttl) at (2,10,\hei) {};
    \node[fill=zx_red,shape=circle,draw=black] (n1\ttl) at (2,6,\hei) {};
    \node[fill=zx_red,shape=circle,draw=black] (n0\ttl) at (2,2,\hei) {};

    \node[fill=zx_red,shape=circle,draw=black] (n9\ttl) at (6,18,\hei) {};
    \node[fill=zx_red,shape=circle,draw=black] (n8\ttl) at (6,14,\hei) {};
    \node[fill=zx_red,shape=circle,draw=black] (n7\ttl) at (6,10,\hei) {};
    \node[fill=zx_red,shape=circle,draw=black] (n6\ttl) at (6,6,\hei) {};
    \node[fill=zx_red,shape=circle,draw=black] (n5\ttl) at (6,2,\hei) {}; 

    \node[fill=zx_red,shape=circle,draw=black] (n14\ttl) at (10,18,\hei) {};
    \node[fill=zx_red,shape=circle,draw=black] (n13\ttl) at (10,14,\hei) {};
    \node[fill=zx_red,shape=circle,draw=black] (n12\ttl) at (10,10,\hei) {};
    \node[fill=zx_red,shape=circle,draw=black] (n11\ttl) at (10,6,\hei) {};
    \node[fill=zx_red,shape=circle,draw=black] (n10\ttl) at (10,2,\hei) {};

    \node[fill=zx_red,shape=circle,draw=black] (n19\ttl) at (14,18,\hei) {};
    \node[fill=zx_red,shape=circle,draw=black] (n18\ttl) at (14,14,\hei) {};
    \node[fill=zx_red,shape=circle,draw=black] (n17\ttl) at (14,10,\hei) {};
    \node[fill=zx_red,shape=circle,draw=black] (n16\ttl) at (14,6,\hei) {};
    \node[fill=zx_red,shape=circle,draw=black] (n15\ttl) at (14,2,\hei) {}; 

    \node[fill=zx_red,shape=circle,draw=black] (n24\ttl) at (18,18,\hei) {};
    \node[fill=zx_red,shape=circle,draw=black] (n23\ttl) at (18,14,\hei) {};
    \node[fill=zx_red,shape=circle,draw=black] (n22\ttl) at (18,10,\hei) {};
    \node[fill=zx_red,shape=circle,draw=black] (n21\ttl) at (18,6,\hei) {};
    \node[fill=zx_red,shape=circle,draw=black] (n20\ttl) at (18,2,\hei) {};
    
    \end{scope}
    \begin{scope}
        \node[fill=zx_green,circle,draw=black] (b0\ttl) at (4,4,\hei) {};
        \node[fill=zx_green,circle,draw=black] (b1\ttl) at (4,12,\hei) {};
        \node[fill=zx_green,circle,draw=black] (b2\ttl) at (4,20,\hei) {};

        \node[fill=zx_green,circle,draw=black] (b3\ttl) at (8,0,\hei) {};
        \node[fill=zx_green,circle,draw=black] (b4\ttl) at (8,8,\hei) {};
        \node[fill=zx_green,circle,draw=black] (b5\ttl) at (8,16,\hei) {};

        \node[fill=zx_green,circle,draw=black] (b6\ttl) at (12,4,\hei) {};
        \node[fill=zx_green,circle,draw=black] (b7\ttl) at (12,12,\hei) {};
        \node[fill=zx_green,circle,draw=black] (b8\ttl) at (12,20,\hei) {};

        \node[fill=zx_green,circle,draw=black] (b9\ttl) at (16,0,\hei) {};
        \node[fill=zx_green,circle,draw=black] (b10\ttl) at (16,8,\hei) {};
        \node[fill=zx_green,circle,draw=black] (b11\ttl) at (16,16,\hei) {};

    \end{scope}
    \begin{scope}
        
    \draw[black,fill=black,opacity=1,line width = 0.05cm] (b0\ttl) to (n0\ttl);
    \draw[black,fill=black,opacity=1,line width = 0.05cm] (b0\ttl) to (n1\ttl);
    \draw[black,fill=black,opacity=1,line width = 0.05cm] (b0\ttl) to (n5\ttl);
    \draw[black,fill=black,opacity=1,line width = 0.05cm] (b0\ttl) to (n6\ttl);

    \draw[black,fill=black,opacity=1,line width = 0.05cm] (b1\ttl) to (n2\ttl);
    \draw[black,fill=black,opacity=1,line width = 0.05cm] (b1\ttl) to (n3\ttl);
    \draw[black,fill=black,opacity=1,line width = 0.05cm] (b1\ttl) to (n7\ttl);
    \draw[black,fill=black,opacity=1,line width = 0.05cm] (b1\ttl) to (n8\ttl);
    
    \draw[black,fill=black,opacity=1,line width = 0.05cm] (b2\ttl) to (n4\ttl);
    \draw[black,fill=black,opacity=1,line width = 0.05cm] (b2\ttl) to (n9\ttl);
    
    \draw[black,fill=black,opacity=1,line width = 0.05cm] (b3\ttl) to (n5\ttl);
    \draw[black,fill=black,opacity=1,line width = 0.05cm] (b3\ttl) to (n10\ttl);

    \draw[black,fill=black,opacity=1,line width = 0.05cm] (b4\ttl) to (n6\ttl);
    \draw[black,fill=black,opacity=1,line width = 0.05cm] (b4\ttl) to (n7\ttl);
    \draw[black,fill=black,opacity=1,line width = 0.05cm] (b4\ttl) to (n11\ttl);
    \draw[black,fill=black,opacity=1,line width = 0.05cm] (b4\ttl) to (n12\ttl);

    \draw[black,fill=black,opacity=1,line width = 0.05cm] (b5\ttl) to (n8\ttl);
    \draw[black,fill=black,opacity=1,line width = 0.05cm] (b5\ttl) to (n9\ttl);
    \draw[black,fill=black,opacity=1,line width = 0.05cm] (b5\ttl) to (n13\ttl);
    \draw[black,fill=black,opacity=1,line width = 0.05cm] (b5\ttl) to (n14\ttl);

    \draw[black,fill=black,opacity=1,line width = 0.05cm] (b6\ttl) to (n10\ttl);
    \draw[black,fill=black,opacity=1,line width = 0.05cm] (b6\ttl) to (n11\ttl);
    \draw[black,fill=black,opacity=1,line width = 0.05cm] (b6\ttl) to (n15\ttl);
    \draw[black,fill=black,opacity=1,line width = 0.05cm] (b6\ttl) to (n16\ttl);

    \draw[black,fill=black,opacity=1,line width = 0.05cm] (b7\ttl) to (n12\ttl);
    \draw[black,fill=black,opacity=1,line width = 0.05cm] (b7\ttl) to (n13\ttl);
    \draw[black,fill=black,opacity=1,line width = 0.05cm] (b7\ttl) to (n17\ttl);
    \draw[black,fill=black,opacity=1,line width = 0.05cm] (b7\ttl) to (n18\ttl);
    
    \draw[black,fill=black,opacity=1,line width = 0.05cm] (b8\ttl) to (n14\ttl);
    \draw[black,fill=black,opacity=1,line width = 0.05cm] (b8\ttl) to (n19\ttl);

    \draw[black,fill=black,opacity=1,line width = 0.05cm] (b9\ttl) to (n15\ttl);
    \draw[black,fill=black,opacity=1,line width = 0.05cm] (b9\ttl) to (n20\ttl);

    \draw[black,fill=black,opacity=1,line width = 0.05cm] (b10\ttl) to (n16\ttl);
    \draw[black,fill=black,opacity=1,line width = 0.05cm] (b10\ttl) to (n17\ttl);
    \draw[black,fill=black,opacity=1,line width = 0.05cm] (b10\ttl) to (n21\ttl);
    \draw[black,fill=black,opacity=1,line width = 0.05cm] (b10\ttl) to (n22\ttl);

    \draw[black,fill=black,opacity=1,line width = 0.05cm] (b11\ttl) to (n18\ttl);
    \draw[black,fill=black,opacity=1,line width = 0.05cm] (b11\ttl) to (n19\ttl);
    \draw[black,fill=black,opacity=1,line width = 0.05cm] (b11\ttl) to (n23\ttl);
    \draw[black,fill=black,opacity=1,line width = 0.05cm] (b11\ttl) to (n24\ttl);

    \end{scope}

    \def\hei{20}
    \def\ttl{t2}
    \begin{scope}
        
    \node[fill=zx_green,shape=circle,draw=black] (n4\ttl) at (2,18,\hei) {};

    \node[fill=zx_green,shape=circle,draw=black] (n3\ttl) at (2,14,\hei) {};
    \node[fill=zx_green,shape=circle,draw=black] (n2\ttl) at (2,10,\hei) {};
    \node[fill=zx_green,shape=circle,draw=black] (n1\ttl) at (2,6,\hei) {};
    \node[fill=zx_green,shape=circle,draw=black] (n0\ttl) at (2,2,\hei) {};

    \node[fill=zx_green,shape=circle,draw=black] (n9\ttl) at (6,18,\hei) {};
    \node[fill=zx_green,shape=circle,draw=black] (n8\ttl) at (6,14,\hei) {};
    \node[fill=zx_green,shape=circle,draw=black] (n7\ttl) at (6,10,\hei) {};
    \node[fill=zx_green,shape=circle,draw=black] (n6\ttl) at (6,6,\hei) {};
    \node[fill=zx_green,shape=circle,draw=black] (n5\ttl) at (6,2,\hei) {}; 

    \node[fill=zx_green,shape=circle,draw=black] (n14\ttl) at (10,18,\hei) {};
    \node[fill=zx_green,shape=circle,draw=black] (n13\ttl) at (10,14,\hei) {};
    \node[fill=zx_green,shape=circle,draw=black] (n12\ttl) at (10,10,\hei) {};
    \node[fill=zx_green,shape=circle,draw=black] (n11\ttl) at (10,6,\hei) {};
    \node[fill=zx_green,shape=circle,draw=black] (n10\ttl) at (10,2,\hei) {};

    \node[fill=zx_green,shape=circle,draw=black] (n19\ttl) at (14,18,\hei) {};
    \node[fill=zx_green,shape=circle,draw=black] (n18\ttl) at (14,14,\hei) {};
    \node[fill=zx_green,shape=circle,draw=black] (n17\ttl) at (14,10,\hei) {};
    \node[fill=zx_green,shape=circle,draw=black] (n16\ttl) at (14,6,\hei) {};
    \node[fill=zx_green,shape=circle,draw=black] (n15\ttl) at (14,2,\hei) {}; 

    \node[fill=zx_green,shape=circle,draw=black] (n24\ttl) at (18,18,\hei) {};
    \node[fill=zx_green,shape=circle,draw=black] (n23\ttl) at (18,14,\hei) {};
    \node[fill=zx_green,shape=circle,draw=black] (n22\ttl) at (18,10,\hei) {};
    \node[fill=zx_green,shape=circle,draw=black] (n21\ttl) at (18,6,\hei) {};
    \node[fill=zx_green,shape=circle,draw=black] (n20\ttl) at (18,2,\hei) {};
    
    \end{scope}
    \begin{scope}

        \node[fill=zx_red,circle,draw=black] (a0\ttl) at (0,4,\hei) {};
        \node[fill=zx_red,circle,draw=black] (a1\ttl) at (0,12,\hei) {};

        \node[fill=zx_red,circle,draw=black] (a2\ttl) at (4,8,\hei) {};
        \node[fill=zx_red,circle,draw=black] (a3\ttl) at (4,16,\hei) {};

        \node[fill=zx_red,circle,draw=black] (a4\ttl) at (8,4,\hei) {};
        \node[fill=zx_red,circle,draw=black] (a5\ttl) at (8,12,\hei) {};

        \node[fill=zx_red,circle,draw=black] (a6\ttl) at (12,8,\hei) {};
        \node[fill=zx_red,circle,draw=black] (a7\ttl) at (12,16,\hei) {};

        \node[fill=zx_red,circle,draw=black] (a8\ttl) at (16,4,\hei) {};
        \node[fill=zx_red,circle,draw=black] (a9\ttl) at (16,12,\hei) {};

        \node[fill=zx_red,circle,draw=black] (a10\ttl) at (20,8,\hei) {};
        \node[fill=zx_red,circle,draw=black] (a11\ttl) at (20,16,\hei) {};
    \end{scope}
    \begin{scope}
        
    \draw[black,fill=black,opacity=1,line width = 0.05cm] (a0\ttl) to (n0\ttl);
    \draw[black,fill=black,opacity=1,line width = 0.05cm] (a0\ttl) to (n1\ttl);

    \draw[black,fill=black,opacity=1,line width = 0.05cm] (a1\ttl) to (n2\ttl);
    \draw[black,fill=black,opacity=1,line width = 0.05cm] (a1\ttl) to (n3\ttl);

    \draw[black,fill=black,opacity=1,line width = 0.05cm] (a2\ttl) to (n1\ttl);
    \draw[black,fill=black,opacity=1,line width = 0.05cm] (a2\ttl) to (n2\ttl);
    \draw[black,fill=black,opacity=1,line width = 0.05cm] (a2\ttl) to (n6\ttl);
    \draw[black,fill=black,opacity=1,line width = 0.05cm] (a2\ttl) to (n7\ttl);

    \draw[black,fill=black,opacity=1,line width = 0.05cm] (a3\ttl) to (n3\ttl);
    \draw[black,fill=black,opacity=1,line width = 0.05cm] (a3\ttl) to (n4\ttl);
    \draw[black,fill=black,opacity=1,line width = 0.05cm] (a3\ttl) to (n8\ttl);
    \draw[black,fill=black,opacity=1,line width = 0.05cm] (a3\ttl) to (n9\ttl);

    \draw[black,fill=black,opacity=1,line width = 0.05cm] (a4\ttl) to (n5\ttl);
    \draw[black,fill=black,opacity=1,line width = 0.05cm] (a4\ttl) to (n6\ttl);
    \draw[black,fill=black,opacity=1,line width = 0.05cm] (a4\ttl) to (n10\ttl);
    \draw[black,fill=black,opacity=1,line width = 0.05cm] (a4\ttl) to (n11\ttl);
    
    \draw[black,fill=black,opacity=1,line width = 0.05cm] (a5\ttl) to (n7\ttl);
    \draw[black,fill=black,opacity=1,line width = 0.05cm] (a5\ttl) to (n8\ttl);
    \draw[black,fill=black,opacity=1,line width = 0.05cm] (a5\ttl) to (n12\ttl);
    \draw[black,fill=black,opacity=1,line width = 0.05cm] (a5\ttl) to (n13\ttl);

    \draw[black,fill=black,opacity=1,line width = 0.05cm] (a6\ttl) to (n11\ttl);
    \draw[black,fill=black,opacity=1,line width = 0.05cm] (a6\ttl) to (n12\ttl);
    \draw[black,fill=black,opacity=1,line width = 0.05cm] (a6\ttl) to (n16\ttl);
    \draw[black,fill=black,opacity=1,line width = 0.05cm] (a6\ttl) to (n17\ttl);

    \draw[black,fill=black,opacity=1,line width = 0.05cm] (a7\ttl) to (n13\ttl);
    \draw[black,fill=black,opacity=1,line width = 0.05cm] (a7\ttl) to (n14\ttl);
    \draw[black,fill=black,opacity=1,line width = 0.05cm] (a7\ttl) to (n18\ttl);
    \draw[black,fill=black,opacity=1,line width = 0.05cm] (a7\ttl) to (n19\ttl);

    \draw[black,fill=black,opacity=1,line width = 0.05cm] (a8\ttl) to (n15\ttl);
    \draw[black,fill=black,opacity=1,line width = 0.05cm] (a8\ttl) to (n16\ttl);
    \draw[black,fill=black,opacity=1,line width = 0.05cm] (a8\ttl) to (n20\ttl);
    \draw[black,fill=black,opacity=1,line width = 0.05cm] (a8\ttl) to (n21\ttl);

    \draw[black,fill=black,opacity=1,line width = 0.05cm] (a9\ttl) to (n17\ttl);
    \draw[black,fill=black,opacity=1,line width = 0.05cm] (a9\ttl) to (n18\ttl);
    \draw[black,fill=black,opacity=1,line width = 0.05cm] (a9\ttl) to (n22\ttl);
    \draw[black,fill=black,opacity=1,line width = 0.05cm] (a9\ttl) to (n23\ttl);
    
    \draw[black,fill=black,opacity=1,line width = 0.05cm] (a10\ttl) to (n21\ttl);
    \draw[black,fill=black,opacity=1,line width = 0.05cm] (a10\ttl) to (n22\ttl);

    \draw[black,fill=black,opacity=1,line width = 0.05cm] (a11\ttl) to (n23\ttl);
    \draw[black,fill=black,opacity=1,line width = 0.05cm] (a11\ttl) to (n24\ttl);

    \end{scope}

    \def\hei{22.5}
    \def\ttl{t3}
    \begin{scope}
        
    \node[fill=none,shape=circle,draw=none] (n4\ttl) at (2,18,\hei) {};

    \node[fill=none,shape=circle,draw=none] (n3\ttl) at (2,14,\hei) {};
    \node[fill=none,shape=circle,draw=none] (n2\ttl) at (2,10,\hei) {};
    \node[fill=none,shape=circle,draw=none] (n1\ttl) at (2,6,\hei) {};
    \node[fill=none,shape=circle,draw=none] (n0\ttl) at (2,2,\hei) {};

    \node[fill=none,shape=circle,draw=none] (n9\ttl) at (6,18,\hei) {};
    \node[fill=none,shape=circle,draw=none] (n8\ttl) at (6,14,\hei) {};
    \node[fill=none,shape=circle,draw=none] (n7\ttl) at (6,10,\hei) {};
    \node[fill=none,shape=circle,draw=none] (n6\ttl) at (6,6,\hei) {};
    \node[fill=none,shape=circle,draw=none] (n5\ttl) at (6,2,\hei) {}; 

    \node[fill=none,shape=circle,draw=none] (n14\ttl) at (10,18,\hei) {};
    \node[fill=none,shape=circle,draw=none] (n13\ttl) at (10,14,\hei) {};
    \node[fill=none,shape=circle,draw=none] (n12\ttl) at (10,10,\hei) {};
    \node[fill=none,shape=circle,draw=none] (n11\ttl) at (10,6,\hei) {};
    \node[fill=none,shape=circle,draw=none] (n10\ttl) at (10,2,\hei) {};

    \node[fill=none,shape=circle,draw=none] (n19\ttl) at (14,18,\hei) {};
    \node[fill=none,shape=circle,draw=none] (n18\ttl) at (14,14,\hei) {};
    \node[fill=none,shape=circle,draw=none] (n17\ttl) at (14,10,\hei) {};
    \node[fill=none,shape=circle,draw=none] (n16\ttl) at (14,6,\hei) {};
    \node[fill=none,shape=circle,draw=none] (n15\ttl) at (14,2,\hei) {}; 

    \node[fill=none,shape=circle,draw=none] (n24\ttl) at (18,18,\hei) {};
    \node[fill=none,shape=circle,draw=none] (n23\ttl) at (18,14,\hei) {};
    \node[fill=none,shape=circle,draw=none] (n22\ttl) at (18,10,\hei) {};
    \node[fill=none,shape=circle,draw=none] (n21\ttl) at (18,6,\hei) {};
    \node[fill=none,shape=circle,draw=none] (n20\ttl) at (18,2,\hei) {};
    
    \end{scope}

    \def\tlow{t0}
    \def\thig{t1}
    \begin{pgfonlayer}{background}
    \foreach \x in {0,...,24}
    {
    \draw[black,fill=black,opacity=1,line width = 0.05cm] (n\x\tlow) to (n\x\thig);
    }
    \end{pgfonlayer} 

    \def\tlow{t1}
    \def\thig{t2}
    \begin{pgfonlayer}{background}
    \foreach \x in {0,...,24}
    {
    \draw[black,fill=black,opacity=1,line width = 0.05cm] (n\x\tlow) to (n\x\thig);
    }
    \end{pgfonlayer}

    \def\tlow{t2}
    \def\thig{t3}
    \begin{pgfonlayer}{background}
    \foreach \x in {0,...,24}
    {
    \draw[black,fill=black,opacity=1,line width = 0.05cm] (n\x\tlow) to (n\x\thig);
    }
    \end{pgfonlayer}

    \def\tlow{t1}

    \def\tlow{t2}

    \def\tlow{t0}
    \def\thig{t1}
    \def\thhig{t2}
    \def\thhhig{t3}
    \begin{pgfonlayer}{background}
    \draw[green,fill=green,opacity=0.8,line width = 0.25cm] (n4\tlow) to (n4\thig);
    \draw[green,fill=green,opacity=0.8,line width = 0.25cm] 
    (n4\thig) to (n4\thhig);
    \draw[green,fill=green,opacity=0.8,line width = 0.25cm] 
    (n4\thig) to (b2\thig);
    \draw[green,fill=green,opacity=0.8,line width = 0.25cm] 
    (n4\thhig) to (n4\thhhig);
    \draw[green,fill=green,opacity=0.8,line width = 0.25cm] 
    (n9\thig) to (b2\thig);
    \draw[green,fill=green,opacity=0.8,line width = 0.25cm] 
    (n9\thig) to (n9\tlow);
     \draw[green,fill=green,opacity=0.8,line width = 0.25cm] 
    (b5\thig) to (n9\thig);
    \draw[green,fill=green,opacity=0.8,line width = 0.25cm] 
    (n9\thhig) to (n9\thig);
    \draw[green,fill=green,opacity=0.8,line width = 0.25cm] 
    (n9\thhig) to (n9\thhhig);
    \draw[green,fill=green,opacity=0.8,line width = 0.25cm] 
    (n14\thig) to (b5\thig);
    \draw[green,fill=green,opacity=0.8,line width = 0.25cm] 
    (n14\thig) to (n14\tlow);
    \draw[green,fill=green,opacity=0.8,line width = 0.25cm] 
    (n14\thig) to (n14\thhig);
    \draw[green,fill=green,opacity=0.8,line width = 0.25cm] 
    (n14\thig) to (b8\thig);
    \draw[green,fill=green,opacity=0.8,line width = 0.25cm] 
    (n14\thhig) to (n14\thhhig);
    \draw[green,fill=green,opacity=0.8,line width = 0.25cm] 
    (n19\thig) to (b8\thig);
    \draw[green,fill=green,opacity=0.8,line width = 0.25cm] 
    (n19\thig) to (n19\tlow);
    \draw[green,fill=green,opacity=0.8,line width = 0.25cm] 
    (n19\thig) to (n19\thhig);
    \draw[green,fill=green,opacity=0.8,line width = 0.25cm] 
    (b11\thig) to (n19\thig);
    \draw[green,fill=green,opacity=0.8,line width = 0.25cm] 
    (n19\thhhig) to (n19\thhig);
    \draw[green,fill=green,opacity=0.8,line width = 0.25cm] 
    (b11\thig) to (n24\thig);
    \draw[green,fill=green,opacity=0.8,line width = 0.25cm] 
    (n24\thig) to (n24\tlow);
    \draw[green,fill=green,opacity=0.8,line width = 0.25cm] 
    (n24\thig) to (n24\thhig);
    \draw[green,fill=green,opacity=0.8,line width = 0.25cm] 
    (n24\thhig) to (n24\thhhig);
    
    \end{pgfonlayer}

    \end{tikzpicture}
    }
    }
    \qquad
    \subfloat[][\label{fig:logical_cor_X}Logical $X$ correlator.]{
    \resizebox{0.4\linewidth}{!}{
   
    \begin{tikzpicture}[tdplot_main_coords,every node/.style={minimum size=1cm}]
     
    \def\hei{7.5}
    \def\ttl{t0}
    \begin{scope}
        
    \node[fill=none,shape=circle,draw=none] (n4\ttl) at (2,18,\hei) {};
    \node[fill=none,shape=circle,draw=none]  (n3\ttl) at (2,14,\hei) {};
    \node[fill=none,shape=circle,draw=none]  (n2\ttl) at (2,10,\hei) {};
    \node[fill=none,shape=circle,draw=none]  (n1\ttl) at (2,6,\hei) {};
   \node[fill=none,shape=circle,draw=none]  (n0\ttl) at (2,2,\hei) {};

   \node[fill=none,shape=circle,draw=none] (n9\ttl) at (6,18,\hei) {};
    \node[fill=none,shape=circle,draw=none]  (n8\ttl) at (6,14,\hei) {};
    \node[fill=none,shape=circle,draw=none]  (n7\ttl) at (6,10,\hei) {};
    \node[fill=none,shape=circle,draw=none]  (n6\ttl) at (6,6,\hei) {};
    \node[fill=none,shape=circle,draw=none] (n5\ttl) at (6,2,\hei) {}; 

   \node[fill=none,shape=circle,draw=none]  (n14\ttl) at (10,18,\hei) {};
   \node[fill=none,shape=circle,draw=none] (n13\ttl) at (10,14,\hei) {};
   \node[fill=none,shape=circle,draw=none]  (n12\ttl) at (10,10,\hei) {};
    \node[fill=none,shape=circle,draw=none]  (n11\ttl) at (10,6,\hei) {};
    \node[fill=none,shape=circle,draw=none]  (n10\ttl) at (10,2,\hei) {};

    \node[fill=none,shape=circle,draw=none] (n19\ttl) at (14,18,\hei) {};
    \node[fill=none,shape=circle,draw=none] (n18\ttl) at (14,14,\hei) {};
    \node[fill=none,shape=circle,draw=none] (n17\ttl) at (14,10,\hei) {};
    \node[fill=none,shape=circle,draw=none]  (n16\ttl) at (14,6,\hei) {};
    \node[fill=none,shape=circle,draw=none]  (n15\ttl) at (14,2,\hei) {}; 

    \node[fill=none,shape=circle,draw=none]  (n24\ttl) at (18,18,\hei) {};
    \node[fill=none,shape=circle,draw=none]  (n23\ttl) at (18,14,\hei) {};
    \node[fill=none,shape=circle,draw=none]  (n22\ttl) at (18,10,\hei) {};
    \node[fill=none,shape=circle,draw=none] (n21\ttl) at (18,6,\hei) {};
    \node[fill=none,shape=circle,draw=none]  (n20\ttl) at (18,2,\hei) {};
    
    \end{scope}

    \def\hei{10}
    \def\ttl{t1}
    \begin{scope}
        
    \node[fill=zx_red,shape=circle,draw=black] (n4\ttl) at (2,18,\hei) {};

    \node[fill=zx_red,shape=circle,draw=black] (n3\ttl) at (2,14,\hei) {};
    \node[fill=zx_red,shape=circle,draw=black] (n2\ttl) at (2,10,\hei) {};
    \node[fill=zx_red,shape=circle,draw=black] (n1\ttl) at (2,6,\hei) {};
    \node[fill=zx_red,shape=circle,draw=black] (n0\ttl) at (2,2,\hei) {};

    \node[fill=zx_red,shape=circle,draw=black] (n9\ttl) at (6,18,\hei) {};
    \node[fill=zx_red,shape=circle,draw=black] (n8\ttl) at (6,14,\hei) {};
    \node[fill=zx_red,shape=circle,draw=black] (n7\ttl) at (6,10,\hei) {};
    \node[fill=zx_red,shape=circle,draw=black] (n6\ttl) at (6,6,\hei) {};
    \node[fill=zx_red,shape=circle,draw=black] (n5\ttl) at (6,2,\hei) {}; 

    \node[fill=zx_red,shape=circle,draw=black] (n14\ttl) at (10,18,\hei) {};
    \node[fill=zx_red,shape=circle,draw=black] (n13\ttl) at (10,14,\hei) {};
    \node[fill=zx_red,shape=circle,draw=black] (n12\ttl) at (10,10,\hei) {};
    \node[fill=zx_red,shape=circle,draw=black] (n11\ttl) at (10,6,\hei) {};
    \node[fill=zx_red,shape=circle,draw=black] (n10\ttl) at (10,2,\hei) {};

    \node[fill=zx_red,shape=circle,draw=black] (n19\ttl) at (14,18,\hei) {};
    \node[fill=zx_red,shape=circle,draw=black] (n18\ttl) at (14,14,\hei) {};
    \node[fill=zx_red,shape=circle,draw=black] (n17\ttl) at (14,10,\hei) {};
    \node[fill=zx_red,shape=circle,draw=black] (n16\ttl) at (14,6,\hei) {};
    \node[fill=zx_red,shape=circle,draw=black] (n15\ttl) at (14,2,\hei) {}; 

    \node[fill=zx_red,shape=circle,draw=black] (n24\ttl) at (18,18,\hei) {};
    \node[fill=zx_red,shape=circle,draw=black] (n23\ttl) at (18,14,\hei) {};
    \node[fill=zx_red,shape=circle,draw=black] (n22\ttl) at (18,10,\hei) {};
    \node[fill=zx_red,shape=circle,draw=black] (n21\ttl) at (18,6,\hei) {};
    \node[fill=zx_red,shape=circle,draw=black] (n20\ttl) at (18,2,\hei) {};
    
    \end{scope}
    \begin{scope}
        \node[fill=zx_green,circle,draw=black] (b0\ttl) at (4,4,\hei) {};
        \node[fill=zx_green,circle,draw=black] (b1\ttl) at (4,12,\hei) {};
        \node[fill=zx_green,circle,draw=black] (b2\ttl) at (4,20,\hei) {};

        \node[fill=zx_green,circle,draw=black] (b3\ttl) at (8,0,\hei) {};
        \node[fill=zx_green,circle,draw=black] (b4\ttl) at (8,8,\hei) {};
        \node[fill=zx_green,circle,draw=black] (b5\ttl) at (8,16,\hei) {};

        \node[fill=zx_green,circle,draw=black] (b6\ttl) at (12,4,\hei) {};
        \node[fill=zx_green,circle,draw=black] (b7\ttl) at (12,12,\hei) {};
        \node[fill=zx_green,circle,draw=black] (b8\ttl) at (12,20,\hei) {};

        \node[fill=zx_green,circle,draw=black] (b9\ttl) at (16,0,\hei) {};
        \node[fill=zx_green,circle,draw=black] (b10\ttl) at (16,8,\hei) {};
        \node[fill=zx_green,circle,draw=black] (b11\ttl) at (16,16,\hei) {};

    \end{scope}
    \begin{scope}
        
    \draw[black,fill=black,opacity=1,line width = 0.05cm] (b0\ttl) to (n0\ttl);
    \draw[black,fill=black,opacity=1,line width = 0.05cm] (b0\ttl) to (n1\ttl);
    \draw[black,fill=black,opacity=1,line width = 0.05cm] (b0\ttl) to (n5\ttl);
    \draw[black,fill=black,opacity=1,line width = 0.05cm] (b0\ttl) to (n6\ttl);

    \draw[black,fill=black,opacity=1,line width = 0.05cm] (b1\ttl) to (n2\ttl);
    \draw[black,fill=black,opacity=1,line width = 0.05cm] (b1\ttl) to (n3\ttl);
    \draw[black,fill=black,opacity=1,line width = 0.05cm] (b1\ttl) to (n7\ttl);
    \draw[black,fill=black,opacity=1,line width = 0.05cm] (b1\ttl) to (n8\ttl);
    
    \draw[black,fill=black,opacity=1,line width = 0.05cm] (b2\ttl) to (n4\ttl);
    \draw[black,fill=black,opacity=1,line width = 0.05cm] (b2\ttl) to (n9\ttl);
    
    \draw[black,fill=black,opacity=1,line width = 0.05cm] (b3\ttl) to (n5\ttl);
    \draw[black,fill=black,opacity=1,line width = 0.05cm] (b3\ttl) to (n10\ttl);

    \draw[black,fill=black,opacity=1,line width = 0.05cm] (b4\ttl) to (n6\ttl);
    \draw[black,fill=black,opacity=1,line width = 0.05cm] (b4\ttl) to (n7\ttl);
    \draw[black,fill=black,opacity=1,line width = 0.05cm] (b4\ttl) to (n11\ttl);
    \draw[black,fill=black,opacity=1,line width = 0.05cm] (b4\ttl) to (n12\ttl);

    \draw[black,fill=black,opacity=1,line width = 0.05cm] (b5\ttl) to (n8\ttl);
    \draw[black,fill=black,opacity=1,line width = 0.05cm] (b5\ttl) to (n9\ttl);
    \draw[black,fill=black,opacity=1,line width = 0.05cm] (b5\ttl) to (n13\ttl);
    \draw[black,fill=black,opacity=1,line width = 0.05cm] (b5\ttl) to (n14\ttl);

    \draw[black,fill=black,opacity=1,line width = 0.05cm] (b6\ttl) to (n10\ttl);
    \draw[black,fill=black,opacity=1,line width = 0.05cm] (b6\ttl) to (n11\ttl);
    \draw[black,fill=black,opacity=1,line width = 0.05cm] (b6\ttl) to (n15\ttl);
    \draw[black,fill=black,opacity=1,line width = 0.05cm] (b6\ttl) to (n16\ttl);

    \draw[black,fill=black,opacity=1,line width = 0.05cm] (b7\ttl) to (n12\ttl);
    \draw[black,fill=black,opacity=1,line width = 0.05cm] (b7\ttl) to (n13\ttl);
    \draw[black,fill=black,opacity=1,line width = 0.05cm] (b7\ttl) to (n17\ttl);
    \draw[black,fill=black,opacity=1,line width = 0.05cm] (b7\ttl) to (n18\ttl);
    
    \draw[black,fill=black,opacity=1,line width = 0.05cm] (b8\ttl) to (n14\ttl);
    \draw[black,fill=black,opacity=1,line width = 0.05cm] (b8\ttl) to (n19\ttl);

    \draw[black,fill=black,opacity=1,line width = 0.05cm] (b9\ttl) to (n15\ttl);
    \draw[black,fill=black,opacity=1,line width = 0.05cm] (b9\ttl) to (n20\ttl);

    \draw[black,fill=black,opacity=1,line width = 0.05cm] (b10\ttl) to (n16\ttl);
    \draw[black,fill=black,opacity=1,line width = 0.05cm] (b10\ttl) to (n17\ttl);
    \draw[black,fill=black,opacity=1,line width = 0.05cm] (b10\ttl) to (n21\ttl);
    \draw[black,fill=black,opacity=1,line width = 0.05cm] (b10\ttl) to (n22\ttl);

    \draw[black,fill=black,opacity=1,line width = 0.05cm] (b11\ttl) to (n18\ttl);
    \draw[black,fill=black,opacity=1,line width = 0.05cm] (b11\ttl) to (n19\ttl);
    \draw[black,fill=black,opacity=1,line width = 0.05cm] (b11\ttl) to (n23\ttl);
    \draw[black,fill=black,opacity=1,line width = 0.05cm] (b11\ttl) to (n24\ttl);

    \end{scope}

    \def\hei{20}
    \def\ttl{t2}
    \begin{scope}
        
    \node[fill=zx_green,shape=circle,draw=black] (n4\ttl) at (2,18,\hei) {};

    \node[fill=zx_green,shape=circle,draw=black] (n3\ttl) at (2,14,\hei) {};
    \node[fill=zx_green,shape=circle,draw=black] (n2\ttl) at (2,10,\hei) {};
    \node[fill=zx_green,shape=circle,draw=black] (n1\ttl) at (2,6,\hei) {};
    \node[fill=zx_green,shape=circle,draw=black] (n0\ttl) at (2,2,\hei) {};

    \node[fill=zx_green,shape=circle,draw=black] (n9\ttl) at (6,18,\hei) {};
    \node[fill=zx_green,shape=circle,draw=black] (n8\ttl) at (6,14,\hei) {};
    \node[fill=zx_green,shape=circle,draw=black] (n7\ttl) at (6,10,\hei) {};
    \node[fill=zx_green,shape=circle,draw=black] (n6\ttl) at (6,6,\hei) {};
    \node[fill=zx_green,shape=circle,draw=black] (n5\ttl) at (6,2,\hei) {}; 

    \node[fill=zx_green,shape=circle,draw=black] (n14\ttl) at (10,18,\hei) {};
    \node[fill=zx_green,shape=circle,draw=black] (n13\ttl) at (10,14,\hei) {};
    \node[fill=zx_green,shape=circle,draw=black] (n12\ttl) at (10,10,\hei) {};
    \node[fill=zx_green,shape=circle,draw=black] (n11\ttl) at (10,6,\hei) {};
    \node[fill=zx_green,shape=circle,draw=black] (n10\ttl) at (10,2,\hei) {};

    \node[fill=zx_green,shape=circle,draw=black] (n19\ttl) at (14,18,\hei) {};
    \node[fill=zx_green,shape=circle,draw=black] (n18\ttl) at (14,14,\hei) {};
    \node[fill=zx_green,shape=circle,draw=black] (n17\ttl) at (14,10,\hei) {};
    \node[fill=zx_green,shape=circle,draw=black] (n16\ttl) at (14,6,\hei) {};
    \node[fill=zx_green,shape=circle,draw=black] (n15\ttl) at (14,2,\hei) {}; 

    \node[fill=zx_green,shape=circle,draw=black] (n24\ttl) at (18,18,\hei) {};
    \node[fill=zx_green,shape=circle,draw=black] (n23\ttl) at (18,14,\hei) {};
    \node[fill=zx_green,shape=circle,draw=black] (n22\ttl) at (18,10,\hei) {};
    \node[fill=zx_green,shape=circle,draw=black] (n21\ttl) at (18,6,\hei) {};
    \node[fill=zx_green,shape=circle,draw=black] (n20\ttl) at (18,2,\hei) {};
    
    \end{scope}
    \begin{scope}

        \node[fill=zx_red,circle,draw=black] (a0\ttl) at (0,4,\hei) {};
        \node[fill=zx_red,circle,draw=black] (a1\ttl) at (0,12,\hei) {};

        \node[fill=zx_red,circle,draw=black] (a2\ttl) at (4,8,\hei) {};
        \node[fill=zx_red,circle,draw=black] (a3\ttl) at (4,16,\hei) {};

        \node[fill=zx_red,circle,draw=black] (a4\ttl) at (8,4,\hei) {};
        \node[fill=zx_red,circle,draw=black] (a5\ttl) at (8,12,\hei) {};

        \node[fill=zx_red,circle,draw=black] (a6\ttl) at (12,8,\hei) {};
        \node[fill=zx_red,circle,draw=black] (a7\ttl) at (12,16,\hei) {};

        \node[fill=zx_red,circle,draw=black] (a8\ttl) at (16,4,\hei) {};
        \node[fill=zx_red,circle,draw=black] (a9\ttl) at (16,12,\hei) {};

        \node[fill=zx_red,circle,draw=black] (a10\ttl) at (20,8,\hei) {};
        \node[fill=zx_red,circle,draw=black] (a11\ttl) at (20,16,\hei) {};
    \end{scope}
    \begin{scope}
        
    \draw[black,fill=black,opacity=1,line width = 0.05cm] (a0\ttl) to (n0\ttl);
    \draw[black,fill=black,opacity=1,line width = 0.05cm] (a0\ttl) to (n1\ttl);

    \draw[black,fill=black,opacity=1,line width = 0.05cm] (a1\ttl) to (n2\ttl);
    \draw[black,fill=black,opacity=1,line width = 0.05cm] (a1\ttl) to (n3\ttl);

    \draw[black,fill=black,opacity=1,line width = 0.05cm] (a2\ttl) to (n1\ttl);
    \draw[black,fill=black,opacity=1,line width = 0.05cm] (a2\ttl) to (n2\ttl);
    \draw[black,fill=black,opacity=1,line width = 0.05cm] (a2\ttl) to (n6\ttl);
    \draw[black,fill=black,opacity=1,line width = 0.05cm] (a2\ttl) to (n7\ttl);

    \draw[black,fill=black,opacity=1,line width = 0.05cm] (a3\ttl) to (n3\ttl);
    \draw[black,fill=black,opacity=1,line width = 0.05cm] (a3\ttl) to (n4\ttl);
    \draw[black,fill=black,opacity=1,line width = 0.05cm] (a3\ttl) to (n8\ttl);
    \draw[black,fill=black,opacity=1,line width = 0.05cm] (a3\ttl) to (n9\ttl);

    \draw[black,fill=black,opacity=1,line width = 0.05cm] (a4\ttl) to (n5\ttl);
    \draw[black,fill=black,opacity=1,line width = 0.05cm] (a4\ttl) to (n6\ttl);
    \draw[black,fill=black,opacity=1,line width = 0.05cm] (a4\ttl) to (n10\ttl);
    \draw[black,fill=black,opacity=1,line width = 0.05cm] (a4\ttl) to (n11\ttl);
    
    \draw[black,fill=black,opacity=1,line width = 0.05cm] (a5\ttl) to (n7\ttl);
    \draw[black,fill=black,opacity=1,line width = 0.05cm] (a5\ttl) to (n8\ttl);
    \draw[black,fill=black,opacity=1,line width = 0.05cm] (a5\ttl) to (n12\ttl);
    \draw[black,fill=black,opacity=1,line width = 0.05cm] (a5\ttl) to (n13\ttl);

    \draw[black,fill=black,opacity=1,line width = 0.05cm] (a6\ttl) to (n11\ttl);
    \draw[black,fill=black,opacity=1,line width = 0.05cm] (a6\ttl) to (n12\ttl);
    \draw[black,fill=black,opacity=1,line width = 0.05cm] (a6\ttl) to (n16\ttl);
    \draw[black,fill=black,opacity=1,line width = 0.05cm] (a6\ttl) to (n17\ttl);

    \draw[black,fill=black,opacity=1,line width = 0.05cm] (a7\ttl) to (n13\ttl);
    \draw[black,fill=black,opacity=1,line width = 0.05cm] (a7\ttl) to (n14\ttl);
    \draw[black,fill=black,opacity=1,line width = 0.05cm] (a7\ttl) to (n18\ttl);
    \draw[black,fill=black,opacity=1,line width = 0.05cm] (a7\ttl) to (n19\ttl);

    \draw[black,fill=black,opacity=1,line width = 0.05cm] (a8\ttl) to (n15\ttl);
    \draw[black,fill=black,opacity=1,line width = 0.05cm] (a8\ttl) to (n16\ttl);
    \draw[black,fill=black,opacity=1,line width = 0.05cm] (a8\ttl) to (n20\ttl);
    \draw[black,fill=black,opacity=1,line width = 0.05cm] (a8\ttl) to (n21\ttl);

    \draw[black,fill=black,opacity=1,line width = 0.05cm] (a9\ttl) to (n17\ttl);
    \draw[black,fill=black,opacity=1,line width = 0.05cm] (a9\ttl) to (n18\ttl);
    \draw[black,fill=black,opacity=1,line width = 0.05cm] (a9\ttl) to (n22\ttl);
    \draw[black,fill=black,opacity=1,line width = 0.05cm] (a9\ttl) to (n23\ttl);
    
    \draw[black,fill=black,opacity=1,line width = 0.05cm] (a10\ttl) to (n21\ttl);
    \draw[black,fill=black,opacity=1,line width = 0.05cm] (a10\ttl) to (n22\ttl);

    \draw[black,fill=black,opacity=1,line width = 0.05cm] (a11\ttl) to (n23\ttl);
    \draw[black,fill=black,opacity=1,line width = 0.05cm] (a11\ttl) to (n24\ttl);

    \end{scope}

    \def\hei{22.5}
    \def\ttl{t3}
    \begin{scope}
        
    \node[fill=none,shape=circle,draw=none] (n4\ttl) at (2,18,\hei) {};

    \node[fill=none,shape=circle,draw=none] (n3\ttl) at (2,14,\hei) {};
    \node[fill=none,shape=circle,draw=none] (n2\ttl) at (2,10,\hei) {};
    \node[fill=none,shape=circle,draw=none] (n1\ttl) at (2,6,\hei) {};
    \node[fill=none,shape=circle,draw=none] (n0\ttl) at (2,2,\hei) {};

    \node[fill=none,shape=circle,draw=none] (n9\ttl) at (6,18,\hei) {};
    \node[fill=none,shape=circle,draw=none] (n8\ttl) at (6,14,\hei) {};
    \node[fill=none,shape=circle,draw=none] (n7\ttl) at (6,10,\hei) {};
    \node[fill=none,shape=circle,draw=none] (n6\ttl) at (6,6,\hei) {};
    \node[fill=none,shape=circle,draw=none] (n5\ttl) at (6,2,\hei) {}; 

    \node[fill=none,shape=circle,draw=none] (n14\ttl) at (10,18,\hei) {};
    \node[fill=none,shape=circle,draw=none] (n13\ttl) at (10,14,\hei) {};
    \node[fill=none,shape=circle,draw=none] (n12\ttl) at (10,10,\hei) {};
    \node[fill=none,shape=circle,draw=none] (n11\ttl) at (10,6,\hei) {};
    \node[fill=none,shape=circle,draw=none] (n10\ttl) at (10,2,\hei) {};

    \node[fill=none,shape=circle,draw=none] (n19\ttl) at (14,18,\hei) {};
    \node[fill=none,shape=circle,draw=none] (n18\ttl) at (14,14,\hei) {};
    \node[fill=none,shape=circle,draw=none] (n17\ttl) at (14,10,\hei) {};
    \node[fill=none,shape=circle,draw=none] (n16\ttl) at (14,6,\hei) {};
    \node[fill=none,shape=circle,draw=none] (n15\ttl) at (14,2,\hei) {}; 

    \node[fill=none,shape=circle,draw=none] (n24\ttl) at (18,18,\hei) {};
    \node[fill=none,shape=circle,draw=none] (n23\ttl) at (18,14,\hei) {};
    \node[fill=none,shape=circle,draw=none] (n22\ttl) at (18,10,\hei) {};
    \node[fill=none,shape=circle,draw=none] (n21\ttl) at (18,6,\hei) {};
    \node[fill=none,shape=circle,draw=none] (n20\ttl) at (18,2,\hei) {};
    
    \end{scope}

    \def\tlow{t0}
    \def\thig{t1}
    \begin{pgfonlayer}{background}
    \foreach \x in {0,...,24}
    {
    \draw[black,fill=black,opacity=1,line width = 0.05cm] (n\x\tlow) to (n\x\thig);
    }
    \end{pgfonlayer} 

    \def\tlow{t1}
    \def\thig{t2}
    \begin{pgfonlayer}{background}
    \foreach \x in {0,...,24}
    {
    \draw[black,fill=black,opacity=1,line width = 0.05cm] (n\x\tlow) to (n\x\thig);
    }
    \end{pgfonlayer}

    \def\tlow{t2}
    \def\thig{t3}
    \begin{pgfonlayer}{background}
    \foreach \x in {0,...,24}
    {
    \draw[black,fill=black,opacity=1,line width = 0.05cm] (n\x\tlow) to (n\x\thig);
    }
    \end{pgfonlayer}

    \def\tlow{t1}

    \def\tlow{t2}

    \def\tlow{t0}
    \def\thig{t1}
    \def\thhig{t2}
    \def\thhhig{t3}
    \begin{pgfonlayer}{background}
    \draw[red,fill=red,opacity=0.6,line width = 0.25cm] (n4\tlow) to (n4\thig);
    
    \draw[red,fill=red,opacity=0.6,line width = 0.25cm] (n4\thhig) to (n4\thig);
    \draw[red,fill=red,opacity=0.6,line width = 0.25cm] (n4\thhig) to (n4\thhhig);
    \draw[red,fill=red,opacity=0.6,line width = 0.25cm] (n4\thhig) to (a3\thhig);
    \draw[red,fill=red,opacity=0.6,line width = 0.25cm] (n3\thhig) to (a3\thhig);
    \draw[red,fill=red,opacity=0.6,line width = 0.25cm] (n3\thhig) to (n3\thhhig);
    \draw[red,fill=red,opacity=0.6,line width = 0.25cm] (n3\thhig) to (n3\thig);
    \draw[red,fill=red,opacity=0.6,line width = 0.25cm] (n3\thhig) to (a1\thhig);
    \draw[red,fill=red,opacity=0.6,line width = 0.25cm] (n2\thhig) to (a1\thhig);
    \draw[red,fill=red,opacity=0.6,line width = 0.25cm] (n2\thhig) to (n2\thhhig);
    \draw[red,fill=red,opacity=0.6,line width = 0.25cm] (n2\thhig) to (n2\thig);
    \draw[red,fill=red,opacity=0.6,line width = 0.25cm] (n2\thhig) to (a2\thhig);
    \draw[red,fill=red,opacity=0.6,line width = 0.25cm] (n1\thhig) to (a2\thhig);
    \draw[red,fill=red,opacity=0.6,line width = 0.25cm] (n1\thhig) to (n1\thhhig);
    \draw[red,fill=red,opacity=0.6,line width = 0.25cm] (n1\thhig) to (n1\thig);
    \draw[red,fill=red,opacity=0.6,line width = 0.25cm] (n1\thhig) to (a0\thhig);
    \draw[red,fill=red,opacity=0.6,line width = 0.25cm] (n0\thhig) to (a0\thhig);
    \draw[red,fill=red,opacity=0.6,line width = 0.25cm] (n0\thhig) to (n0\thhhig);
    \draw[red,fill=red,opacity=0.6,line width = 0.25cm] (n0\thhig) to (n0\thig);
    \draw[red,fill=red,opacity=0.6,line width = 0.25cm] (n0\tlow) to (n0\thig);
    \draw[red,fill=red,opacity=0.6,line width = 0.25cm] (n1\tlow) to (n1\thig);
    \draw[red,fill=red,opacity=0.6,line width = 0.25cm] (n2\tlow) to (n2\thig);
    \draw[red,fill=red,opacity=0.6,line width = 0.25cm] (n3\tlow) to (n3\thig);

    \end{pgfonlayer}

    \end{tikzpicture}
    }
    }
    \caption{\label{fig:logical_cor}These are the logical $Z$ and $X$ correlators, linking the logical ($Z$ and $X$ respectively) operators of the surface code from the input (bottom legs) to the output (top legs).}
\end{figure}

The surface code encoder have logical $Z$ or $X$ correlators \cite{Bombin_2024} that connects the $Z$ or $X$ logical operators from the input to output of the ZX-diagram (bottom to top). In figure \ref{fig:logical_cor}, the logical $Z$ correlator is the green Pauli web (figure \ref{fig:logical_cor_Z}), similarly, the logical $X$ correlator is the red Pauli web (figure \ref{fig:logical_cor_X}). Note that the Pauli web colouring convention is the exact opposite of \cite{Bombin_2024}, following the colour map from \cite{rodatz2024floquetifyingstabilisercodesdistancepreserving}.

\subsection{Combined ZX-diagram}

\begin{figure}[!h]
    \centering     
    \tdplotsetmaincoords{70}{23}
    \resizebox{0.45\linewidth}{!}{
   
    \begin{tikzpicture}[tdplot_main_coords,every node/.style={minimum size=1cm}]
    \def\hei{0}
    \def\ttl{t0}
    \begin{scope}
        
    \node[fill=zx_green,shape=circle,draw=black] (n4\ttl) at (2,18,\hei) {$\pi/2$};
    \node[fill=zx_green,shape=circle,draw=black] (n3\ttl) at (2,14,\hei) {};
    \node[fill=zx_green,shape=circle,draw=black] (n2\ttl) at (2,10,\hei) {};
    \node[fill=zx_green,shape=circle,draw=black] (n1\ttl) at (2,6,\hei) {};
    \node[fill=zx_green,shape=circle,draw=black] (n0\ttl) at (2,2,\hei) {};

    \node[fill=zx_red,shape=circle,draw=black] (n9\ttl) at (6,18,\hei) {};
    \node[fill=zx_green,shape=circle,draw=black] (n8\ttl) at (6,14,\hei) {};
    \node[fill=zx_green,shape=circle,draw=black] (n7\ttl) at (6,10,\hei) {};
    \node[fill=zx_green,shape=circle,draw=black] (n6\ttl) at (6,6,\hei) {};
    \node[fill=zx_green,shape=circle,draw=black] (n5\ttl) at (6,2,\hei) {}; 

    \node[fill=zx_red,shape=circle,draw=black] (n14\ttl) at (10,18,\hei) {};
    \node[fill=zx_red,shape=circle,draw=black] (n13\ttl) at (10,14,\hei) {};
    \node[fill=zx_green,shape=circle,draw=black] (n12\ttl) at (10,10,\hei) {};
    \node[fill=zx_green,shape=circle,draw=black] (n11\ttl) at (10,6,\hei) {};
    \node[fill=zx_green,shape=circle,draw=black] (n10\ttl) at (10,2,\hei) {};

    \node[fill=zx_red,shape=circle,draw=black] (n19\ttl) at (14,18,\hei) {};
    \node[fill=zx_red,shape=circle,draw=black] (n18\ttl) at (14,14,\hei) {};
    \node[fill=zx_red,shape=circle,draw=black] (n17\ttl) at (14,10,\hei) {};
    \node[fill=zx_green,shape=circle,draw=black] (n16\ttl) at (14,6,\hei) {};
    \node[fill=zx_green,shape=circle,draw=black] (n15\ttl) at (14,2,\hei) {}; 

    \node[fill=zx_red,shape=circle,draw=black] (n24\ttl) at (18,18,\hei) {};
    \node[fill=zx_red,shape=circle,draw=black] (n23\ttl) at (18,14,\hei) {};
    \node[fill=zx_red,shape=circle,draw=black] (n22\ttl) at (18,10,\hei) {};
    \node[fill=zx_red,shape=circle,draw=black] (n21\ttl) at (18,6,\hei) {};
    \node[fill=zx_green,shape=circle,draw=black] (n20\ttl) at (18,2,\hei) {};
    
    \end{scope} 

    \def\hei{10}
    \def\ttl{t1}
    \begin{scope}
        
    \node[fill=zx_red,shape=circle,draw=black] (n4\ttl) at (2,18,\hei) {};

    \node[fill=zx_red,shape=circle,draw=black] (n3\ttl) at (2,14,\hei) {};
    \node[fill=zx_red,shape=circle,draw=black] (n2\ttl) at (2,10,\hei) {};
    \node[fill=zx_red,shape=circle,draw=black] (n1\ttl) at (2,6,\hei) {};
    \node[fill=zx_red,shape=circle,draw=black] (n0\ttl) at (2,2,\hei) {};

    \node[fill=zx_red,shape=circle,draw=black] (n9\ttl) at (6,18,\hei) {};
    \node[fill=zx_red,shape=circle,draw=black] (n8\ttl) at (6,14,\hei) {};
    \node[fill=zx_red,shape=circle,draw=black] (n7\ttl) at (6,10,\hei) {};
    \node[fill=zx_red,shape=circle,draw=black] (n6\ttl) at (6,6,\hei) {};
    \node[fill=zx_red,shape=circle,draw=black] (n5\ttl) at (6,2,\hei) {}; 

    \node[fill=zx_red,shape=circle,draw=black] (n14\ttl) at (10,18,\hei) {};
    \node[fill=zx_red,shape=circle,draw=black] (n13\ttl) at (10,14,\hei) {};
    \node[fill=zx_red,shape=circle,draw=black] (n12\ttl) at (10,10,\hei) {};
    \node[fill=zx_red,shape=circle,draw=black] (n11\ttl) at (10,6,\hei) {};
    \node[fill=zx_red,shape=circle,draw=black] (n10\ttl) at (10,2,\hei) {};

    \node[fill=zx_red,shape=circle,draw=black] (n19\ttl) at (14,18,\hei) {};
    \node[fill=zx_red,shape=circle,draw=black] (n18\ttl) at (14,14,\hei) {};
    \node[fill=zx_red,shape=circle,draw=black] (n17\ttl) at (14,10,\hei) {};
    \node[fill=zx_red,shape=circle,draw=black] (n16\ttl) at (14,6,\hei) {};
    \node[fill=zx_red,shape=circle,draw=black] (n15\ttl) at (14,2,\hei) {}; 

    \node[fill=zx_red,shape=circle,draw=black] (n24\ttl) at (18,18,\hei) {};
    \node[fill=zx_red,shape=circle,draw=black] (n23\ttl) at (18,14,\hei) {};
    \node[fill=zx_red,shape=circle,draw=black] (n22\ttl) at (18,10,\hei) {};
    \node[fill=zx_red,shape=circle,draw=black] (n21\ttl) at (18,6,\hei) {};
    \node[fill=zx_red,shape=circle,draw=black] (n20\ttl) at (18,2,\hei) {};
    
    \end{scope}
    \begin{scope}
        \node[fill=zx_green,circle,draw=black] (b0\ttl) at (4,4,\hei) {};
        \node[fill=zx_green,circle,draw=black] (b1\ttl) at (4,12,\hei) {};
        \node[fill=zx_green,circle,draw=black] (b2\ttl) at (4,20,\hei) {};

        \node[fill=zx_green,circle,draw=black] (b3\ttl) at (8,0,\hei) {};
        \node[fill=zx_green,circle,draw=black] (b4\ttl) at (8,8,\hei) {};
        \node[fill=zx_green,circle,draw=black] (b5\ttl) at (8,16,\hei) {};

        \node[fill=zx_green,circle,draw=black] (b6\ttl) at (12,4,\hei) {};
        \node[fill=zx_green,circle,draw=black] (b7\ttl) at (12,12,\hei) {};
        \node[fill=zx_green,circle,draw=black] (b8\ttl) at (12,20,\hei) {};

        \node[fill=zx_green,circle,draw=black] (b9\ttl) at (16,0,\hei) {};
        \node[fill=zx_green,circle,draw=black] (b10\ttl) at (16,8,\hei) {};
        \node[fill=zx_green,circle,draw=black] (b11\ttl) at (16,16,\hei) {};

    \end{scope}
    \begin{scope}
        
    \draw[black,fill=black,opacity=1,line width = 0.05cm] (b0\ttl) to (n0\ttl);
    \draw[black,fill=black,opacity=1,line width = 0.05cm] (b0\ttl) to (n1\ttl);
    \draw[black,fill=black,opacity=1,line width = 0.05cm] (b0\ttl) to (n5\ttl);
    \draw[black,fill=black,opacity=1,line width = 0.05cm] (b0\ttl) to (n6\ttl);

    \draw[black,fill=black,opacity=1,line width = 0.05cm] (b1\ttl) to (n2\ttl);
    \draw[black,fill=black,opacity=1,line width = 0.05cm] (b1\ttl) to (n3\ttl);
    \draw[black,fill=black,opacity=1,line width = 0.05cm] (b1\ttl) to (n7\ttl);
    \draw[black,fill=black,opacity=1,line width = 0.05cm] (b1\ttl) to (n8\ttl);
    
    \draw[black,fill=black,opacity=1,line width = 0.05cm] (b2\ttl) to (n4\ttl);
    \draw[black,fill=black,opacity=1,line width = 0.05cm] (b2\ttl) to (n9\ttl);
    
    \draw[black,fill=black,opacity=1,line width = 0.05cm] (b3\ttl) to (n5\ttl);
    \draw[black,fill=black,opacity=1,line width = 0.05cm] (b3\ttl) to (n10\ttl);

    \draw[black,fill=black,opacity=1,line width = 0.05cm] (b4\ttl) to (n6\ttl);
    \draw[black,fill=black,opacity=1,line width = 0.05cm] (b4\ttl) to (n7\ttl);
    \draw[black,fill=black,opacity=1,line width = 0.05cm] (b4\ttl) to (n11\ttl);
    \draw[black,fill=black,opacity=1,line width = 0.05cm] (b4\ttl) to (n12\ttl);

    \draw[black,fill=black,opacity=1,line width = 0.05cm] (b5\ttl) to (n8\ttl);
    \draw[black,fill=black,opacity=1,line width = 0.05cm] (b5\ttl) to (n9\ttl);
    \draw[black,fill=black,opacity=1,line width = 0.05cm] (b5\ttl) to (n13\ttl);
    \draw[black,fill=black,opacity=1,line width = 0.05cm] (b5\ttl) to (n14\ttl);

    \draw[black,fill=black,opacity=1,line width = 0.05cm] (b6\ttl) to (n10\ttl);
    \draw[black,fill=black,opacity=1,line width = 0.05cm] (b6\ttl) to (n11\ttl);
    \draw[black,fill=black,opacity=1,line width = 0.05cm] (b6\ttl) to (n15\ttl);
    \draw[black,fill=black,opacity=1,line width = 0.05cm] (b6\ttl) to (n16\ttl);

    \draw[black,fill=black,opacity=1,line width = 0.05cm] (b7\ttl) to (n12\ttl);
    \draw[black,fill=black,opacity=1,line width = 0.05cm] (b7\ttl) to (n13\ttl);
    \draw[black,fill=black,opacity=1,line width = 0.05cm] (b7\ttl) to (n17\ttl);
    \draw[black,fill=black,opacity=1,line width = 0.05cm] (b7\ttl) to (n18\ttl);
    
    \draw[black,fill=black,opacity=1,line width = 0.05cm] (b8\ttl) to (n14\ttl);
    \draw[black,fill=black,opacity=1,line width = 0.05cm] (b8\ttl) to (n19\ttl);

    \draw[black,fill=black,opacity=1,line width = 0.05cm] (b9\ttl) to (n15\ttl);
    \draw[black,fill=black,opacity=1,line width = 0.05cm] (b9\ttl) to (n20\ttl);

    \draw[black,fill=black,opacity=1,line width = 0.05cm] (b10\ttl) to (n16\ttl);
    \draw[black,fill=black,opacity=1,line width = 0.05cm] (b10\ttl) to (n17\ttl);
    \draw[black,fill=black,opacity=1,line width = 0.05cm] (b10\ttl) to (n21\ttl);
    \draw[black,fill=black,opacity=1,line width = 0.05cm] (b10\ttl) to (n22\ttl);

    \draw[black,fill=black,opacity=1,line width = 0.05cm] (b11\ttl) to (n18\ttl);
    \draw[black,fill=black,opacity=1,line width = 0.05cm] (b11\ttl) to (n19\ttl);
    \draw[black,fill=black,opacity=1,line width = 0.05cm] (b11\ttl) to (n23\ttl);
    \draw[black,fill=black,opacity=1,line width = 0.05cm] (b11\ttl) to (n24\ttl);

    \end{scope}

    \def\hei{20}
    \def\ttl{t2}
    \begin{scope}
        
    \node[fill=zx_green,shape=circle,draw=black] (n4\ttl) at (2,18,\hei) {};

    \node[fill=zx_green,shape=circle,draw=black] (n3\ttl) at (2,14,\hei) {};
    \node[fill=zx_green,shape=circle,draw=black] (n2\ttl) at (2,10,\hei) {};
    \node[fill=zx_green,shape=circle,draw=black] (n1\ttl) at (2,6,\hei) {};
    \node[fill=zx_green,shape=circle,draw=black] (n0\ttl) at (2,2,\hei) {};

    \node[fill=zx_green,shape=circle,draw=black] (n9\ttl) at (6,18,\hei) {};
    \node[fill=zx_green,shape=circle,draw=black] (n8\ttl) at (6,14,\hei) {};
    \node[fill=zx_green,shape=circle,draw=black] (n7\ttl) at (6,10,\hei) {};
    \node[fill=zx_green,shape=circle,draw=black] (n6\ttl) at (6,6,\hei) {};
    \node[fill=zx_green,shape=circle,draw=black] (n5\ttl) at (6,2,\hei) {}; 

    \node[fill=zx_green,shape=circle,draw=black] (n14\ttl) at (10,18,\hei) {};
    \node[fill=zx_green,shape=circle,draw=black] (n13\ttl) at (10,14,\hei) {};
    \node[fill=zx_green,shape=circle,draw=black] (n12\ttl) at (10,10,\hei) {};
    \node[fill=zx_green,shape=circle,draw=black] (n11\ttl) at (10,6,\hei) {};
    \node[fill=zx_green,shape=circle,draw=black] (n10\ttl) at (10,2,\hei) {};

    \node[fill=zx_green,shape=circle,draw=black] (n19\ttl) at (14,18,\hei) {};
    \node[fill=zx_green,shape=circle,draw=black] (n18\ttl) at (14,14,\hei) {};
    \node[fill=zx_green,shape=circle,draw=black] (n17\ttl) at (14,10,\hei) {};
    \node[fill=zx_green,shape=circle,draw=black] (n16\ttl) at (14,6,\hei) {};
    \node[fill=zx_green,shape=circle,draw=black] (n15\ttl) at (14,2,\hei) {}; 

    \node[fill=zx_green,shape=circle,draw=black] (n24\ttl) at (18,18,\hei) {};
    \node[fill=zx_green,shape=circle,draw=black] (n23\ttl) at (18,14,\hei) {};
    \node[fill=zx_green,shape=circle,draw=black] (n22\ttl) at (18,10,\hei) {};
    \node[fill=zx_green,shape=circle,draw=black] (n21\ttl) at (18,6,\hei) {};
    \node[fill=zx_green,shape=circle,draw=black] (n20\ttl) at (18,2,\hei) {};
    
    \end{scope}
    \begin{scope}

        \node[fill=zx_red,circle,draw=black] (a0\ttl) at (0,4,\hei) {};
        \node[fill=zx_red,circle,draw=black] (a1\ttl) at (0,12,\hei) {};

        \node[fill=zx_red,circle,draw=black] (a2\ttl) at (4,8,\hei) {};
        \node[fill=zx_red,circle,draw=black] (a3\ttl) at (4,16,\hei) {};

        \node[fill=zx_red,circle,draw=black] (a4\ttl) at (8,4,\hei) {};
        \node[fill=zx_red,circle,draw=black] (a5\ttl) at (8,12,\hei) {};

        \node[fill=zx_red,circle,draw=black] (a6\ttl) at (12,8,\hei) {};
        \node[fill=zx_red,circle,draw=black] (a7\ttl) at (12,16,\hei) {};

        \node[fill=zx_red,circle,draw=black] (a8\ttl) at (16,4,\hei) {};
        \node[fill=zx_red,circle,draw=black] (a9\ttl) at (16,12,\hei) {};

        \node[fill=zx_red,circle,draw=black] (a10\ttl) at (20,8,\hei) {};
        \node[fill=zx_red,circle,draw=black] (a11\ttl) at (20,16,\hei) {};
    \end{scope}
    \begin{scope}
        
    \draw[black,fill=black,opacity=1,line width = 0.05cm] (a0\ttl) to (n0\ttl);
    \draw[black,fill=black,opacity=1,line width = 0.05cm] (a0\ttl) to (n1\ttl);

    \draw[black,fill=black,opacity=1,line width = 0.05cm] (a1\ttl) to (n2\ttl);
    \draw[black,fill=black,opacity=1,line width = 0.05cm] (a1\ttl) to (n3\ttl);

    \draw[black,fill=black,opacity=1,line width = 0.05cm] (a2\ttl) to (n1\ttl);
    \draw[black,fill=black,opacity=1,line width = 0.05cm] (a2\ttl) to (n2\ttl);
    \draw[black,fill=black,opacity=1,line width = 0.05cm] (a2\ttl) to (n6\ttl);
    \draw[black,fill=black,opacity=1,line width = 0.05cm] (a2\ttl) to (n7\ttl);

    \draw[black,fill=black,opacity=1,line width = 0.05cm] (a3\ttl) to (n3\ttl);
    \draw[black,fill=black,opacity=1,line width = 0.05cm] (a3\ttl) to (n4\ttl);
    \draw[black,fill=black,opacity=1,line width = 0.05cm] (a3\ttl) to (n8\ttl);
    \draw[black,fill=black,opacity=1,line width = 0.05cm] (a3\ttl) to (n9\ttl);

    \draw[black,fill=black,opacity=1,line width = 0.05cm] (a4\ttl) to (n5\ttl);
    \draw[black,fill=black,opacity=1,line width = 0.05cm] (a4\ttl) to (n6\ttl);
    \draw[black,fill=black,opacity=1,line width = 0.05cm] (a4\ttl) to (n10\ttl);
    \draw[black,fill=black,opacity=1,line width = 0.05cm] (a4\ttl) to (n11\ttl);
    
    \draw[black,fill=black,opacity=1,line width = 0.05cm] (a5\ttl) to (n7\ttl);
    \draw[black,fill=black,opacity=1,line width = 0.05cm] (a5\ttl) to (n8\ttl);
    \draw[black,fill=black,opacity=1,line width = 0.05cm] (a5\ttl) to (n12\ttl);
    \draw[black,fill=black,opacity=1,line width = 0.05cm] (a5\ttl) to (n13\ttl);

    \draw[black,fill=black,opacity=1,line width = 0.05cm] (a6\ttl) to (n11\ttl);
    \draw[black,fill=black,opacity=1,line width = 0.05cm] (a6\ttl) to (n12\ttl);
    \draw[black,fill=black,opacity=1,line width = 0.05cm] (a6\ttl) to (n16\ttl);
    \draw[black,fill=black,opacity=1,line width = 0.05cm] (a6\ttl) to (n17\ttl);

    \draw[black,fill=black,opacity=1,line width = 0.05cm] (a7\ttl) to (n13\ttl);
    \draw[black,fill=black,opacity=1,line width = 0.05cm] (a7\ttl) to (n14\ttl);
    \draw[black,fill=black,opacity=1,line width = 0.05cm] (a7\ttl) to (n18\ttl);
    \draw[black,fill=black,opacity=1,line width = 0.05cm] (a7\ttl) to (n19\ttl);

    \draw[black,fill=black,opacity=1,line width = 0.05cm] (a8\ttl) to (n15\ttl);
    \draw[black,fill=black,opacity=1,line width = 0.05cm] (a8\ttl) to (n16\ttl);
    \draw[black,fill=black,opacity=1,line width = 0.05cm] (a8\ttl) to (n20\ttl);
    \draw[black,fill=black,opacity=1,line width = 0.05cm] (a8\ttl) to (n21\ttl);

    \draw[black,fill=black,opacity=1,line width = 0.05cm] (a9\ttl) to (n17\ttl);
    \draw[black,fill=black,opacity=1,line width = 0.05cm] (a9\ttl) to (n18\ttl);
    \draw[black,fill=black,opacity=1,line width = 0.05cm] (a9\ttl) to (n22\ttl);
    \draw[black,fill=black,opacity=1,line width = 0.05cm] (a9\ttl) to (n23\ttl);
    
    \draw[black,fill=black,opacity=1,line width = 0.05cm] (a10\ttl) to (n21\ttl);
    \draw[black,fill=black,opacity=1,line width = 0.05cm] (a10\ttl) to (n22\ttl);

    \draw[black,fill=black,opacity=1,line width = 0.05cm] (a11\ttl) to (n23\ttl);
    \draw[black,fill=black,opacity=1,line width = 0.05cm] (a11\ttl) to (n24\ttl);

    \end{scope}

    \def\hei{22.5}
    \def\ttl{t3}
    \begin{scope}
        
    \node[fill=none,shape=circle,draw=none] (n4\ttl) at (2,18,\hei) {};

    \node[fill=none,shape=circle,draw=none] (n3\ttl) at (2,14,\hei) {};
    \node[fill=none,shape=circle,draw=none] (n2\ttl) at (2,10,\hei) {};
    \node[fill=none,shape=circle,draw=none] (n1\ttl) at (2,6,\hei) {};
    \node[fill=none,shape=circle,draw=none] (n0\ttl) at (2,2,\hei) {};

    \node[fill=none,shape=circle,draw=none] (n9\ttl) at (6,18,\hei) {};
    \node[fill=none,shape=circle,draw=none] (n8\ttl) at (6,14,\hei) {};
    \node[fill=none,shape=circle,draw=none] (n7\ttl) at (6,10,\hei) {};
    \node[fill=none,shape=circle,draw=none] (n6\ttl) at (6,6,\hei) {};
    \node[fill=none,shape=circle,draw=none] (n5\ttl) at (6,2,\hei) {}; 

    \node[fill=none,shape=circle,draw=none] (n14\ttl) at (10,18,\hei) {};
    \node[fill=none,shape=circle,draw=none] (n13\ttl) at (10,14,\hei) {};
    \node[fill=none,shape=circle,draw=none] (n12\ttl) at (10,10,\hei) {};
    \node[fill=none,shape=circle,draw=none] (n11\ttl) at (10,6,\hei) {};
    \node[fill=none,shape=circle,draw=none] (n10\ttl) at (10,2,\hei) {};

    \node[fill=none,shape=circle,draw=none] (n19\ttl) at (14,18,\hei) {};
    \node[fill=none,shape=circle,draw=none] (n18\ttl) at (14,14,\hei) {};
    \node[fill=none,shape=circle,draw=none] (n17\ttl) at (14,10,\hei) {};
    \node[fill=none,shape=circle,draw=none] (n16\ttl) at (14,6,\hei) {};
    \node[fill=none,shape=circle,draw=none] (n15\ttl) at (14,2,\hei) {}; 

    \node[fill=none,shape=circle,draw=none] (n24\ttl) at (18,18,\hei) {};
    \node[fill=none,shape=circle,draw=none] (n23\ttl) at (18,14,\hei) {};
    \node[fill=none,shape=circle,draw=none] (n22\ttl) at (18,10,\hei) {};
    \node[fill=none,shape=circle,draw=none] (n21\ttl) at (18,6,\hei) {};
    \node[fill=none,shape=circle,draw=none] (n20\ttl) at (18,2,\hei) {};
    
    \end{scope}

    \def\tlow{t0}
    \def\thig{t1}
    \begin{pgfonlayer}{background}
    \foreach \x in {0,...,24}
    {
    \draw[black,fill=black,opacity=1,line width = 0.05cm] (n\x\tlow) to (n\x\thig);
    }
    \end{pgfonlayer} 

    \def\tlow{t1}
    \def\thig{t2}
    \begin{pgfonlayer}{background}
    \foreach \x in {0,...,24}
    {
    \draw[black,fill=black,opacity=1,line width = 0.05cm] (n\x\tlow) to (n\x\thig);
    }
    \end{pgfonlayer}

    \def\tlow{t2}
    \def\thig{t3}
    \begin{pgfonlayer}{background}
    \foreach \x in {0,...,24}
    {
    \draw[black,fill=black,opacity=1,line width = 0.05cm] (n\x\tlow) to (n\x\thig);
    }
    \end{pgfonlayer}

    \def\tlow{t1}

    \def\tlow{t2}

    \end{tikzpicture}
    }
    \caption{\label{fig:y_injection_init_encoder}The surface code encoder circuit in figure \ref{fig:inj_encoder} applied to the initial states (figure \ref{fig:inj_1}).}
\end{figure}

The combined ZX-diagram with initialisation and encoder circuit is shown in figure \ref{fig:y_injection_init_encoder}. Multiple rounds of parity measurement circuit should be stacked on top repeatedly in time to perform multi-round syndrome extractions. For simplicity, we shall ignore the $2$ full rounds of parity measurements or the post-selection and distance growth specified in the Li/Lao--Criger schemes \cite{Li_2015,lao_criger_magic}, as a single round of parity measurement captures the essence when a Pauli web is constructed in an error-free circuit.

\section{Pauli web of the $\ket{Y}$ state injection scheme}
\label{sec:pauliweb_y}
Following the Pauli web notation of \cite{rodatz2024floquetifyingstabilisercodesdistancepreserving}, we start with the $\pi/2$ phase Z-spider in the ZX-diagram (figure \ref{fig:y_injection_init_encoder}) and choose definition 2.7 from \cite{rodatz2024floquetifyingstabilisercodesdistancepreserving}:

\begin{webrulebox}
`a spider with $\pm\frac{\pi}{2}$ phase can have ... an odd number of legs highlighted in its own colour \textbf{and} all legs highlighted in the
opposite colour'. 
\end{webrulebox}

\noindent For simplicity, we post-select onto the $+1$ parity measurement outcome throughout, and the bolded classical measurement wires of \cite{Bombin_2024} are not drawn: a measurement effect is a one-legged spider of the same colour as its ancilla, so it fuses into that ancilla and vanishes.
\\

\noindent This leads to a multi-colour Pauli web decorated ZX-diagram as shown in figure \ref{fig:y_injection_first_pauli_web}. 

\begin{figure}[!h]
    \centering     
    \tdplotsetmaincoords{70}{23}
    \subfloat[][\label{fig:y_injection_first_pauli_web}The first red and green decorated edge in the ZX-diagram.]{
    \resizebox{0.4\linewidth}{!}{
   
    \begin{tikzpicture}[tdplot_main_coords,every node/.style={minimum size=1cm}]
    \def\hei{0}
    \def\ttl{t0}
    \begin{scope}
        
    \node[fill=zx_green,shape=circle,draw=black] (n4\ttl) at (2,18,\hei) {$\pi/2$};
    \node[fill=zx_green,shape=circle,draw=black] (n3\ttl) at (2,14,\hei) {};
    \node[fill=zx_green,shape=circle,draw=black] (n2\ttl) at (2,10,\hei) {};
    \node[fill=zx_green,shape=circle,draw=black] (n1\ttl) at (2,6,\hei) {};
    \node[fill=zx_green,shape=circle,draw=black] (n0\ttl) at (2,2,\hei) {};

    \node[fill=zx_red,shape=circle,draw=black] (n9\ttl) at (6,18,\hei) {};
    \node[fill=zx_green,shape=circle,draw=black] (n8\ttl) at (6,14,\hei) {};
    \node[fill=zx_green,shape=circle,draw=black] (n7\ttl) at (6,10,\hei) {};
    \node[fill=zx_green,shape=circle,draw=black] (n6\ttl) at (6,6,\hei) {};
    \node[fill=zx_green,shape=circle,draw=black] (n5\ttl) at (6,2,\hei) {}; 

    \node[fill=zx_red,shape=circle,draw=black] (n14\ttl) at (10,18,\hei) {};
    \node[fill=zx_red,shape=circle,draw=black] (n13\ttl) at (10,14,\hei) {};
    \node[fill=zx_green,shape=circle,draw=black] (n12\ttl) at (10,10,\hei) {};
    \node[fill=zx_green,shape=circle,draw=black] (n11\ttl) at (10,6,\hei) {};
    \node[fill=zx_green,shape=circle,draw=black] (n10\ttl) at (10,2,\hei) {};

    \node[fill=zx_red,shape=circle,draw=black] (n19\ttl) at (14,18,\hei) {};
    \node[fill=zx_red,shape=circle,draw=black] (n18\ttl) at (14,14,\hei) {};
    \node[fill=zx_red,shape=circle,draw=black] (n17\ttl) at (14,10,\hei) {};
    \node[fill=zx_green,shape=circle,draw=black] (n16\ttl) at (14,6,\hei) {};
    \node[fill=zx_green,shape=circle,draw=black] (n15\ttl) at (14,2,\hei) {}; 

    \node[fill=zx_red,shape=circle,draw=black] (n24\ttl) at (18,18,\hei) {};
    \node[fill=zx_red,shape=circle,draw=black] (n23\ttl) at (18,14,\hei) {};
    \node[fill=zx_red,shape=circle,draw=black] (n22\ttl) at (18,10,\hei) {};
    \node[fill=zx_red,shape=circle,draw=black] (n21\ttl) at (18,6,\hei) {};
    \node[fill=zx_green,shape=circle,draw=black] (n20\ttl) at (18,2,\hei) {};
    
    \end{scope} 

    \def\hei{10}
    \def\ttl{t1}
    \begin{scope}
        
    \node[fill=zx_red,shape=circle,draw=black] (n4\ttl) at (2,18,\hei) {};

    \node[fill=zx_red,shape=circle,draw=black] (n3\ttl) at (2,14,\hei) {};
    \node[fill=zx_red,shape=circle,draw=black] (n2\ttl) at (2,10,\hei) {};
    \node[fill=zx_red,shape=circle,draw=black] (n1\ttl) at (2,6,\hei) {};
    \node[fill=zx_red,shape=circle,draw=black] (n0\ttl) at (2,2,\hei) {};

    \node[fill=zx_red,shape=circle,draw=black] (n9\ttl) at (6,18,\hei) {};
    \node[fill=zx_red,shape=circle,draw=black] (n8\ttl) at (6,14,\hei) {};
    \node[fill=zx_red,shape=circle,draw=black] (n7\ttl) at (6,10,\hei) {};
    \node[fill=zx_red,shape=circle,draw=black] (n6\ttl) at (6,6,\hei) {};
    \node[fill=zx_red,shape=circle,draw=black] (n5\ttl) at (6,2,\hei) {}; 

    \node[fill=zx_red,shape=circle,draw=black] (n14\ttl) at (10,18,\hei) {};
    \node[fill=zx_red,shape=circle,draw=black] (n13\ttl) at (10,14,\hei) {};
    \node[fill=zx_red,shape=circle,draw=black] (n12\ttl) at (10,10,\hei) {};
    \node[fill=zx_red,shape=circle,draw=black] (n11\ttl) at (10,6,\hei) {};
    \node[fill=zx_red,shape=circle,draw=black] (n10\ttl) at (10,2,\hei) {};

    \node[fill=zx_red,shape=circle,draw=black] (n19\ttl) at (14,18,\hei) {};
    \node[fill=zx_red,shape=circle,draw=black] (n18\ttl) at (14,14,\hei) {};
    \node[fill=zx_red,shape=circle,draw=black] (n17\ttl) at (14,10,\hei) {};
    \node[fill=zx_red,shape=circle,draw=black] (n16\ttl) at (14,6,\hei) {};
    \node[fill=zx_red,shape=circle,draw=black] (n15\ttl) at (14,2,\hei) {}; 

    \node[fill=zx_red,shape=circle,draw=black] (n24\ttl) at (18,18,\hei) {};
    \node[fill=zx_red,shape=circle,draw=black] (n23\ttl) at (18,14,\hei) {};
    \node[fill=zx_red,shape=circle,draw=black] (n22\ttl) at (18,10,\hei) {};
    \node[fill=zx_red,shape=circle,draw=black] (n21\ttl) at (18,6,\hei) {};
    \node[fill=zx_red,shape=circle,draw=black] (n20\ttl) at (18,2,\hei) {};
    
    \end{scope}
    \begin{scope}
        \node[fill=zx_green,circle,draw=black] (b0\ttl) at (4,4,\hei) {};
        \node[fill=zx_green,circle,draw=black] (b1\ttl) at (4,12,\hei) {};
        \node[fill=zx_green,circle,draw=black] (b2\ttl) at (4,20,\hei) {};

        \node[fill=zx_green,circle,draw=black] (b3\ttl) at (8,0,\hei) {};
        \node[fill=zx_green,circle,draw=black] (b4\ttl) at (8,8,\hei) {};
        \node[fill=zx_green,circle,draw=black] (b5\ttl) at (8,16,\hei) {};

        \node[fill=zx_green,circle,draw=black] (b6\ttl) at (12,4,\hei) {};
        \node[fill=zx_green,circle,draw=black] (b7\ttl) at (12,12,\hei) {};
        \node[fill=zx_green,circle,draw=black] (b8\ttl) at (12,20,\hei) {};

        \node[fill=zx_green,circle,draw=black] (b9\ttl) at (16,0,\hei) {};
        \node[fill=zx_green,circle,draw=black] (b10\ttl) at (16,8,\hei) {};
        \node[fill=zx_green,circle,draw=black] (b11\ttl) at (16,16,\hei) {};

    \end{scope}
    \begin{scope}
        
    \draw[black,fill=black,opacity=1,line width = 0.05cm] (b0\ttl) to (n0\ttl);
    \draw[black,fill=black,opacity=1,line width = 0.05cm] (b0\ttl) to (n1\ttl);
    \draw[black,fill=black,opacity=1,line width = 0.05cm] (b0\ttl) to (n5\ttl);
    \draw[black,fill=black,opacity=1,line width = 0.05cm] (b0\ttl) to (n6\ttl);

    \draw[black,fill=black,opacity=1,line width = 0.05cm] (b1\ttl) to (n2\ttl);
    \draw[black,fill=black,opacity=1,line width = 0.05cm] (b1\ttl) to (n3\ttl);
    \draw[black,fill=black,opacity=1,line width = 0.05cm] (b1\ttl) to (n7\ttl);
    \draw[black,fill=black,opacity=1,line width = 0.05cm] (b1\ttl) to (n8\ttl);
    
    \draw[black,fill=black,opacity=1,line width = 0.05cm] (b2\ttl) to (n4\ttl);
    \draw[black,fill=black,opacity=1,line width = 0.05cm] (b2\ttl) to (n9\ttl);
    
    \draw[black,fill=black,opacity=1,line width = 0.05cm] (b3\ttl) to (n5\ttl);
    \draw[black,fill=black,opacity=1,line width = 0.05cm] (b3\ttl) to (n10\ttl);

    \draw[black,fill=black,opacity=1,line width = 0.05cm] (b4\ttl) to (n6\ttl);
    \draw[black,fill=black,opacity=1,line width = 0.05cm] (b4\ttl) to (n7\ttl);
    \draw[black,fill=black,opacity=1,line width = 0.05cm] (b4\ttl) to (n11\ttl);
    \draw[black,fill=black,opacity=1,line width = 0.05cm] (b4\ttl) to (n12\ttl);

    \draw[black,fill=black,opacity=1,line width = 0.05cm] (b5\ttl) to (n8\ttl);
    \draw[black,fill=black,opacity=1,line width = 0.05cm] (b5\ttl) to (n9\ttl);
    \draw[black,fill=black,opacity=1,line width = 0.05cm] (b5\ttl) to (n13\ttl);
    \draw[black,fill=black,opacity=1,line width = 0.05cm] (b5\ttl) to (n14\ttl);

    \draw[black,fill=black,opacity=1,line width = 0.05cm] (b6\ttl) to (n10\ttl);
    \draw[black,fill=black,opacity=1,line width = 0.05cm] (b6\ttl) to (n11\ttl);
    \draw[black,fill=black,opacity=1,line width = 0.05cm] (b6\ttl) to (n15\ttl);
    \draw[black,fill=black,opacity=1,line width = 0.05cm] (b6\ttl) to (n16\ttl);

    \draw[black,fill=black,opacity=1,line width = 0.05cm] (b7\ttl) to (n12\ttl);
    \draw[black,fill=black,opacity=1,line width = 0.05cm] (b7\ttl) to (n13\ttl);
    \draw[black,fill=black,opacity=1,line width = 0.05cm] (b7\ttl) to (n17\ttl);
    \draw[black,fill=black,opacity=1,line width = 0.05cm] (b7\ttl) to (n18\ttl);
    
    \draw[black,fill=black,opacity=1,line width = 0.05cm] (b8\ttl) to (n14\ttl);
    \draw[black,fill=black,opacity=1,line width = 0.05cm] (b8\ttl) to (n19\ttl);

    \draw[black,fill=black,opacity=1,line width = 0.05cm] (b9\ttl) to (n15\ttl);
    \draw[black,fill=black,opacity=1,line width = 0.05cm] (b9\ttl) to (n20\ttl);

    \draw[black,fill=black,opacity=1,line width = 0.05cm] (b10\ttl) to (n16\ttl);
    \draw[black,fill=black,opacity=1,line width = 0.05cm] (b10\ttl) to (n17\ttl);
    \draw[black,fill=black,opacity=1,line width = 0.05cm] (b10\ttl) to (n21\ttl);
    \draw[black,fill=black,opacity=1,line width = 0.05cm] (b10\ttl) to (n22\ttl);

    \draw[black,fill=black,opacity=1,line width = 0.05cm] (b11\ttl) to (n18\ttl);
    \draw[black,fill=black,opacity=1,line width = 0.05cm] (b11\ttl) to (n19\ttl);
    \draw[black,fill=black,opacity=1,line width = 0.05cm] (b11\ttl) to (n23\ttl);
    \draw[black,fill=black,opacity=1,line width = 0.05cm] (b11\ttl) to (n24\ttl);

    \end{scope}

    \def\hei{20}
    \def\ttl{t2}
    \begin{scope}
        
    \node[fill=zx_green,shape=circle,draw=black] (n4\ttl) at (2,18,\hei) {};

    \node[fill=zx_green,shape=circle,draw=black] (n3\ttl) at (2,14,\hei) {};
    \node[fill=zx_green,shape=circle,draw=black] (n2\ttl) at (2,10,\hei) {};
    \node[fill=zx_green,shape=circle,draw=black] (n1\ttl) at (2,6,\hei) {};
    \node[fill=zx_green,shape=circle,draw=black] (n0\ttl) at (2,2,\hei) {};

    \node[fill=zx_green,shape=circle,draw=black] (n9\ttl) at (6,18,\hei) {};
    \node[fill=zx_green,shape=circle,draw=black] (n8\ttl) at (6,14,\hei) {};
    \node[fill=zx_green,shape=circle,draw=black] (n7\ttl) at (6,10,\hei) {};
    \node[fill=zx_green,shape=circle,draw=black] (n6\ttl) at (6,6,\hei) {};
    \node[fill=zx_green,shape=circle,draw=black] (n5\ttl) at (6,2,\hei) {}; 

    \node[fill=zx_green,shape=circle,draw=black] (n14\ttl) at (10,18,\hei) {};
    \node[fill=zx_green,shape=circle,draw=black] (n13\ttl) at (10,14,\hei) {};
    \node[fill=zx_green,shape=circle,draw=black] (n12\ttl) at (10,10,\hei) {};
    \node[fill=zx_green,shape=circle,draw=black] (n11\ttl) at (10,6,\hei) {};
    \node[fill=zx_green,shape=circle,draw=black] (n10\ttl) at (10,2,\hei) {};

    \node[fill=zx_green,shape=circle,draw=black] (n19\ttl) at (14,18,\hei) {};
    \node[fill=zx_green,shape=circle,draw=black] (n18\ttl) at (14,14,\hei) {};
    \node[fill=zx_green,shape=circle,draw=black] (n17\ttl) at (14,10,\hei) {};
    \node[fill=zx_green,shape=circle,draw=black] (n16\ttl) at (14,6,\hei) {};
    \node[fill=zx_green,shape=circle,draw=black] (n15\ttl) at (14,2,\hei) {}; 

    \node[fill=zx_green,shape=circle,draw=black] (n24\ttl) at (18,18,\hei) {};
    \node[fill=zx_green,shape=circle,draw=black] (n23\ttl) at (18,14,\hei) {};
    \node[fill=zx_green,shape=circle,draw=black] (n22\ttl) at (18,10,\hei) {};
    \node[fill=zx_green,shape=circle,draw=black] (n21\ttl) at (18,6,\hei) {};
    \node[fill=zx_green,shape=circle,draw=black] (n20\ttl) at (18,2,\hei) {};
    
    \end{scope}
    \begin{scope}

        \node[fill=zx_red,circle,draw=black] (a0\ttl) at (0,4,\hei) {};
        \node[fill=zx_red,circle,draw=black] (a1\ttl) at (0,12,\hei) {};

        \node[fill=zx_red,circle,draw=black] (a2\ttl) at (4,8,\hei) {};
        \node[fill=zx_red,circle,draw=black] (a3\ttl) at (4,16,\hei) {};

        \node[fill=zx_red,circle,draw=black] (a4\ttl) at (8,4,\hei) {};
        \node[fill=zx_red,circle,draw=black] (a5\ttl) at (8,12,\hei) {};

        \node[fill=zx_red,circle,draw=black] (a6\ttl) at (12,8,\hei) {};
        \node[fill=zx_red,circle,draw=black] (a7\ttl) at (12,16,\hei) {};

        \node[fill=zx_red,circle,draw=black] (a8\ttl) at (16,4,\hei) {};
        \node[fill=zx_red,circle,draw=black] (a9\ttl) at (16,12,\hei) {};

        \node[fill=zx_red,circle,draw=black] (a10\ttl) at (20,8,\hei) {};
        \node[fill=zx_red,circle,draw=black] (a11\ttl) at (20,16,\hei) {};
    \end{scope}
    \begin{scope}
        
    \draw[black,fill=black,opacity=1,line width = 0.05cm] (a0\ttl) to (n0\ttl);
    \draw[black,fill=black,opacity=1,line width = 0.05cm] (a0\ttl) to (n1\ttl);

    \draw[black,fill=black,opacity=1,line width = 0.05cm] (a1\ttl) to (n2\ttl);
    \draw[black,fill=black,opacity=1,line width = 0.05cm] (a1\ttl) to (n3\ttl);

    \draw[black,fill=black,opacity=1,line width = 0.05cm] (a2\ttl) to (n1\ttl);
    \draw[black,fill=black,opacity=1,line width = 0.05cm] (a2\ttl) to (n2\ttl);
    \draw[black,fill=black,opacity=1,line width = 0.05cm] (a2\ttl) to (n6\ttl);
    \draw[black,fill=black,opacity=1,line width = 0.05cm] (a2\ttl) to (n7\ttl);

    \draw[black,fill=black,opacity=1,line width = 0.05cm] (a3\ttl) to (n3\ttl);
    \draw[black,fill=black,opacity=1,line width = 0.05cm] (a3\ttl) to (n4\ttl);
    \draw[black,fill=black,opacity=1,line width = 0.05cm] (a3\ttl) to (n8\ttl);
    \draw[black,fill=black,opacity=1,line width = 0.05cm] (a3\ttl) to (n9\ttl);

    \draw[black,fill=black,opacity=1,line width = 0.05cm] (a4\ttl) to (n5\ttl);
    \draw[black,fill=black,opacity=1,line width = 0.05cm] (a4\ttl) to (n6\ttl);
    \draw[black,fill=black,opacity=1,line width = 0.05cm] (a4\ttl) to (n10\ttl);
    \draw[black,fill=black,opacity=1,line width = 0.05cm] (a4\ttl) to (n11\ttl);
    
    \draw[black,fill=black,opacity=1,line width = 0.05cm] (a5\ttl) to (n7\ttl);
    \draw[black,fill=black,opacity=1,line width = 0.05cm] (a5\ttl) to (n8\ttl);
    \draw[black,fill=black,opacity=1,line width = 0.05cm] (a5\ttl) to (n12\ttl);
    \draw[black,fill=black,opacity=1,line width = 0.05cm] (a5\ttl) to (n13\ttl);

    \draw[black,fill=black,opacity=1,line width = 0.05cm] (a6\ttl) to (n11\ttl);
    \draw[black,fill=black,opacity=1,line width = 0.05cm] (a6\ttl) to (n12\ttl);
    \draw[black,fill=black,opacity=1,line width = 0.05cm] (a6\ttl) to (n16\ttl);
    \draw[black,fill=black,opacity=1,line width = 0.05cm] (a6\ttl) to (n17\ttl);

    \draw[black,fill=black,opacity=1,line width = 0.05cm] (a7\ttl) to (n13\ttl);
    \draw[black,fill=black,opacity=1,line width = 0.05cm] (a7\ttl) to (n14\ttl);
    \draw[black,fill=black,opacity=1,line width = 0.05cm] (a7\ttl) to (n18\ttl);
    \draw[black,fill=black,opacity=1,line width = 0.05cm] (a7\ttl) to (n19\ttl);

    \draw[black,fill=black,opacity=1,line width = 0.05cm] (a8\ttl) to (n15\ttl);
    \draw[black,fill=black,opacity=1,line width = 0.05cm] (a8\ttl) to (n16\ttl);
    \draw[black,fill=black,opacity=1,line width = 0.05cm] (a8\ttl) to (n20\ttl);
    \draw[black,fill=black,opacity=1,line width = 0.05cm] (a8\ttl) to (n21\ttl);

    \draw[black,fill=black,opacity=1,line width = 0.05cm] (a9\ttl) to (n17\ttl);
    \draw[black,fill=black,opacity=1,line width = 0.05cm] (a9\ttl) to (n18\ttl);
    \draw[black,fill=black,opacity=1,line width = 0.05cm] (a9\ttl) to (n22\ttl);
    \draw[black,fill=black,opacity=1,line width = 0.05cm] (a9\ttl) to (n23\ttl);
    
    \draw[black,fill=black,opacity=1,line width = 0.05cm] (a10\ttl) to (n21\ttl);
    \draw[black,fill=black,opacity=1,line width = 0.05cm] (a10\ttl) to (n22\ttl);

    \draw[black,fill=black,opacity=1,line width = 0.05cm] (a11\ttl) to (n23\ttl);
    \draw[black,fill=black,opacity=1,line width = 0.05cm] (a11\ttl) to (n24\ttl);

    \end{scope}

    \def\hei{22.5}
    \def\ttl{t3}
    \begin{scope}
        
    \node[fill=none,shape=circle,draw=none] (n4\ttl) at (2,18,\hei) {};

    \node[fill=none,shape=circle,draw=none] (n3\ttl) at (2,14,\hei) {};
    \node[fill=none,shape=circle,draw=none] (n2\ttl) at (2,10,\hei) {};
    \node[fill=none,shape=circle,draw=none] (n1\ttl) at (2,6,\hei) {};
    \node[fill=none,shape=circle,draw=none] (n0\ttl) at (2,2,\hei) {};

    \node[fill=none,shape=circle,draw=none] (n9\ttl) at (6,18,\hei) {};
    \node[fill=none,shape=circle,draw=none] (n8\ttl) at (6,14,\hei) {};
    \node[fill=none,shape=circle,draw=none] (n7\ttl) at (6,10,\hei) {};
    \node[fill=none,shape=circle,draw=none] (n6\ttl) at (6,6,\hei) {};
    \node[fill=none,shape=circle,draw=none] (n5\ttl) at (6,2,\hei) {}; 

    \node[fill=none,shape=circle,draw=none] (n14\ttl) at (10,18,\hei) {};
    \node[fill=none,shape=circle,draw=none] (n13\ttl) at (10,14,\hei) {};
    \node[fill=none,shape=circle,draw=none] (n12\ttl) at (10,10,\hei) {};
    \node[fill=none,shape=circle,draw=none] (n11\ttl) at (10,6,\hei) {};
    \node[fill=none,shape=circle,draw=none] (n10\ttl) at (10,2,\hei) {};

    \node[fill=none,shape=circle,draw=none] (n19\ttl) at (14,18,\hei) {};
    \node[fill=none,shape=circle,draw=none] (n18\ttl) at (14,14,\hei) {};
    \node[fill=none,shape=circle,draw=none] (n17\ttl) at (14,10,\hei) {};
    \node[fill=none,shape=circle,draw=none] (n16\ttl) at (14,6,\hei) {};
    \node[fill=none,shape=circle,draw=none] (n15\ttl) at (14,2,\hei) {}; 

    \node[fill=none,shape=circle,draw=none] (n24\ttl) at (18,18,\hei) {};
    \node[fill=none,shape=circle,draw=none] (n23\ttl) at (18,14,\hei) {};
    \node[fill=none,shape=circle,draw=none] (n22\ttl) at (18,10,\hei) {};
    \node[fill=none,shape=circle,draw=none] (n21\ttl) at (18,6,\hei) {};
    \node[fill=none,shape=circle,draw=none] (n20\ttl) at (18,2,\hei) {};
    
    \end{scope}

    \def\tlow{t0}
    \def\thig{t1}
    \begin{pgfonlayer}{background}
    \foreach \x in {0,...,24}
    {
    \draw[black,fill=black,opacity=1,line width = 0.05cm] (n\x\tlow) to (n\x\thig);
    }
    \end{pgfonlayer} 

    \def\tlow{t1}
    \def\thig{t2}
    \begin{pgfonlayer}{background}
    \foreach \x in {0,...,24}
    {
    \draw[black,fill=black,opacity=1,line width = 0.05cm] (n\x\tlow) to (n\x\thig);
    }
    \end{pgfonlayer}

    \def\tlow{t2}
    \def\thig{t3}
    \begin{pgfonlayer}{background}
    \foreach \x in {0,...,24}
    {
    \draw[black,fill=black,opacity=1,line width = 0.05cm] (n\x\tlow) to (n\x\thig);
    }
    \end{pgfonlayer}

    \def\tlow{t1}

    \def\tlow{t2}

    \def\tlow{t0}
    \def\thig{t1}
    \def\thhig{t2}
    \def\thhhig{t3}
    \begin{pgfonlayer}{background}
    \draw[green,fill=green,opacity=0.8,line width = 0.45cm] (n4\tlow) to (n4\thig);
    \draw[red,fill=red,opacity=0.6,line width = 0.25cm] (n4\tlow) to (n4\thig);
    
    \end{pgfonlayer}

    \end{tikzpicture}
    }
    }
    \qquad
    \subfloat[][\label{fig:y_injection_first_pauli_web_green}All the green decorated Pauli web shown.]{
        \resizebox{0.4\linewidth}{!}{
   
    \begin{tikzpicture}[tdplot_main_coords,every node/.style={minimum size=1cm}]
    \def\hei{0}
    \def\ttl{t0}
    \begin{scope}
        
    \node[fill=zx_green,shape=circle,draw=black] (n4\ttl) at (2,18,\hei) {$\pi/2$};
    \node[fill=zx_green,shape=circle,draw=black] (n3\ttl) at (2,14,\hei) {};
    \node[fill=zx_green,shape=circle,draw=black] (n2\ttl) at (2,10,\hei) {};
    \node[fill=zx_green,shape=circle,draw=black] (n1\ttl) at (2,6,\hei) {};
    \node[fill=zx_green,shape=circle,draw=black] (n0\ttl) at (2,2,\hei) {};

    \node[fill=zx_red,shape=circle,draw=black] (n9\ttl) at (6,18,\hei) {};
    \node[fill=zx_green,shape=circle,draw=black] (n8\ttl) at (6,14,\hei) {};
    \node[fill=zx_green,shape=circle,draw=black] (n7\ttl) at (6,10,\hei) {};
    \node[fill=zx_green,shape=circle,draw=black] (n6\ttl) at (6,6,\hei) {};
    \node[fill=zx_green,shape=circle,draw=black] (n5\ttl) at (6,2,\hei) {}; 

    \node[fill=zx_red,shape=circle,draw=black] (n14\ttl) at (10,18,\hei) {};
    \node[fill=zx_red,shape=circle,draw=black] (n13\ttl) at (10,14,\hei) {};
    \node[fill=zx_green,shape=circle,draw=black] (n12\ttl) at (10,10,\hei) {};
    \node[fill=zx_green,shape=circle,draw=black] (n11\ttl) at (10,6,\hei) {};
    \node[fill=zx_green,shape=circle,draw=black] (n10\ttl) at (10,2,\hei) {};

    \node[fill=zx_red,shape=circle,draw=black] (n19\ttl) at (14,18,\hei) {};
    \node[fill=zx_red,shape=circle,draw=black] (n18\ttl) at (14,14,\hei) {};
    \node[fill=zx_red,shape=circle,draw=black] (n17\ttl) at (14,10,\hei) {};
    \node[fill=zx_green,shape=circle,draw=black] (n16\ttl) at (14,6,\hei) {};
    \node[fill=zx_green,shape=circle,draw=black] (n15\ttl) at (14,2,\hei) {}; 

    \node[fill=zx_red,shape=circle,draw=black] (n24\ttl) at (18,18,\hei) {};
    \node[fill=zx_red,shape=circle,draw=black] (n23\ttl) at (18,14,\hei) {};
    \node[fill=zx_red,shape=circle,draw=black] (n22\ttl) at (18,10,\hei) {};
    \node[fill=zx_red,shape=circle,draw=black] (n21\ttl) at (18,6,\hei) {};
    \node[fill=zx_green,shape=circle,draw=black] (n20\ttl) at (18,2,\hei) {};
    
    \end{scope} 

    \def\hei{10}
    \def\ttl{t1}
    \begin{scope}
        
    \node[fill=zx_red,shape=circle,draw=black] (n4\ttl) at (2,18,\hei) {};

    \node[fill=zx_red,shape=circle,draw=black] (n3\ttl) at (2,14,\hei) {};
    \node[fill=zx_red,shape=circle,draw=black] (n2\ttl) at (2,10,\hei) {};
    \node[fill=zx_red,shape=circle,draw=black] (n1\ttl) at (2,6,\hei) {};
    \node[fill=zx_red,shape=circle,draw=black] (n0\ttl) at (2,2,\hei) {};

    \node[fill=zx_red,shape=circle,draw=black] (n9\ttl) at (6,18,\hei) {};
    \node[fill=zx_red,shape=circle,draw=black] (n8\ttl) at (6,14,\hei) {};
    \node[fill=zx_red,shape=circle,draw=black] (n7\ttl) at (6,10,\hei) {};
    \node[fill=zx_red,shape=circle,draw=black] (n6\ttl) at (6,6,\hei) {};
    \node[fill=zx_red,shape=circle,draw=black] (n5\ttl) at (6,2,\hei) {}; 

    \node[fill=zx_red,shape=circle,draw=black] (n14\ttl) at (10,18,\hei) {};
    \node[fill=zx_red,shape=circle,draw=black] (n13\ttl) at (10,14,\hei) {};
    \node[fill=zx_red,shape=circle,draw=black] (n12\ttl) at (10,10,\hei) {};
    \node[fill=zx_red,shape=circle,draw=black] (n11\ttl) at (10,6,\hei) {};
    \node[fill=zx_red,shape=circle,draw=black] (n10\ttl) at (10,2,\hei) {};

    \node[fill=zx_red,shape=circle,draw=black] (n19\ttl) at (14,18,\hei) {};
    \node[fill=zx_red,shape=circle,draw=black] (n18\ttl) at (14,14,\hei) {};
    \node[fill=zx_red,shape=circle,draw=black] (n17\ttl) at (14,10,\hei) {};
    \node[fill=zx_red,shape=circle,draw=black] (n16\ttl) at (14,6,\hei) {};
    \node[fill=zx_red,shape=circle,draw=black] (n15\ttl) at (14,2,\hei) {}; 

    \node[fill=zx_red,shape=circle,draw=black] (n24\ttl) at (18,18,\hei) {};
    \node[fill=zx_red,shape=circle,draw=black] (n23\ttl) at (18,14,\hei) {};
    \node[fill=zx_red,shape=circle,draw=black] (n22\ttl) at (18,10,\hei) {};
    \node[fill=zx_red,shape=circle,draw=black] (n21\ttl) at (18,6,\hei) {};
    \node[fill=zx_red,shape=circle,draw=black] (n20\ttl) at (18,2,\hei) {};
    
    \end{scope}
    \begin{scope}
        \node[fill=zx_green,circle,draw=black] (b0\ttl) at (4,4,\hei) {};
        \node[fill=zx_green,circle,draw=black] (b1\ttl) at (4,12,\hei) {};
        \node[fill=zx_green,circle,draw=black] (b2\ttl) at (4,20,\hei) {};

        \node[fill=zx_green,circle,draw=black] (b3\ttl) at (8,0,\hei) {};
        \node[fill=zx_green,circle,draw=black] (b4\ttl) at (8,8,\hei) {};
        \node[fill=zx_green,circle,draw=black] (b5\ttl) at (8,16,\hei) {};

        \node[fill=zx_green,circle,draw=black] (b6\ttl) at (12,4,\hei) {};
        \node[fill=zx_green,circle,draw=black] (b7\ttl) at (12,12,\hei) {};
        \node[fill=zx_green,circle,draw=black] (b8\ttl) at (12,20,\hei) {};

        \node[fill=zx_green,circle,draw=black] (b9\ttl) at (16,0,\hei) {};
        \node[fill=zx_green,circle,draw=black] (b10\ttl) at (16,8,\hei) {};
        \node[fill=zx_green,circle,draw=black] (b11\ttl) at (16,16,\hei) {};

    \end{scope}
    \begin{scope}
        
    \draw[black,fill=black,opacity=1,line width = 0.05cm] (b0\ttl) to (n0\ttl);
    \draw[black,fill=black,opacity=1,line width = 0.05cm] (b0\ttl) to (n1\ttl);
    \draw[black,fill=black,opacity=1,line width = 0.05cm] (b0\ttl) to (n5\ttl);
    \draw[black,fill=black,opacity=1,line width = 0.05cm] (b0\ttl) to (n6\ttl);

    \draw[black,fill=black,opacity=1,line width = 0.05cm] (b1\ttl) to (n2\ttl);
    \draw[black,fill=black,opacity=1,line width = 0.05cm] (b1\ttl) to (n3\ttl);
    \draw[black,fill=black,opacity=1,line width = 0.05cm] (b1\ttl) to (n7\ttl);
    \draw[black,fill=black,opacity=1,line width = 0.05cm] (b1\ttl) to (n8\ttl);
    
    \draw[black,fill=black,opacity=1,line width = 0.05cm] (b2\ttl) to (n4\ttl);
    \draw[black,fill=black,opacity=1,line width = 0.05cm] (b2\ttl) to (n9\ttl);
    
    \draw[black,fill=black,opacity=1,line width = 0.05cm] (b3\ttl) to (n5\ttl);
    \draw[black,fill=black,opacity=1,line width = 0.05cm] (b3\ttl) to (n10\ttl);

    \draw[black,fill=black,opacity=1,line width = 0.05cm] (b4\ttl) to (n6\ttl);
    \draw[black,fill=black,opacity=1,line width = 0.05cm] (b4\ttl) to (n7\ttl);
    \draw[black,fill=black,opacity=1,line width = 0.05cm] (b4\ttl) to (n11\ttl);
    \draw[black,fill=black,opacity=1,line width = 0.05cm] (b4\ttl) to (n12\ttl);

    \draw[black,fill=black,opacity=1,line width = 0.05cm] (b5\ttl) to (n8\ttl);
    \draw[black,fill=black,opacity=1,line width = 0.05cm] (b5\ttl) to (n9\ttl);
    \draw[black,fill=black,opacity=1,line width = 0.05cm] (b5\ttl) to (n13\ttl);
    \draw[black,fill=black,opacity=1,line width = 0.05cm] (b5\ttl) to (n14\ttl);

    \draw[black,fill=black,opacity=1,line width = 0.05cm] (b6\ttl) to (n10\ttl);
    \draw[black,fill=black,opacity=1,line width = 0.05cm] (b6\ttl) to (n11\ttl);
    \draw[black,fill=black,opacity=1,line width = 0.05cm] (b6\ttl) to (n15\ttl);
    \draw[black,fill=black,opacity=1,line width = 0.05cm] (b6\ttl) to (n16\ttl);

    \draw[black,fill=black,opacity=1,line width = 0.05cm] (b7\ttl) to (n12\ttl);
    \draw[black,fill=black,opacity=1,line width = 0.05cm] (b7\ttl) to (n13\ttl);
    \draw[black,fill=black,opacity=1,line width = 0.05cm] (b7\ttl) to (n17\ttl);
    \draw[black,fill=black,opacity=1,line width = 0.05cm] (b7\ttl) to (n18\ttl);
    
    \draw[black,fill=black,opacity=1,line width = 0.05cm] (b8\ttl) to (n14\ttl);
    \draw[black,fill=black,opacity=1,line width = 0.05cm] (b8\ttl) to (n19\ttl);

    \draw[black,fill=black,opacity=1,line width = 0.05cm] (b9\ttl) to (n15\ttl);
    \draw[black,fill=black,opacity=1,line width = 0.05cm] (b9\ttl) to (n20\ttl);

    \draw[black,fill=black,opacity=1,line width = 0.05cm] (b10\ttl) to (n16\ttl);
    \draw[black,fill=black,opacity=1,line width = 0.05cm] (b10\ttl) to (n17\ttl);
    \draw[black,fill=black,opacity=1,line width = 0.05cm] (b10\ttl) to (n21\ttl);
    \draw[black,fill=black,opacity=1,line width = 0.05cm] (b10\ttl) to (n22\ttl);

    \draw[black,fill=black,opacity=1,line width = 0.05cm] (b11\ttl) to (n18\ttl);
    \draw[black,fill=black,opacity=1,line width = 0.05cm] (b11\ttl) to (n19\ttl);
    \draw[black,fill=black,opacity=1,line width = 0.05cm] (b11\ttl) to (n23\ttl);
    \draw[black,fill=black,opacity=1,line width = 0.05cm] (b11\ttl) to (n24\ttl);

    \end{scope}

    \def\hei{20}
    \def\ttl{t2}
    \begin{scope}
        
    \node[fill=zx_green,shape=circle,draw=black] (n4\ttl) at (2,18,\hei) {};

    \node[fill=zx_green,shape=circle,draw=black] (n3\ttl) at (2,14,\hei) {};
    \node[fill=zx_green,shape=circle,draw=black] (n2\ttl) at (2,10,\hei) {};
    \node[fill=zx_green,shape=circle,draw=black] (n1\ttl) at (2,6,\hei) {};
    \node[fill=zx_green,shape=circle,draw=black] (n0\ttl) at (2,2,\hei) {};

    \node[fill=zx_green,shape=circle,draw=black] (n9\ttl) at (6,18,\hei) {};
    \node[fill=zx_green,shape=circle,draw=black] (n8\ttl) at (6,14,\hei) {};
    \node[fill=zx_green,shape=circle,draw=black] (n7\ttl) at (6,10,\hei) {};
    \node[fill=zx_green,shape=circle,draw=black] (n6\ttl) at (6,6,\hei) {};
    \node[fill=zx_green,shape=circle,draw=black] (n5\ttl) at (6,2,\hei) {}; 

    \node[fill=zx_green,shape=circle,draw=black] (n14\ttl) at (10,18,\hei) {};
    \node[fill=zx_green,shape=circle,draw=black] (n13\ttl) at (10,14,\hei) {};
    \node[fill=zx_green,shape=circle,draw=black] (n12\ttl) at (10,10,\hei) {};
    \node[fill=zx_green,shape=circle,draw=black] (n11\ttl) at (10,6,\hei) {};
    \node[fill=zx_green,shape=circle,draw=black] (n10\ttl) at (10,2,\hei) {};

    \node[fill=zx_green,shape=circle,draw=black] (n19\ttl) at (14,18,\hei) {};
    \node[fill=zx_green,shape=circle,draw=black] (n18\ttl) at (14,14,\hei) {};
    \node[fill=zx_green,shape=circle,draw=black] (n17\ttl) at (14,10,\hei) {};
    \node[fill=zx_green,shape=circle,draw=black] (n16\ttl) at (14,6,\hei) {};
    \node[fill=zx_green,shape=circle,draw=black] (n15\ttl) at (14,2,\hei) {}; 

    \node[fill=zx_green,shape=circle,draw=black] (n24\ttl) at (18,18,\hei) {};
    \node[fill=zx_green,shape=circle,draw=black] (n23\ttl) at (18,14,\hei) {};
    \node[fill=zx_green,shape=circle,draw=black] (n22\ttl) at (18,10,\hei) {};
    \node[fill=zx_green,shape=circle,draw=black] (n21\ttl) at (18,6,\hei) {};
    \node[fill=zx_green,shape=circle,draw=black] (n20\ttl) at (18,2,\hei) {};
    
    \end{scope}
    \begin{scope}

        \node[fill=zx_red,circle,draw=black] (a0\ttl) at (0,4,\hei) {};
        \node[fill=zx_red,circle,draw=black] (a1\ttl) at (0,12,\hei) {};

        \node[fill=zx_red,circle,draw=black] (a2\ttl) at (4,8,\hei) {};
        \node[fill=zx_red,circle,draw=black] (a3\ttl) at (4,16,\hei) {};

        \node[fill=zx_red,circle,draw=black] (a4\ttl) at (8,4,\hei) {};
        \node[fill=zx_red,circle,draw=black] (a5\ttl) at (8,12,\hei) {};

        \node[fill=zx_red,circle,draw=black] (a6\ttl) at (12,8,\hei) {};
        \node[fill=zx_red,circle,draw=black] (a7\ttl) at (12,16,\hei) {};

        \node[fill=zx_red,circle,draw=black] (a8\ttl) at (16,4,\hei) {};
        \node[fill=zx_red,circle,draw=black] (a9\ttl) at (16,12,\hei) {};

        \node[fill=zx_red,circle,draw=black] (a10\ttl) at (20,8,\hei) {};
        \node[fill=zx_red,circle,draw=black] (a11\ttl) at (20,16,\hei) {};
    \end{scope}
    \begin{scope}
        
    \draw[black,fill=black,opacity=1,line width = 0.05cm] (a0\ttl) to (n0\ttl);
    \draw[black,fill=black,opacity=1,line width = 0.05cm] (a0\ttl) to (n1\ttl);

    \draw[black,fill=black,opacity=1,line width = 0.05cm] (a1\ttl) to (n2\ttl);
    \draw[black,fill=black,opacity=1,line width = 0.05cm] (a1\ttl) to (n3\ttl);

    \draw[black,fill=black,opacity=1,line width = 0.05cm] (a2\ttl) to (n1\ttl);
    \draw[black,fill=black,opacity=1,line width = 0.05cm] (a2\ttl) to (n2\ttl);
    \draw[black,fill=black,opacity=1,line width = 0.05cm] (a2\ttl) to (n6\ttl);
    \draw[black,fill=black,opacity=1,line width = 0.05cm] (a2\ttl) to (n7\ttl);

    \draw[black,fill=black,opacity=1,line width = 0.05cm] (a3\ttl) to (n3\ttl);
    \draw[black,fill=black,opacity=1,line width = 0.05cm] (a3\ttl) to (n4\ttl);
    \draw[black,fill=black,opacity=1,line width = 0.05cm] (a3\ttl) to (n8\ttl);
    \draw[black,fill=black,opacity=1,line width = 0.05cm] (a3\ttl) to (n9\ttl);

    \draw[black,fill=black,opacity=1,line width = 0.05cm] (a4\ttl) to (n5\ttl);
    \draw[black,fill=black,opacity=1,line width = 0.05cm] (a4\ttl) to (n6\ttl);
    \draw[black,fill=black,opacity=1,line width = 0.05cm] (a4\ttl) to (n10\ttl);
    \draw[black,fill=black,opacity=1,line width = 0.05cm] (a4\ttl) to (n11\ttl);
    
    \draw[black,fill=black,opacity=1,line width = 0.05cm] (a5\ttl) to (n7\ttl);
    \draw[black,fill=black,opacity=1,line width = 0.05cm] (a5\ttl) to (n8\ttl);
    \draw[black,fill=black,opacity=1,line width = 0.05cm] (a5\ttl) to (n12\ttl);
    \draw[black,fill=black,opacity=1,line width = 0.05cm] (a5\ttl) to (n13\ttl);

    \draw[black,fill=black,opacity=1,line width = 0.05cm] (a6\ttl) to (n11\ttl);
    \draw[black,fill=black,opacity=1,line width = 0.05cm] (a6\ttl) to (n12\ttl);
    \draw[black,fill=black,opacity=1,line width = 0.05cm] (a6\ttl) to (n16\ttl);
    \draw[black,fill=black,opacity=1,line width = 0.05cm] (a6\ttl) to (n17\ttl);

    \draw[black,fill=black,opacity=1,line width = 0.05cm] (a7\ttl) to (n13\ttl);
    \draw[black,fill=black,opacity=1,line width = 0.05cm] (a7\ttl) to (n14\ttl);
    \draw[black,fill=black,opacity=1,line width = 0.05cm] (a7\ttl) to (n18\ttl);
    \draw[black,fill=black,opacity=1,line width = 0.05cm] (a7\ttl) to (n19\ttl);

    \draw[black,fill=black,opacity=1,line width = 0.05cm] (a8\ttl) to (n15\ttl);
    \draw[black,fill=black,opacity=1,line width = 0.05cm] (a8\ttl) to (n16\ttl);
    \draw[black,fill=black,opacity=1,line width = 0.05cm] (a8\ttl) to (n20\ttl);
    \draw[black,fill=black,opacity=1,line width = 0.05cm] (a8\ttl) to (n21\ttl);

    \draw[black,fill=black,opacity=1,line width = 0.05cm] (a9\ttl) to (n17\ttl);
    \draw[black,fill=black,opacity=1,line width = 0.05cm] (a9\ttl) to (n18\ttl);
    \draw[black,fill=black,opacity=1,line width = 0.05cm] (a9\ttl) to (n22\ttl);
    \draw[black,fill=black,opacity=1,line width = 0.05cm] (a9\ttl) to (n23\ttl);
    
    \draw[black,fill=black,opacity=1,line width = 0.05cm] (a10\ttl) to (n21\ttl);
    \draw[black,fill=black,opacity=1,line width = 0.05cm] (a10\ttl) to (n22\ttl);

    \draw[black,fill=black,opacity=1,line width = 0.05cm] (a11\ttl) to (n23\ttl);
    \draw[black,fill=black,opacity=1,line width = 0.05cm] (a11\ttl) to (n24\ttl);

    \end{scope}

    \def\hei{22.5}
    \def\ttl{t3}
    \begin{scope}
        
    \node[fill=none,shape=circle,draw=none] (n4\ttl) at (2,18,\hei) {};

    \node[fill=none,shape=circle,draw=none] (n3\ttl) at (2,14,\hei) {};
    \node[fill=none,shape=circle,draw=none] (n2\ttl) at (2,10,\hei) {};
    \node[fill=none,shape=circle,draw=none] (n1\ttl) at (2,6,\hei) {};
    \node[fill=none,shape=circle,draw=none] (n0\ttl) at (2,2,\hei) {};

    \node[fill=none,shape=circle,draw=none] (n9\ttl) at (6,18,\hei) {};
    \node[fill=none,shape=circle,draw=none] (n8\ttl) at (6,14,\hei) {};
    \node[fill=none,shape=circle,draw=none] (n7\ttl) at (6,10,\hei) {};
    \node[fill=none,shape=circle,draw=none] (n6\ttl) at (6,6,\hei) {};
    \node[fill=none,shape=circle,draw=none] (n5\ttl) at (6,2,\hei) {}; 

    \node[fill=none,shape=circle,draw=none] (n14\ttl) at (10,18,\hei) {};
    \node[fill=none,shape=circle,draw=none] (n13\ttl) at (10,14,\hei) {};
    \node[fill=none,shape=circle,draw=none] (n12\ttl) at (10,10,\hei) {};
    \node[fill=none,shape=circle,draw=none] (n11\ttl) at (10,6,\hei) {};
    \node[fill=none,shape=circle,draw=none] (n10\ttl) at (10,2,\hei) {};

    \node[fill=none,shape=circle,draw=none] (n19\ttl) at (14,18,\hei) {};
    \node[fill=none,shape=circle,draw=none] (n18\ttl) at (14,14,\hei) {};
    \node[fill=none,shape=circle,draw=none] (n17\ttl) at (14,10,\hei) {};
    \node[fill=none,shape=circle,draw=none] (n16\ttl) at (14,6,\hei) {};
    \node[fill=none,shape=circle,draw=none] (n15\ttl) at (14,2,\hei) {}; 

    \node[fill=none,shape=circle,draw=none] (n24\ttl) at (18,18,\hei) {};
    \node[fill=none,shape=circle,draw=none] (n23\ttl) at (18,14,\hei) {};
    \node[fill=none,shape=circle,draw=none] (n22\ttl) at (18,10,\hei) {};
    \node[fill=none,shape=circle,draw=none] (n21\ttl) at (18,6,\hei) {};
    \node[fill=none,shape=circle,draw=none] (n20\ttl) at (18,2,\hei) {};
    
    \end{scope}

    \def\tlow{t0}
    \def\thig{t1}
    \begin{pgfonlayer}{background}
    \foreach \x in {0,...,24}
    {
    \draw[black,fill=black,opacity=1,line width = 0.05cm] (n\x\tlow) to (n\x\thig);
    }
    \end{pgfonlayer} 

    \def\tlow{t1}
    \def\thig{t2}
    \begin{pgfonlayer}{background}
    \foreach \x in {0,...,24}
    {
    \draw[black,fill=black,opacity=1,line width = 0.05cm] (n\x\tlow) to (n\x\thig);
    }
    \end{pgfonlayer}

    \def\tlow{t2}
    \def\thig{t3}
    \begin{pgfonlayer}{background}
    \foreach \x in {0,...,24}
    {
    \draw[black,fill=black,opacity=1,line width = 0.05cm] (n\x\tlow) to (n\x\thig);
    }
    \end{pgfonlayer}

    \def\tlow{t1}

    \def\tlow{t2}

    \def\tlow{t0}
    \def\thig{t1}
    \def\thhig{t2}
    \def\thhhig{t3}
    \begin{pgfonlayer}{background}
    \draw[green,fill=green,opacity=0.8,line width = 0.45cm] (n4\tlow) to (n4\thig);
    \draw[red,fill=red,opacity=0.6,line width = 0.25cm] (n4\tlow) to (n4\thig);
    \draw[green,fill=green,opacity=0.8,line width = 0.45cm] 
    (n4\thig) to (n4\thhig);
    \draw[green,fill=green,opacity=0.8,line width = 0.45cm] 
    (n4\thig) to (b2\thig);
    \draw[green,fill=green,opacity=0.8,line width = 0.45cm] 
    (n4\thhig) to (n4\thhhig);
    \draw[green,fill=green,opacity=0.8,line width = 0.45cm] 
    (n9\thig) to (b2\thig);
    \draw[green,fill=green,opacity=0.8,line width = 0.45cm] 
    (n9\thig) to (n9\tlow);
     \draw[green,fill=green,opacity=0.8,line width = 0.45cm] 
    (b5\thig) to (n9\thig);
    \draw[green,fill=green,opacity=0.8,line width = 0.45cm] 
    (n9\thhig) to (n9\thig);
    \draw[green,fill=green,opacity=0.8,line width = 0.45cm] 
    (n9\thhig) to (n9\thhhig);
    \draw[green,fill=green,opacity=0.8,line width = 0.45cm] 
    (n14\thig) to (b5\thig);
    \draw[green,fill=green,opacity=0.8,line width = 0.45cm] 
    (n14\thig) to (n14\tlow);
    \draw[green,fill=green,opacity=0.8,line width = 0.45cm] 
    (n14\thig) to (n14\thhig);
    \draw[green,fill=green,opacity=0.8,line width = 0.45cm] 
    (n14\thig) to (b8\thig);
    \draw[green,fill=green,opacity=0.8,line width = 0.45cm] 
    (n14\thhig) to (n14\thhhig);
    \draw[green,fill=green,opacity=0.8,line width = 0.45cm] 
    (n19\thig) to (b8\thig);
    \draw[green,fill=green,opacity=0.8,line width = 0.45cm] 
    (n19\thig) to (n19\tlow);
    \draw[green,fill=green,opacity=0.8,line width = 0.45cm] 
    (n19\thig) to (n19\thhig);
    \draw[green,fill=green,opacity=0.8,line width = 0.45cm] 
    (b11\thig) to (n19\thig);
    \draw[green,fill=green,opacity=0.8,line width = 0.45cm] 
    (n19\thhhig) to (n19\thhig);
    \draw[green,fill=green,opacity=0.8,line width = 0.45cm] 
    (b11\thig) to (n24\thig);
    \draw[green,fill=green,opacity=0.8,line width = 0.45cm] 
    (n24\thig) to (n24\tlow);
    \draw[green,fill=green,opacity=0.8,line width = 0.45cm] 
    (n24\thig) to (n24\thhig);
    \draw[green,fill=green,opacity=0.8,line width = 0.45cm] 
    (n24\thhig) to (n24\thhhig);
    
    \end{pgfonlayer}

    \end{tikzpicture}
    }

    }
    \caption{}
\end{figure}

\noindent We shall then apply the other rules from definition 2.7 in \cite{rodatz2024floquetifyingstabilisercodesdistancepreserving}:

\begin{webrulebox}
`a spider with $k\pi$ phase can have:
\begin{itemize}
    \item an even number of legs highlighted in its own colour, and\sout{/or}
    \item all or none of its legs highlighted in the opposite colour'
\end{itemize}
\end{webrulebox}

\noindent to find the green Pauli web first, by decorating all the phase-less spiders next, resulting in the ZX-diagram in figure \ref{fig:y_injection_first_pauli_web_green}. Finally, we can work out the red decorated Pauli web, resulting in figure \ref{fig:y_injection_pauli_web} using the same phase-less ($k\pi = 0$) spider Pauli web rules. Figure \ref{fig:y_injection_pauli_web} shows the logical correlator for a distance $5$ surface code after $1$ full round of error-free parity measurement. The resulting Pauli web coincides with a logical $Y$ correlator which is a combination of a logical $X$ and logical $Z$ correlator\footnote{Since $iY=ZX$.} as shown in figure \ref{fig:logical_cor}.

\begin{figure}[!h]
    \centering     
    \tdplotsetmaincoords{70}{23}
    \resizebox{0.45\linewidth}{!}{
   

    }
    \caption{\label{fig:y_injection_pauli_web}All the red and green decorated edge Pauli web shown. This recovers the logical $Y$ correlators of the rotated surface code, which is the combination of the logical $X$ and $Z$ correlators.}
\end{figure}

\section{Pauli web analysis of the Lao--Criger corner scheme}
  The ``propagation'' of the Pauli web to future rounds of parity measurement rounds and larger distances is straightforward. We suspect that similar approaches can be applied to replace the injected Z-spiders from $\pi/2 \rightarrow \pi/4$ under appropriate Pauli web rule modifications \cite{Kissinger_2022} to diagrammatically visualise the $\ket{T}$ state injection protocol \cite{Li_2015,lao_criger_magic}. 

  When we first read Li's scheme \cite{Li_2015} years ago, we were puzzled by the strange yet structured initialised qubit pattern. With ZX-calculus and Pauli web, we can see that if any of the states in the yellowed highlighted rectangles (figure \ref{fig:inj_exp}) are any different or of the opposite colour, the logical $Y$ correlator Pauli web will not terminate properly. This will result in the incorrect logical $Y$ correlator. Furthermore, the upper and lower triangular separated state initialisation supports the logical $Y$ correlator and enables plaquette parity measurements needed for post-selecting clean injected states, as detailed in \cite{Li_2015}. We shall provide a brief sketch on how to interpret the post-selection procedure at a superficial level in the next few subsections.

\begin{figure}[!h]
    \centering     
    \tdplotsetmaincoords{70}{23}
    \resizebox{0.5\linewidth}{!}{
   
    \begin{tikzpicture}[tdplot_main_coords,every node/.style={minimum size=1cm}]
    \def\hei{0}
    \def\ttl{t0}
    \begin{scope}
        
    \node[fill=zx_green,shape=circle,draw=black] (n4\ttl) at (2,18,\hei) {$\pi/2$};
    \node[fill=zx_green,shape=circle,draw=black] (n3\ttl) at (2,14,\hei) {};
    \node[fill=zx_green,shape=circle,draw=black] (n2\ttl) at (2,10,\hei) {};
    \node[fill=zx_green,shape=circle,draw=black] (n1\ttl) at (2,6,\hei) {};
    \node[fill=zx_green,shape=circle,draw=black] (n0\ttl) at (2,2,\hei) {};

    \node[fill=zx_red,shape=circle,draw=black] (n9\ttl) at (6,18,\hei) {};
    \node[fill=zx_green,shape=circle,draw=black] (n8\ttl) at (6,14,\hei) {};
    \node[fill=zx_green,shape=circle,draw=black] (n7\ttl) at (6,10,\hei) {};
    \node[fill=zx_green,shape=circle,draw=black] (n6\ttl) at (6,6,\hei) {};
    \node[fill=zx_green,shape=circle,draw=black] (n5\ttl) at (6,2,\hei) {}; 

    \node[fill=zx_red,shape=circle,draw=black] (n14\ttl) at (10,18,\hei) {};
    \node[fill=zx_red,shape=circle,draw=black] (n13\ttl) at (10,14,\hei) {};
    \node[fill=zx_green,shape=circle,draw=black] (n12\ttl) at (10,10,\hei) {};
    \node[fill=zx_green,shape=circle,draw=black] (n11\ttl) at (10,6,\hei) {};
    \node[fill=zx_green,shape=circle,draw=black] (n10\ttl) at (10,2,\hei) {};

    \node[fill=zx_red,shape=circle,draw=black] (n19\ttl) at (14,18,\hei) {};
    \node[fill=zx_red,shape=circle,draw=black] (n18\ttl) at (14,14,\hei) {};
    \node[fill=zx_red,shape=circle,draw=black] (n17\ttl) at (14,10,\hei) {};
    \node[fill=zx_green,shape=circle,draw=black] (n16\ttl) at (14,6,\hei) {};
    \node[fill=zx_green,shape=circle,draw=black] (n15\ttl) at (14,2,\hei) {}; 

    \node[fill=zx_red,shape=circle,draw=black] (n24\ttl) at (18,18,\hei) {};
    \node[fill=zx_red,shape=circle,draw=black] (n23\ttl) at (18,14,\hei) {};
    \node[fill=zx_red,shape=circle,draw=black] (n22\ttl) at (18,10,\hei) {};
    \node[fill=zx_red,shape=circle,draw=black] (n21\ttl) at (18,6,\hei) {};
    \node[fill=zx_green,shape=circle,draw=black] (n20\ttl) at (18,2,\hei) {};
    
    \end{scope}

    \draw[yellow!50,fill=yellow!50,opacity=0.5,line width = 0.05cm,rounded corners=10pt] ($(n0\ttl)+(-2,-2,0)$) -- ($(n3\ttl)+(-2,2,0)$) -- ($(n3\ttl)+(2,2,0)$) -- ($(n0\ttl)+(2,-2,0)$) -- cycle;

    \draw[yellow!50,fill=yellow!50,opacity=0.5,line width = 0.05cm,rounded corners=10pt] ($(n9\ttl)+(-2,2,0)$) -- ($(n24\ttl)+(2,2,0)$) -- ($(n24\ttl)+(2,-2,0)$) -- ($(n9\ttl)+(-2,-2,0)$) -- cycle;

    \end{tikzpicture}
    }
    \caption{\label{fig:inj_exp}The initial states inside the yellow coloured rectangles must be green or red respectively in order to have the correct logical $Y$ correlator \cite{Bombin_2024}.}
\end{figure}

\subsection{Parity measurements and syndrome extractions}
In order to examine how to post-select in Li/Lao--Criger's schemes, let's review how to perform error correction when a logical $\ket{0}$ state is initialised (figure \ref{fig:Z_mem}). In the literature this is known as a ($\ket{\bar{0}}$ initial state) memory experiment \cite{Fowler_2012,higgott2021pymatchingpythonpackagedecoding}. For a distance $d$ ($=5$ here) rotated surface code, $n = d^2$ data qubits in the $\ket{0}^{\otimes n}$ state will be initialised in the very first time slice. Following that, the encoder circuit from figure \ref{fig:inj_encoder} is repeatedly applied. In figure \ref{fig:Z_mem}, we have applied $1.5$ encoder circuit with the final half full-round of parity measurements omitted for simplicity. Figure \ref{fig:Z_mem} shows one initial time $Z$-type \textit{syndrome}/\textit{parity check} to the $\ket{0}$ memory experiment as a green decorated Pauli web \cite{Bombin_2024}. Parity check is the product between consecutive rounds of parity measurements of the same type (i.e $Z$ then $Z$ or $X$ then $X$). If a Pauli $X$ error occurred within the \textit{parity check cube} in figure \ref{fig:Z_init_check}, the product in parity measurements ($1\times -1 = -1$) will be non-trivial and hence we can potentially detect this error\footnote{Assuming appropriate decoding \cite{Fowler_2012}.}.
\begin{figure}[!h]
    \centering
    \tdplotsetmaincoords{70}{23}
    \subfloat[\label{fig:Z_mem}]{
    \resizebox{0.34\linewidth}{!}{
   

    }
    }
    \caption{a) ZX-diagram for a logical $\ket{0}$ initialised state on the surface code. b) The {\color{green}} Pauli web decorated $Z$-type initial parity check cube from figure \ref{fig:Z_mem}. The $X$-type ancillas are drawn with only the legs internal to the cube, the omitted legs carry no Pauli web.}
\end{figure}

The curious reader will notice that the $Z$-type parity check cube in figure \ref{fig:Z_init_check} is different\footnote{Aside from the opposite Pauli web colouring convention.} from the one shown in \cite{Bombin_2024} as this parity check cube comes from the initial round of checks. Hence, the $Z$-type parity measurements will always be $1$ if no errors occurred in the data qubits at the lowest time slice and $-1$ if an error occurred on any data qubit above the lowest time slice\footnote{This is also known as initialisation error.}, shown as a $\pi$ X-spider in figure \ref{fig:Z_mem_before_prop}. As a result, we do not need to perform a $ZZZZ$ measurement on the very first time slice because we know $ZZZZ = +1$ if no $X$ errors occurred and $-1$ otherwise. As it happens, $\ket{\bar{0}/\bar{1}},\ket{\bar{+}/\bar{-}}$ logical states can be fault-tolerantly initialised on the surface code. Starting from $n=d^2$ data qubits in the $\ket{0/1}^{\otimes n}$ (or $\ket{+/-}^{\otimes n}$) separable state, measuring all the plaquette parities lead to a logical encoded $X$ or $Z$ eigenstate.

\subsection{Data qubit error}
Consider the consequence of a single data qubit $X$ error after the first time slice on data qubit labelled {\color{cyan}$9$} in cyan. All data qubit labels are written in cyan, {\color{cyan}$A$} for example, and do not denote a spider with phase $A$.
\begin{figure}[!h]
    \centering
    \tdplotsetmaincoords{70}{23}
    \subfloat[\label{fig:Z_mem_before_prop}]{
    \resizebox{0.34\linewidth}{!}{
   

    }
    }
    \caption{a) ZX-diagram for a logical $\ket{0}$ initialised state on the surface code with an $X$ error ($\pi$ phase X-spider) at a particular data qubit on the first layer. b) ZX-diagram for a logical $\ket{0}$ initialised state on the surface code with an $X$ error. The $\pi$ phase X-spider error is pushed along in time and flips a particular parity measurement (labelled with another $\pi$ phase X-spider).}
\end{figure}
The $\pi$ phase X-spider (Pauli $X$ error) from figure \ref{fig:Z_mem_before_prop} right after the first state initialisation time step on data qubit {\color{cyan}$9$} can be pushed into future time layers with ZX-calculus. In figure \ref{fig:Z_mem_before_prop}, the $Z$ parity check at the third time slice picks up an $X$ error ($\pi$ phase X-spider) after pushing the $\pi$ phase spider through. This means the parity check cube syndrome result will return $-1$ in parity (or $1$ in binary). On data qubit {\color{cyan}$9$}, the Pauli frame is modified and the $\pi$ phase X-spider persists in its world-line in time. If we perform minimum-weight perfect matching \cite{Fowler_2012,higgott2021pymatchingpythonpackagedecoding,higgott2025sparseblossomcorrectingmillion} to decode this syndrome, the syndrome\footnote{Or a detector in PyMatching \cite{higgott2021pymatchingpythonpackagedecoding}.} will be matched to the boundary of the decoding graph \cite{higgott2021pymatchingpythonpackagedecoding}, hence we can infer that data qubit {\color{cyan}$9$} or {\color{cyan}$14$} (see figure \ref{fig:Z_mem_after_prop}) had experienced an initialisation $X$ error. The Pauli-frame will be modified accordingly.   

\subsection{$\ket{Y}$ state injection and post-selection}
Consider two neighbouring parity check cubes in the $\ket{Y}$ state injection protocol (figure \ref{fig:Y_inj_two_cell_both}).
\begin{figure}[!h]
    \centering
    \tdplotsetmaincoords{70}{23}
    \subfloat[\label{fig:Y_inj_two_cell}]{
    \resizebox{0.35\linewidth}{!}{
   

    }
    }  
    \caption{a) Two parity check cubes highlighted from the $\ket{Y}$ state injection protocol associated with qubit {\color{cyan}$13$}. The {\color{orange}orange} Pauli web is also a `{\color{green}green}' Z-type Pauli web. It coloured {\color{orange}orange} to make clear that it is a invalid Pauli web, as the {\color{orange}orange} Pauli web terminate on the initial states in an forbidden manner. b) The only valid Pauli web associated with the left orange (invalid) check cube. This Pauli web terminates on no initial state, it only runs up to the outputs of the four data qubits. Hence the left check cube does not check \mbox{$ZZZZ$} against the initial states, it projects \mbox{$ZZZZ$} into the future with a random sign. This is why its first round parity measurement cannot be used as a syndrome. c) The right parity check cube from figure \ref{fig:Y_inj_two_cell}.}
    \label{fig:Y_inj_two_cell_both}
\end{figure}

We draw a left Z-type parity check cube (highlighted in {\color{orange}orange} instead of {\color{green}green} for illustration and differentiation purposes) in figure \ref{fig:Y_inj_two_cell}. This orange parity check cube is invalid as the initial green Z-spiders from the first time slices associated with it does not have the correct X-type, red Pauli web terminating on them. This invalid orange check cube has three $\ket{+}$ states and one $\ket{0}$ state in the first time slice dictated by the injection protocol. This implies that in the first time slice, the left orange check cube has $ZZZZ = \pm 1$, with either of the $-1$ or $+1$ value occurring with probability $1/2$ when measured\footnote{In stabiliser notation: if we were to measure $Z_aZ_bZ_cZ_d$ in a system stabilised by $\langle X_a,X_b,X_c,Z_d\rangle$, these three stabiliser generators: $X_a,X_b,X_c$ each anti-commutes with the measurement basis: $Z_aZ_bZ_cZ_d$. Hence, the resultant stabiliser generators post measurement is: $\langle (-1)^{m}Z_aZ_bZ_cZ_d,X_aX_b,X_aX_c,Z_d\rangle$, with the measured $Z_aZ_bZ_cZ_d$ binary result, $m\in\{0,1\}$, happening uniformly at random.}. This means the measurement results associated with the left orange check cube cannot be reliably used to construct syndromes and infer errors when data qubit {\color{cyan}$14$} (coloured {\color{cyan}cyan} in figure \ref{fig:Y_inj_two_cell_both}) had experienced an initialisation error. Suppose the true initialisation $X$ error is on data qubit {\color{cyan}$14$}. The right parity check cube will return a parity value of $-1$. However, the random parity measurement outcome in the left parity measurement could lead to the left parity check cube also returning a parity value of $-1$, hence we would have decoded and misidentified a $X$ initial error on data qubit {\color{cyan}$13$}. This in turn leads to a mis-calculated Pauli-frame concerning the logical $Y$ correlator. The remedy is to post-select $+1$ parity measurement results in all the plaquettes labelled `$+1$' in figure \ref{fig:surface_code_Y_inj_post_selection}, where we have a green $X$-type and red $Z$-type plaquettes rotated surface code \cite{Fowler_2012}. This reduces the probability of misidentifying the Pauli-frame for the logical $Y$ correlator, which leads to the high fidelity logical state injection, as shown in the Li \cite{Li_2015} and Lao--Criger \cite{lao_criger_magic} schemes.

The set of plaquettes to post-select on can also be read off from the Pauli webs directly. Out of the $24$ plaquettes in a distance $5$ code, only $10$ have a first round web that terminates on the initial states: the $6$ $X$-type plaquettes whose data qubits all start in $\ket{+}$, and the $4$ $Z$-type plaquettes whose data qubits all start in $\ket{0}$. These are the same $10$ plaquettes labelled `$+1$' in figure \ref{fig:surface_code_Y_inj_post_selection}. The other $14$ sit across the boundary between the $\ket{+}$ and $\ket{0}$ regions, and none of them has a Pauli web that reaches back to the first time slice. Each of the $14$ still has a Pauli web, but it runs only forward in time, so its first round parity measurement projects a new stabiliser instead of checking one that is already there. We draw both sets in figure \ref{fig:web_mesh}. Each panel there is not one big Pauli web. It is all $10$, or all $14$, of the individual Pauli webs drawn on top of one another, so the edges that several Pauli webs share are drawn more than once. This is also why two full rounds of parity measurement are quoted in the Li \cite{Li_2015} and Lao--Criger \cite{lao_criger_magic} schemes. The Pauli webs in figure \ref{fig:web_mesh} were computed and drawn with \cite{injection_webs_code}, using the 3D ZX-diagram viewer of \cite{pyzx_3d_viewer}.

\begin{figure}[!h]
    \centering
    \resizebox{0.45\linewidth}{!}{
    \begin{tikzpicture}[scale=.7,every node/.style={minimum size=0.5cm},on grid]

    \node[fill=white,shape=circle,draw=black] (n4) at (2,18) {${\color{blue}\ket{Y}}$};
    \draw[blue, very thick] (2,18) circle (0.85);
    \node[fill=white,shape=circle,draw=black] (n3) at (2,14) {${\ket{+}}$};
    \node[fill=white,shape=circle,draw=black] (n2) at (2,10) {${\ket{+}}$};
    \node[fill=white,shape=circle,draw=black] (n1) at (2,6) {${\ket{+}}$};
    \node[fill=white,shape=circle,draw=black] (n0) at (2,2) {${\ket{+}}$};

    \node[fill=white,shape=circle,draw=black] (n9) at (6,18) {${\ket{0}}$};
    \node[fill=white,shape=circle,draw=black] (n8) at (6,14) {${\ket{+}}$};
    \node[fill=white,shape=circle,draw=black] (n7) at (6,10) {${\ket{+}}$};
    \node[fill=white,shape=circle,draw=black] (n6) at (6,6) {${\ket{+}}$};
    \node[fill=white,shape=circle,draw=black] (n5) at (6,2) {${\ket{+}}$}; 

    \node[fill=white,shape=circle,draw=black] (n14) at (10,18) {${\ket{0}}$};
    \node[fill=white,shape=circle,draw=black] (n13) at (10,14) {${\ket{0}}$};
    \node[fill=white,shape=circle,draw=black] (n12) at (10,10) {${\ket{+}}$};
    \node[fill=white,shape=circle,draw=black] (n11) at (10,6) {${\ket{+}}$};
    \node[fill=white,shape=circle,draw=black] (n10) at (10,2) {${\ket{+}}$};

    \node[fill=white,shape=circle,draw=black] (n19) at (14,18) {${\ket{0}}$};
    \node[fill=white,shape=circle,draw=black] (n18) at (14,14) {${\ket{0}}$};
    \node[fill=white,shape=circle,draw=black] (n17) at (14,10) {${\ket{0}}$};
    \node[fill=white,shape=circle,draw=black] (n16) at (14,6) {${\ket{+}}$};
    \node[fill=white,shape=circle,draw=black] (n15) at (14,2) {${\ket{+}}$}; 

    \node[fill=white,shape=circle,draw=black] (n24) at (18,18) {${\ket{0}}$};
    \node[fill=white,shape=circle,draw=black] (n23) at (18,14) {${\ket{0}}$};
    \node[fill=white,shape=circle,draw=black] (n22) at (18,10) {${\ket{0}}$};
    \node[fill=white,shape=circle,draw=black] (n21) at (18,6) {${\ket{0}}$};
    \node[fill=white,shape=circle,draw=black] (n20) at (18,2) {${\ket{+}}$};

    \begin{pgfonlayer}{background}

    \draw[zx_red,fill=zx_red,opacity=0.65](1.8,2.2) to[curve through={(0,4)}](1.8,5.8);
    \filldraw[fill=zx_green, opacity=0.65, draw=zx_green] (2.2,2.2) rectangle (5.8,5.8);
    \filldraw[fill=zx_red, opacity=0.65, draw=zx_red] (2.2,6.2) rectangle (5.8,9.8);
    \filldraw[fill=zx_green, opacity=0.65, draw=zx_green] (2.2,10.2) rectangle (5.8,13.8);
    \draw[zx_red,fill=zx_red,opacity=0.65](1.8,10.2) to[curve through={(0,12)}](1.8,13.8);
    \filldraw[fill=zx_red, opacity=0.65, draw=zx_red] (2.2,14.2) rectangle (5.8,17.8);
    \draw[zx_green,fill=zx_green,opacity=0.65](2.2,18.2) to[curve through={(4,20)}](5.8,18.2);

    \draw[zx_green,fill=zx_green,opacity=0.65](9.8,1.8) to[curve through={(8,0)}](6.2,1.8);
    \filldraw[fill=zx_red, opacity=0.65, draw=zx_red] (6.2,2.2) rectangle (9.8,5.8);
    \filldraw[fill=zx_green, opacity=0.65, draw=zx_green] (6.2,6.2) rectangle (9.8,9.8);
    \filldraw[fill=zx_red, opacity=0.65, draw=zx_red] (6.2,10.2) rectangle (9.8,13.8);
    \filldraw[fill=zx_green, opacity=0.65, draw=zx_green] (6.2,14.2) rectangle (9.8,17.8);

    \filldraw[fill=zx_green, opacity=0.65, draw=zx_green] (10.2,2.2) rectangle (13.8,5.8);
    \filldraw[fill=zx_red, opacity=0.65, draw=zx_red] (10.2,6.2) rectangle (13.8,9.8);
    \filldraw[fill=zx_green, opacity=0.65, draw=zx_green] (10.2,10.2) rectangle (13.8,13.8);
    \filldraw[fill=zx_red, opacity=0.65, draw=zx_red] (10.2,14.2) rectangle (13.8,17.8);
    \draw[zx_green,fill=zx_green,opacity=0.65](10.2,18.2) to[curve through={(12,20)}](13.8,18.2);
    
    \draw[zx_green,fill=zx_green,opacity=0.65](17.8,1.8) to[curve through={(16,0)}](14.2,1.8);
    \filldraw[fill=zx_red, opacity=0.65, draw=zx_red] (14.2,2.2) rectangle (17.8,5.8);
    \filldraw[fill=zx_green, opacity=0.65, draw=zx_green] (14.2,6.2) rectangle (17.8,9.8);
    \draw[zx_red,fill=zx_red,opacity=0.65](18.2,9.8) to[curve through={(20,8)}](18.2,6.2);
    \filldraw[fill=zx_red, opacity=0.65, draw=zx_red] (14.2,10.2) rectangle (17.8,13.8);
    \filldraw[fill=zx_green, opacity=0.65, draw=zx_green] (14.2,14.2) rectangle (17.8,17.8);
    \draw[zx_red,fill=zx_red,opacity=0.65](18.2,17.8) to[curve through={(20,16)}](18.2,14.2);
    \end{pgfonlayer}

    \begin{pgfonlayer}{background}
        \path [-,line width=0.1cm,black,opacity=1] (n0) edge node {} (n20);
        \path [-,line width=0.1cm,black,opacity=1] (n1) edge node {} (n21);
        \path [-,line width=0.1cm,black,opacity=1] (n2) edge node {} (n22);
        \path [-,line width=0.1cm,black,opacity=1] (n3) edge node {} (n23);
        \path [-,line width=0.1cm,black,opacity=1] (n4) edge node {} (n24);

        \path [-,line width=0.1cm,black,opacity=1] (n0) edge node {} (n4);
        \path [-,line width=0.1cm,black,opacity=1] (n5) edge node {} (n9);
        \path [-,line width=0.1cm,black,opacity=1] (n10) edge node {} (n14);
        \path [-,line width=0.1cm,black,opacity=1] (n15) edge node {} (n19);
        \path [-,line width=0.1cm,black,opacity=1] (n20) edge node {} (n24);

        \node[fill=black,circle,draw=black,text =white,scale = 1.5] (b0) at (4,4) {$+1$};
        \node[fill=black,circle,draw=black,text =white,scale = 1.5] (b1) at (4,12) {$+1$};
        \node[fill=black,circle,draw=black,scale = 0.25] (b2) at (4,20) {};

        \node[fill=black,circle,draw=black,scale = 1.5, text=white] (b3) at (8,0) {$+1$};
        \node[fill=black,circle,draw=black,scale = 1.5, text=white] (b4) at (8,8) {$+1$};
        \node[fill=black,circle,draw=black,scale = 0.25] (b5) at (8,16) {};

        \node[fill=black,circle,draw=black,scale = 1.5, text=white] (b6) at (12,4) {$+1$};
        \node[fill=black,circle,draw=black,scale = 0.25] (b7) at (12,12) {};
        \node[fill=black,circle,draw=black,scale = 0.25] (b8) at (12,20) {};

        \node[fill=black,circle,draw=black,scale = 1.5, text=white] (b9) at (16,0) {$+1$};
        \node[fill=black,circle,draw=black,scale = 0.25] (b10) at (16,8) {};
        \node[fill=black,circle,draw=black,scale = 0.25] (b11) at (16,16) {};

        \node[fill=white,circle,draw=black,scale = 0.25] (a0) at (0,4) {};
        \node[fill=white,circle,draw=black,scale = 0.25] (a1) at (0,12) {};

        \node[fill=white,circle,draw=black,scale = 0.25] (a2) at (4,8) {};
        \node[fill=white,circle,draw=black,scale = 0.25] (a3) at (4,16) {};

        \node[fill=white,circle,draw=black,scale = 0.25] (a4) at (8,4) {};
        \node[fill=white,circle,draw=black,scale = 0.25] (a5) at (8,12) {};

        \node[fill=white,circle,draw=black,scale = 0.25] (a6) at (12,8) {};
        \node[fill=white,circle,draw=black,scale = 1.5, text=black] (a7) at (12,16) {$+1$};

        \node[fill=white,circle,draw=black,scale = 0.25] (a8) at (16,4) {};
        \node[fill=white,circle,draw=black,scale = 1.5, text=black] (a9) at (16,12) {$+1$};

        \node[fill=white,circle,draw=black,scale = 1.5, text=black] (a10) at (20,8) {$+1$};
        \node[fill=white,circle,draw=black,scale = 1.5, text=black] (a11) at (20,16) {$+1$};

    \end{pgfonlayer}

\end{tikzpicture}
    }
    \caption{A distance $d=5$ rotated surface code with green $X$-type and red $Z$-type plaquettes. In the injection scheme, the $+1$ labelled plaquettes are the parity measurements that one post-select on given a $+1$ parity value (see detailed discussion in \cite{Li_2015,lao_criger_magic}).}
    \label{fig:surface_code_Y_inj_post_selection}
\end{figure}

\begin{figure}[!h]
    \centering
    \tdplotsetmaincoords{70}{23}
    \subfloat[\label{fig:webs_postselect}The $10$ plaquettes that do admit a first round
    Pauli web, all drawn together. Every one of these Pauli webs terminates on the initial states at
    the bottom. They are the same plaquettes labelled `$+1$' in figure
    \ref{fig:surface_code_Y_inj_post_selection}.]{
    \resizebox{0.4\linewidth}{!}{

    }
    }
    \quad
    \subfloat[\label{fig:webs_nodetector}The $14$ plaquettes that do not, again all
    drawn together. These Pauli webs run only forward in time to the last time slice and reach none
    of the initial states.]{
    \resizebox{0.4\linewidth}{!}{
%
    }
    }
    \caption{\label{fig:web_mesh}The first round Pauli webs of the $\ket{Y}$ injection
    protocol. Neither panel is a single Pauli web. Each one shows all of the individual Pauli webs
    superposed, so an edge that several Pauli webs share is drawn several times over. Green edges
    carry a $Z$-type decoration and red edges an $X$-type one.}
\end{figure}

\section{Other injection schemes}
\label{sec:otherinjections}

The corner placement analysed so far is one member of a family of injection schemes. Technically the injection qubit can be placed anywhere on the $d\times d$ grid of the rotated surface code. Figure 7 of reference~\cite{strikis2023quantumcomputingscalableplanar} even shows the preferred injection point pushed off-centre by fabrication defects. On a defect-free grid the two distinguished placements are the corner and the centre, and both are analysed in this manuscript. In the
sections (or appendix) that follow we run the other injection schemes for the surface code through the same Pauli web analysis. Firstly with the injected qubit placed in the central via Lao--Criger's scheme (section~\ref{app:injection}), its behaviour
under circuit-level noise (section~\ref{app:circuitnoise}), ZZ injection
(appendix~\ref{app:zz}), hook injection (appendix~\ref{app:hook}) and
unitary-encoder injection (appendix~\ref{app:unitary}). For the Lao--Criger central qubit
placement scheme, we aim to find out:
\begin{itemize} 
\item which plaquette parity measurements admit a first round web that terminates on the
initial states, and 
\item which initialisation errors remain invisible to every such Pauli web, 
\item while flipping the logical correlator.
\end{itemize}

Before drawing the Pauli webs, we fix two pieces of notation (collected in
figure~\ref{fig:blue_notation}) that hold from now onwards. Purely for
visualisation, a green spider with phase $\pi/2$, which is to be
injected as a $\ket{Y}$ state, will be drawn as a blue
node without a phase label. A Hadamard gate on a wire is drawn either as
a yellow box or, equivalently, as a blue dashed edge (a Hadamard edge).
Since an $X$-type spider is a $Z$-type spider conjugated by Hadamards, a
red spider with phase $\pi/2$ is the same diagram as a blue node sitting
between two Hadamard edges (this appears in appendix~\ref{app:hook}).

\begin{figure}[!h]
    \centering
    \resizebox{0.42\linewidth}{!}{
    \begin{tikzpicture}[scale=.7,every node/.style={minimum size=0.55cm},on grid]
    \node[fill=zx_green,shape=circle,draw=black] (g) at (0,6) {$\frac{\pi}{2}$};
    \draw[black,line width=0.05cm] (g) -- (1.6,6);
    \node[draw=none,fill=none] at (2.6,6) {$=$};
    \node[fill=zx_blue,shape=circle,draw=black] (bl) at (3.9,6) {};
    \draw[black,line width=0.05cm] (bl) -- (5.5,6);

    \draw[black,line width=0.05cm] (-0.7,4) -- (3.1,4);
    \node[fill=yellow,draw=black,shape=rectangle,minimum size=0.3cm,
          inner sep=0pt] (h) at (1.2,4) {};
    \node[draw=none,fill=none] at (4.0,4) {$=$};
    \draw[-, dashed, dash pattern=on 2pt off 1.2pt, line width=0.06cm,
          draw=zx_hadamard] (5.0,4) -- (7.2,4);

    \node[fill=zx_red,shape=circle,draw=black] (r) at (0,2) {$\frac{\pi}{2}$};
    \draw[black,line width=0.05cm] (-1.4,2) -- (r);
    \draw[black,line width=0.05cm] (r) -- (1.6,2);
    \node[draw=none,fill=none] at (2.6,2) {$=$};
    \node[fill=zx_blue,shape=circle,draw=black] (b2) at (5.1,2) {};
    \draw[-, dashed, dash pattern=on 2pt off 1.2pt, line width=0.06cm,
          draw=zx_hadamard] (3.5,2) -- (b2);
    \draw[-, dashed, dash pattern=on 2pt off 1.2pt, line width=0.06cm,
          draw=zx_hadamard] (b2) -- (6.7,2);

    \draw[-, dashed, dash pattern=on 2pt off 1.2pt, line width=0.06cm,
          draw=zx_hadamard] (-0.2,0) -- (3.1,0);
    \draw[red,fill=red,opacity=0.8,line width = 0.25cm] (-0.2,0) -- (1.45,0);
    \draw[green,fill=green,opacity=0.8,line width = 0.25cm] (1.45,0) -- (3.1,0);
\end{tikzpicture}
    }
    \caption{Notation used from now onwards. First row: a green spider
    with phase $\pi/2$ is drawn as a blue node. Second row: a Hadamard
    gate is drawn as a yellow box or, equivalently, as a blue dashed
    Hadamard edge, and an edge is either plain or Hadamard in full. Third
    row: a red spider with phase $\pi/2$ equals a blue node whose edges
    are Hadamard edges. Fourth row: a Pauli web decoration changes
    colour when it crosses a Hadamard edge, so a decorated Hadamard edge
    is drawn half in each colour ($X$-type on one side, $Z$-type on the
    other).}
    \label{fig:blue_notation}
\end{figure}

\section{The Lao--Criger central-qubit injection}
\label{app:injection}

The corner placement analysed in the previous sections is not the sole injection
scheme in \cite{lao_criger_magic}. The injected qubit can also sit in the
middle of the patch. In this section we study the central-qubit scheme
through the same Pauli web analysis (initialisation,
first round of parity measurements, Pauli webs, post-selection) and use the Pauli webs
to fix the choice of the initial $\ket{+}$ and $\ket{0}$
states. At $d = 5$, this results in a central-qubit initialisation with
12 post-selectable plaquettes instead of the corner scheme's 10, and
with only two undetectable malicious initialisation errors,
initialisation errors that no post-selected check can detect,
instead of four. This scheme coincides with Lao and Criger's middle placement in the position of the injected qubit, the initialisation pattern (identically to their figure 3(b) at $d=3$, generalising it at larger distances) and the post-selected plaquettes, but not in the syndrome extraction schedule: under circuit-level noise we replace their Tomita--Svore ordering by the optimised serialised schedule of section~\ref{sub:mc}, improving an injected logical error rate of $32p/15$
(section~\ref{sub:budget}) to $28p/15$. We work with $d = 5$ throughout for visualisation purposes.

\subsection{Initialisation}

For a distance $d$ surface code, a square grid of $d \times d$ physical
(data) qubits needs to be initialised: the bare physical $\ket{Y}$ state
now sits on the central data qubit ({\color{cyan}$12$} at $d=5$), and the
remaining $d^2 - 1$ data qubits are initialised in $\ket{+}$ or $\ket{0}$.
Webs decide the initialisation pattern of the data qubits. For each candidate initialisation pattern,
\begin{itemize}
\item different plaquette parity measurements admit a first round web that
terminates on the initial states (these are post-selectable), while
others admit none, so their first round parities are random\footnote{For the Lao--Criger central qubit scheme, these parity measurements are the `$+1$'-labelled
plaquettes one may post-select on, exactly as in
figure~\ref{fig:surface_code_Y_inj_post_selection}.}, and 
\item different initialisation errors remain invisible to every such plaquette while
flipping the logical $Y$ correlator. 
\end{itemize}
We call the latter `malicious'
initialisation errors: they corrupt the injected state and no amount of
post-selection can catch them. Searching exhaustively over all
$180^{\circ}$-symmetric central initialisation patterns at $d = 5$ gives
the placement of figure~\ref{fig:central_injection}: the row directly
above the injected qubit holds $\ket{+}$ on the two data qubits
{\color{cyan}$8$} and {\color{cyan}$13$}. The top row holds $\ket{+}$ on
its four leftmost data qubits {\color{cyan}$4$}, {\color{cyan}$9$},
{\color{cyan}$14$}, {\color{cyan}$19$}. The lower half of the grid is the
$180^{\circ}$ rotation of the upper half, and every remaining data qubit
starts in $\ket{0}$. This pattern coincides
with the initialisation Lao and Criger draw for their own middle
placement (figure 3(b) of~\cite{lao_criger_magic}).

\begin{figure}[!h]
    \centering
    \resizebox{0.50\linewidth}{!}{
    \begin{tikzpicture}[scale=.7,every node/.style={minimum size=0.5cm},on grid]
    \node[fill=white,shape=circle,draw=black] (n4) at (2,18) {${\ket{+}}$};
    \node[fill=white,shape=circle,draw=black] (n3) at (2,14) {${\ket{0}}$};
    \node[fill=white,shape=circle,draw=black] (n2) at (2,10) {${\ket{0}}$};
    \node[fill=white,shape=circle,draw=black] (n1) at (2,6) {${\ket{0}}$};
    \node[fill=white,shape=circle,draw=black] (n0) at (2,2) {${\ket{0}}$};

    \node[fill=white,shape=circle,draw=black] (n9) at (6,18) {${\ket{+}}$};
    \node[fill=white,shape=circle,draw=black] (n8) at (6,14) {${\ket{+}}$};
    \node[fill=white,shape=circle,draw=black] (n7) at (6,10) {${\ket{0}}$};
    \node[fill=white,shape=circle,draw=black] (n6) at (6,6) {${\ket{0}}$};
    \node[fill=white,shape=circle,draw=black] (n5) at (6,2) {${\ket{+}}$};

    \node[fill=white,shape=circle,draw=black] (n14) at (10,18) {${\ket{+}}$};
    \node[fill=white,shape=circle,draw=black] (n13) at (10,14) {${\ket{+}}$};
    \node[fill=white,shape=circle,draw=black] (n12) at (10,10) {${\color{blue}\ket{Y}}$};
    \draw[blue, very thick] (10,10) circle (0.85);
    \node[fill=white,shape=circle,draw=black] (n11) at (10,6) {${\ket{+}}$};
    \node[fill=white,shape=circle,draw=black] (n10) at (10,2) {${\ket{+}}$};

    \node[fill=white,shape=circle,draw=black] (n19) at (14,18) {${\ket{+}}$};
    \node[fill=white,shape=circle,draw=black] (n18) at (14,14) {${\ket{0}}$};
    \node[fill=white,shape=circle,draw=black] (n17) at (14,10) {${\ket{0}}$};
    \node[fill=white,shape=circle,draw=black] (n16) at (14,6) {${\ket{+}}$};
    \node[fill=white,shape=circle,draw=black] (n15) at (14,2) {${\ket{+}}$} ;

    \node[fill=white,shape=circle,draw=black] (n24) at (18,18) {${\ket{0}}$};
    \node[fill=white,shape=circle,draw=black] (n23) at (18,14) {${\ket{0}}$};
    \node[fill=white,shape=circle,draw=black] (n22) at (18,10) {${\ket{0}}$};
    \node[fill=white,shape=circle,draw=black] (n21) at (18,6) {${\ket{0}}$};
    \node[fill=white,shape=circle,draw=black] (n20) at (18,2) {${\ket{+}}$} ;

    \begin{pgfonlayer}{background}
    \draw[zx_red,fill=zx_red,opacity=0.65](1.8,2.2) to[curve through={(0,4)}](1.8,5.8);
    \filldraw[fill=zx_green, opacity=0.65, draw=zx_green] (2.2,2.2) rectangle (5.8,5.8);
    \filldraw[fill=zx_red, opacity=0.65, draw=zx_red] (2.2,6.2) rectangle (5.8,9.8);
    \filldraw[fill=zx_green, opacity=0.65, draw=zx_green] (2.2,10.2) rectangle (5.8,13.8);
    \draw[zx_red,fill=zx_red,opacity=0.65](1.8,10.2) to[curve through={(0,12)}](1.8,13.8);
    \filldraw[fill=zx_red, opacity=0.65, draw=zx_red] (2.2,14.2) rectangle (5.8,17.8);
    \draw[zx_green,fill=zx_green,opacity=0.65](2.2,18.2) to[curve through={(4,20)}](5.8,18.2);

    \draw[zx_green,fill=zx_green,opacity=0.65](9.8,1.8) to[curve through={(8,0)}](6.2,1.8);
    \filldraw[fill=zx_red, opacity=0.65, draw=zx_red] (6.2,2.2) rectangle (9.8,5.8);
    \filldraw[fill=zx_green, opacity=0.65, draw=zx_green] (6.2,6.2) rectangle (9.8,9.8);
    \filldraw[fill=zx_red, opacity=0.65, draw=zx_red] (6.2,10.2) rectangle (9.8,13.8);
    \filldraw[fill=zx_green, opacity=0.65, draw=zx_green] (6.2,14.2) rectangle (9.8,17.8);

    \filldraw[fill=zx_green, opacity=0.65, draw=zx_green] (10.2,2.2) rectangle (13.8,5.8);
    \filldraw[fill=zx_red, opacity=0.65, draw=zx_red] (10.2,6.2) rectangle (13.8,9.8);
    \filldraw[fill=zx_green, opacity=0.65, draw=zx_green] (10.2,10.2) rectangle (13.8,13.8);
    \filldraw[fill=zx_red, opacity=0.65, draw=zx_red] (10.2,14.2) rectangle (13.8,17.8);
    \draw[zx_green,fill=zx_green,opacity=0.65](10.2,18.2) to[curve through={(12,20)}](13.8,18.2);

    \draw[zx_green,fill=zx_green,opacity=0.65](17.8,1.8) to[curve through={(16,0)}](14.2,1.8);
    \filldraw[fill=zx_red, opacity=0.65, draw=zx_red] (14.2,2.2) rectangle (17.8,5.8);
    \filldraw[fill=zx_green, opacity=0.65, draw=zx_green] (14.2,6.2) rectangle (17.8,9.8);
    \draw[zx_red,fill=zx_red,opacity=0.65](18.2,9.8) to[curve through={(20,8)}](18.2,6.2);
    \filldraw[fill=zx_red, opacity=0.65, draw=zx_red] (14.2,10.2) rectangle (17.8,13.8);
    \filldraw[fill=zx_green, opacity=0.65, draw=zx_green] (14.2,14.2) rectangle (17.8,17.8);
    \draw[zx_red,fill=zx_red,opacity=0.65](18.2,17.8) to[curve through={(20,16)}](18.2,14.2);
    \end{pgfonlayer}

    \begin{pgfonlayer}{background}
        \path [-,line width=0.1cm,black,opacity=1] (n0) edge node {} (n20);
        \path [-,line width=0.1cm,black,opacity=1] (n1) edge node {} (n21);
        \path [-,line width=0.1cm,black,opacity=1] (n2) edge node {} (n22);
        \path [-,line width=0.1cm,black,opacity=1] (n3) edge node {} (n23);
        \path [-,line width=0.1cm,black,opacity=1] (n4) edge node {} (n24);

        \path [-,line width=0.1cm,black,opacity=1] (n0) edge node {} (n4);
        \path [-,line width=0.1cm,black,opacity=1] (n5) edge node {} (n9);
        \path [-,line width=0.1cm,black,opacity=1] (n10) edge node {} (n14);
        \path [-,line width=0.1cm,black,opacity=1] (n15) edge node {} (n19);
        \path [-,line width=0.1cm,black,opacity=1] (n20) edge node {} (n24);

        \node[fill=black,circle,draw=black,scale = 0.25] (b0) at (4,4) {};
        \node[fill=black,circle,draw=black,scale = 0.25] (b1) at (4,12) {};
        \node[fill=black,circle,draw=black,text =white,scale = 1.5] (b2) at (4,20) {$+1$};

        \node[fill=black,circle,draw=black,scale = 1.5, text=white] (b3) at (8,0) {$+1$};
        \node[fill=black,circle,draw=black,scale = 0.25] (b4) at (8,8) {};
        \node[fill=black,circle,draw=black,scale = 1.5, text=white] (b5) at (8,16) {$+1$};

        \node[fill=black,circle,draw=black,scale = 1.5, text=white] (b6) at (12,4) {$+1$};
        \node[fill=black,circle,draw=black,scale = 0.25] (b7) at (12,12) {};
        \node[fill=black,circle,draw=black,scale = 1.5, text=white] (b8) at (12,20) {$+1$};

        \node[fill=black,circle,draw=black,scale = 1.5, text=white] (b9) at (16,0) {$+1$};
        \node[fill=black,circle,draw=black,scale = 0.25] (b10) at (16,8) {};
        \node[fill=black,circle,draw=black,scale = 0.25] (b11) at (16,16) {};

        \node[fill=white,circle,draw=black,scale = 1.5, text=black] (a0) at (0,4) {$+1$};
        \node[fill=white,circle,draw=black,scale = 1.5, text=black] (a1) at (0,12) {$+1$};

        \node[fill=white,circle,draw=black,scale = 1.5, text=black] (a2) at (4,8) {$+1$};
        \node[fill=white,circle,draw=black,scale = 0.25] (a3) at (4,16) {};

        \node[fill=white,circle,draw=black,scale = 0.25] (a4) at (8,4) {};
        \node[fill=white,circle,draw=black,scale = 0.25] (a5) at (8,12) {};

        \node[fill=white,circle,draw=black,scale = 0.25] (a6) at (12,8) {};
        \node[fill=white,circle,draw=black,scale = 0.25] (a7) at (12,16) {};

        \node[fill=white,circle,draw=black,scale = 0.25] (a8) at (16,4) {};
        \node[fill=white,circle,draw=black,scale = 1.5, text=black] (a9) at (16,12) {$+1$};

        \node[fill=white,circle,draw=black,scale = 1.5, text=black] (a10) at (20,8) {$+1$};
        \node[fill=white,circle,draw=black,scale = 1.5, text=black] (a11) at (20,16) {$+1$};
    \end{pgfonlayer}
\end{tikzpicture}
    }
    \caption{The Lao--Criger central-qubit injection on a
    distance $d=5$ rotated surface code with green $X$-type and red
    $Z$-type plaquettes, drawn in the conventions of
    figure~\ref{fig:surface_code_Y_inj_post_selection}. The bare
    $\ket{Y}$ state sits on the central data qubit ({\color{cyan}$12$}).
    The $12$ plaquettes labelled `$+1$' are the parity measurements to
    post-select on: the $6$ green $X$-type plaquettes whose data qubits
    all start in $\ket{+}$ and the $6$ red $Z$-type plaquettes whose data
    qubits all start in $\ket{0}$.}
    \label{fig:central_injection}
\end{figure}

\subsection{First round parity measurements and their Pauli webs}

The first full round of parity measurements follows: an $X$-check layer,
then a $Z$-check layer, exactly as in the encoder circuit of the main
text. Consider first a single green $X$-type plaquette whose four data
qubits all start in $\ket{+}$, for instance the plaquette just above the
injected qubit, covering data qubits {\color{cyan}$8$},
{\color{cyan}$9$}, {\color{cyan}$13$} and {\color{cyan}$14$}. Its first
round parity measurement admits the Pauli web drawn in
figure~\ref{fig:central_web_single}: a red $X$-type web that terminates on
the four initial $\ket{+}$ states. As in the main text, this means the
$XXXX$ parity is already fixed to $+1$ by the initial product state, so
the measurement checks an existing stabiliser rather than projecting a
new one. A $-1$ parity outcome implies an initialisation error on one of the
four qubits, in particular on {\color{cyan}$13$}, the neighbour directly
above the injected qubit. The same reasoning applies to every plaquette
whose data qubits share a basis: at $d=5$ there are $6$ such green
$X$-type plaquettes (all $\ket{+}$) and $6$ such red $Z$-type plaquettes
(all $\ket{0}$).

\begin{figure}[!h]
    \centering
    \tdplotsetmaincoords{70}{23}
    \resizebox{0.46\linewidth}{!}{
    \input{fig_central_web_single}
    }
    \caption{The first round Pauli web of one post-selectable plaquette of
    the central-qubit injection: the green $X$-type plaquette covering
    data qubits {\color{cyan}$8$}, {\color{cyan}$9$}, {\color{cyan}$13$},
    {\color{cyan}$14$} (all initialised in $\ket{+}$, labelled in cyan on
    the first time slice). The red $X$-type web terminates on the four
    initial states, so the plaquette's first round parity is
    deterministic and can be post-selected. The blue node on the first
    time slice is the injected $\ket{Y}$ state, a green spider with phase
    $\pi/2$. Drawn with the same pipeline as
    figure~\ref{fig:web_mesh}~\cite{injection_webs_code,
    pyzx_3d_viewer}.}
    \label{fig:central_web_single}
\end{figure}

Figure~\ref{fig:central_web_mesh}(a) overlays the Pauli webs of all $12$
post-selectable plaquettes. Every one terminates on the initial states at
the bottom time slice. The other $12$ plaquettes sit across the boundary
between the $\ket{+}$ and $\ket{0}$ regions, or contain the injected
qubit itself. Their Pauli webs, overlaid in
figure~\ref{fig:central_web_mesh}(b), run only forward in time and reach
none of the initial states: their first round parities are uniformly
random and project new stabilisers instead of checking existing ones, the
same phenomenon as the orange check cube of
figure~\ref{fig:Y_inj_two_cell}. As in the main text, the first round is
therefore followed by a second full round of parity measurements, and one
post-selects the `$+1$' plaquettes of
figure~\ref{fig:central_injection} together with agreement between the
two rounds.

\begin{figure}[!h]
    \centering
    \tdplotsetmaincoords{70}{23}
    \subfloat[\label{fig:central_webs_postselect}The $12$ plaquettes that
    do admit a first round Pauli web, all drawn together. Every one of
    these Pauli webs terminates on the initial states at the bottom. They are
    the plaquettes labelled `$+1$' in
    figure~\ref{fig:central_injection}.]{
    \resizebox{0.4\linewidth}{!}{
    \input{fig_central_webs_det}
    }
    }
    \quad
    \subfloat[\label{fig:central_webs_nodetector}The $12$ plaquettes that
    do not, again all drawn together. These Pauli webs run only forward in time
    to the last time slice and reach none of the initial states. Their
    first round parities are random projections.]{
    \resizebox{0.4\linewidth}{!}{
    \input{fig_central_webs_nodet}
    }
    }
    \caption{The first round Pauli webs of the central-qubit injection at
    $d=5$, in the style of figure~\ref{fig:web_mesh}. Each panel is not
    one big Pauli web but all $12$ individual Pauli webs drawn on top of one
    another. Edges shared by several Pauli webs are drawn several times over.
    Green edges carry a $Z$-type decoration and red edges an $X$-type
    one. The blue node on the first time slice is the injected $\ket{Y}$
    state, a green spider with phase $\pi/2$. Computed and drawn
    with~\cite{injection_webs_code}, using the 3D ZX-diagram viewer
    of~\cite{pyzx_3d_viewer}.}
    \label{fig:central_web_mesh}
\end{figure}

\subsection{What post-selection can and cannot catch}

The central scheme has $12$ post-selectable plaquettes against the corner
scheme's $10$, but the more important difference is where they sit. In
the corner scheme, the four malicious initialisation errors are Pauli $X$
and $Z$ on the injected corner qubit ({\color{cyan}$4$}) and Pauli $X$
and $Y$ on its $\ket{0}$ neighbour ({\color{cyan}$9$}). The neighbour
touches one weight-two boundary check and two bulk plaquettes, and
each of these either contains the (unknown) injected qubit or crosses
the $\ket{+}$/$\ket{0}$ boundary, so none of them is post-selectable. In
the central scheme every neighbour of the injected qubit belongs to four
plaquettes, only two of which contain the injected qubit, and the
pattern of figure~\ref{fig:central_injection} gives each neighbour a
post-selectable plaquette. Figure~\ref{fig:central_web_single} showed the
one checking {\color{cyan}$13$}. The only malicious initialisation errors
left are Pauli $X$ and $Z$ on the injected qubit itself.
Figure~\ref{fig:malicious_2panel} draws the comparison.

\begin{figure}[H]
    \centering
    \resizebox{0.8\linewidth}{!}{\begin{tabular}{cc}
\begin{tikzpicture}[scale=0.68]
\fill[green!14] (0,0) arc[start angle=-90.0, end angle=-270.0, radius=0.500] -- cycle;
\fill[green!14] (0,2) arc[start angle=-90.0, end angle=-270.0, radius=0.500] -- cycle;
\fill[red!12] (0.0,0.0) rectangle (1.0,1.0);
\fill[green!14] (0.0,1.0) rectangle (1.0,2.0);
\fill[red!12] (0.0,2.0) rectangle (1.0,3.0);
\fill[green!14] (0.0,3.0) rectangle (1.0,4.0);
\fill[red!12] (0,4) arc[start angle=180.0, end angle=0.0, radius=0.500] -- cycle;
\fill[red!12] (1,0) arc[start angle=180.0, end angle=360.0, radius=0.500] -- cycle;
\fill[green!14] (1.0,0.0) rectangle (2.0,1.0);
\fill[red!12] (1.0,1.0) rectangle (2.0,2.0);
\fill[green!14] (1.0,2.0) rectangle (2.0,3.0);
\fill[red!12] (1.0,3.0) rectangle (2.0,4.0);
\fill[red!12] (2.0,0.0) rectangle (3.0,1.0);
\fill[green!14] (2.0,1.0) rectangle (3.0,2.0);
\fill[red!12] (2.0,2.0) rectangle (3.0,3.0);
\fill[green!14] (2.0,3.0) rectangle (3.0,4.0);
\fill[red!12] (2,4) arc[start angle=180.0, end angle=0.0, radius=0.500] -- cycle;
\fill[red!12] (3,0) arc[start angle=180.0, end angle=360.0, radius=0.500] -- cycle;
\fill[green!14] (3.0,0.0) rectangle (4.0,1.0);
\fill[red!12] (3.0,1.0) rectangle (4.0,2.0);
\fill[green!14] (3.0,2.0) rectangle (4.0,3.0);
\fill[red!12] (3.0,3.0) rectangle (4.0,4.0);
\fill[green!14] (4,1) arc[start angle=-90.0, end angle=90.0, radius=0.500] -- cycle;
\fill[green!14] (4,3) arc[start angle=-90.0, end angle=90.0, radius=0.500] -- cycle;
\node[draw=none,fill=none,scale=0.55] at (0.5,0.5) {$+1$};
\node[draw=none,fill=none,scale=0.55] at (0.5,2.5) {$+1$};
\node[draw=none,fill=none,scale=0.55] at (1.5,-0.27) {$+1$};
\node[draw=none,fill=none,scale=0.55] at (1.5,1.5) {$+1$};
\node[draw=none,fill=none,scale=0.55] at (2.5,0.5) {$+1$};
\node[draw=none,fill=none,scale=0.55] at (2.5,3.5) {$+1$};
\node[draw=none,fill=none,scale=0.55] at (3.5,-0.27) {$+1$};
\node[draw=none,fill=none,scale=0.55] at (3.5,2.5) {$+1$};
\node[draw=none,fill=none,scale=0.55] at (4.27,1.5) {$+1$};
\node[draw=none,fill=none,scale=0.55] at (4.27,3.5) {$+1$};
\draw[black, dashed, very thick] (0.0,3.0) rectangle (1.0,4.0);
\draw[black, dashed, very thick] (0,4) arc[start angle=180.0, end angle=0.0, radius=0.500] -- cycle;
\draw[black, dashed, very thick] (1.0,3.0) rectangle (2.0,4.0);
\draw[black, fill=zx_green] (0,0) circle (0.14);
\draw[black, fill=zx_green] (0,1) circle (0.14);
\draw[black, fill=zx_green] (0,2) circle (0.14);
\draw[black, fill=zx_green] (0,3) circle (0.14);
\draw[black, fill=blue!40] (0,4) circle (0.14);
\draw[black, fill=zx_green] (1,0) circle (0.14);
\draw[black, fill=zx_green] (1,1) circle (0.14);
\draw[black, fill=zx_green] (1,2) circle (0.14);
\draw[black, fill=zx_green] (1,3) circle (0.14);
\draw[black, fill=zx_red] (1,4) circle (0.14);
\draw[black, fill=zx_green] (2,0) circle (0.14);
\draw[black, fill=zx_green] (2,1) circle (0.14);
\draw[black, fill=zx_green] (2,2) circle (0.14);
\draw[black, fill=zx_red] (2,3) circle (0.14);
\draw[black, fill=zx_red] (2,4) circle (0.14);
\draw[black, fill=zx_green] (3,0) circle (0.14);
\draw[black, fill=zx_green] (3,1) circle (0.14);
\draw[black, fill=zx_red] (3,2) circle (0.14);
\draw[black, fill=zx_red] (3,3) circle (0.14);
\draw[black, fill=zx_red] (3,4) circle (0.14);
\draw[black, fill=zx_green] (4,0) circle (0.14);
\draw[black, fill=zx_red] (4,1) circle (0.14);
\draw[black, fill=zx_red] (4,2) circle (0.14);
\draw[black, fill=zx_red] (4,3) circle (0.14);
\draw[black, fill=zx_red] (4,4) circle (0.14);
\draw[blue, very thick] (0, 4) circle (0.34);
\node[draw=none,fill=none,text=blue,scale=0.75] at (-0.05,3.38) {$X, Z$};
\draw[violet, very thick] (1, 4) circle (0.34);
\node[draw=none,fill=none,text=violet,scale=0.75] at (1.05,4.62) {$X, Y$};
\node[draw=none,fill=none,text=cyan,scale=0.7] at (1.42,3.7) {9};
\end{tikzpicture} &
\begin{tikzpicture}[scale=0.68]
\fill[green!14] (0,0) arc[start angle=-90.0, end angle=-270.0, radius=0.500] -- cycle;
\fill[green!14] (0,2) arc[start angle=-90.0, end angle=-270.0, radius=0.500] -- cycle;
\fill[red!12] (0.0,0.0) rectangle (1.0,1.0);
\fill[green!14] (0.0,1.0) rectangle (1.0,2.0);
\fill[red!12] (0.0,2.0) rectangle (1.0,3.0);
\fill[green!14] (0.0,3.0) rectangle (1.0,4.0);
\fill[red!12] (0,4) arc[start angle=180.0, end angle=0.0, radius=0.500] -- cycle;
\fill[red!12] (1,0) arc[start angle=180.0, end angle=360.0, radius=0.500] -- cycle;
\fill[green!14] (1.0,0.0) rectangle (2.0,1.0);
\fill[red!12] (1.0,1.0) rectangle (2.0,2.0);
\fill[green!14] (1.0,2.0) rectangle (2.0,3.0);
\fill[red!12] (1.0,3.0) rectangle (2.0,4.0);
\fill[red!12] (2.0,0.0) rectangle (3.0,1.0);
\fill[green!14] (2.0,1.0) rectangle (3.0,2.0);
\fill[red!12] (2.0,2.0) rectangle (3.0,3.0);
\fill[green!14] (2.0,3.0) rectangle (3.0,4.0);
\fill[red!12] (2,4) arc[start angle=180.0, end angle=0.0, radius=0.500] -- cycle;
\fill[red!12] (3,0) arc[start angle=180.0, end angle=360.0, radius=0.500] -- cycle;
\fill[green!14] (3.0,0.0) rectangle (4.0,1.0);
\fill[red!12] (3.0,1.0) rectangle (4.0,2.0);
\fill[green!14] (3.0,2.0) rectangle (4.0,3.0);
\fill[red!12] (3.0,3.0) rectangle (4.0,4.0);
\fill[green!14] (4,1) arc[start angle=-90.0, end angle=90.0, radius=0.500] -- cycle;
\fill[green!14] (4,3) arc[start angle=-90.0, end angle=90.0, radius=0.500] -- cycle;
\node[draw=none,fill=none,scale=0.55] at (-0.27,0.5) {$+1$};
\node[draw=none,fill=none,scale=0.55] at (-0.27,2.5) {$+1$};
\node[draw=none,fill=none,scale=0.55] at (0.5,1.5) {$+1$};
\node[draw=none,fill=none,scale=0.55] at (0.5,4.27) {$+1$};
\node[draw=none,fill=none,scale=0.55] at (1.5,-0.27) {$+1$};
\node[draw=none,fill=none,scale=0.55] at (1.5,3.5) {$+1$};
\node[draw=none,fill=none,scale=0.55] at (2.5,0.5) {$+1$};
\node[draw=none,fill=none,scale=0.55] at (2.5,4.27) {$+1$};
\node[draw=none,fill=none,scale=0.55] at (3.5,-0.27) {$+1$};
\node[draw=none,fill=none,scale=0.55] at (3.5,2.5) {$+1$};
\node[draw=none,fill=none,scale=0.55] at (4.27,1.5) {$+1$};
\node[draw=none,fill=none,scale=0.55] at (4.27,3.5) {$+1$};
\draw[black, dashed, very thick] (1.0,2.0) rectangle (2.0,3.0);
\draw[black, dashed, very thick] (1.0,3.0) rectangle (2.0,4.0);
\draw[black, dashed, very thick] (2.0,2.0) rectangle (3.0,3.0);
\draw[black, dashed, very thick] (2.0,3.0) rectangle (3.0,4.0);
\draw[black, fill=zx_red] (0,0) circle (0.14);
\draw[black, fill=zx_red] (0,1) circle (0.14);
\draw[black, fill=zx_red] (0,2) circle (0.14);
\draw[black, fill=zx_red] (0,3) circle (0.14);
\draw[black, fill=zx_green] (0,4) circle (0.14);
\draw[black, fill=zx_green] (1,0) circle (0.14);
\draw[black, fill=zx_red] (1,1) circle (0.14);
\draw[black, fill=zx_red] (1,2) circle (0.14);
\draw[black, fill=zx_green] (1,3) circle (0.14);
\draw[black, fill=zx_green] (1,4) circle (0.14);
\draw[black, fill=zx_green] (2,0) circle (0.14);
\draw[black, fill=zx_green] (2,1) circle (0.14);
\draw[black, fill=blue!40] (2,2) circle (0.14);
\draw[black, fill=zx_green] (2,3) circle (0.14);
\draw[black, fill=zx_green] (2,4) circle (0.14);
\draw[black, fill=zx_green] (3,0) circle (0.14);
\draw[black, fill=zx_green] (3,1) circle (0.14);
\draw[black, fill=zx_red] (3,2) circle (0.14);
\draw[black, fill=zx_red] (3,3) circle (0.14);
\draw[black, fill=zx_green] (3,4) circle (0.14);
\draw[black, fill=zx_green] (4,0) circle (0.14);
\draw[black, fill=zx_red] (4,1) circle (0.14);
\draw[black, fill=zx_red] (4,2) circle (0.14);
\draw[black, fill=zx_red] (4,3) circle (0.14);
\draw[black, fill=zx_red] (4,4) circle (0.14);
\draw[blue, very thick] (2, 2) circle (0.34);
\node[draw=none,fill=none,text=blue,scale=0.75] at (1.95,1.38) {$X, Z$};
\draw[violet, very thick] (2, 3) circle (0.34);
\node[draw=none,fill=none,text=cyan,scale=0.7] at (2.42,2.7) {13};
\end{tikzpicture}\\
{\footnotesize (a) corner scheme} & {\footnotesize (b) central scheme}
\end{tabular}}
    \caption{Where the malicious initialisation errors live, at
    $d = 5$. Data qubits are coloured by their initial state (green
    $\ket{+}$, red $\ket{0}$, blue the injected $\ket{Y}$), the
    verified post-selectable plaquettes are marked `$+1$', and the
    dashed outlines are the plaquettes containing the ringed
    neighbour. (a) Corner scheme: the injected qubit (blue ring,
    malicious $X, Z$) and its $\ket{0}$ neighbour
    ({\color{cyan}$9$}, violet ring, malicious $X, Y$). Every dashed
    plaquette either contains the injected qubit or crosses the
    $\ket{+}$/$\ket{0}$ boundary, so none carries a `$+1$'. (b)
    Central scheme: every neighbour of the injected qubit, such as
    {\color{cyan}$13$}, has a post-selectable dashed plaquette, and
    only $X, Z$ on the injected qubit itself remain
    malicious~\cite{injection_webs_code}.}
    \label{fig:malicious_2panel}
\end{figure}

At larger distances the pattern extends by the same quadrant rule,
drawn at $d = 9$ in figure~\ref{fig:pinwheel_d9}: the $\ket{+}$
qubits fill a wedge above the centre, its edges stepped like a
staircase (the row $k$ steps up holds $2k$ of them, beginning $k$
columns left of centre), the wedge
below is its $180^{\circ}$ rotation, and the remaining left and right
wedges start in $\ket{0}$: four quadrants pinwheeling around the
injected qubit. The same wedges keep growing at larger
distances (dotted continuations in the figure). The Pauli webs
certify,
at every odd distance checked ($d = 3, 5, 7, 9$): the post-selection
set consists of $(d^2-1)/2$ individual plaquettes, the only malicious initialisation errors are
$X$ and $Z$ on the injected qubit, and all three logical classes are
transported from the injected qubit to the encoded logical qubit, so
the placement injects arbitrary states, not only $\ket{Y}$. We return
to the two schemes under circuit-level noise in
section~\ref{app:circuitnoise}.

\begin{figure}[tbp]
    \centering
    \resizebox{0.52\linewidth}{!}{\input{fig_pinwheel_d9}}
    \caption{The central-qubit initialisation pattern at $d = 9$:
    green data qubits start in $\ket{+}$, red in $\ket{0}$, the blue
    ringed centre carries the injected state. The dashed outlines are
    the two $\ket{+}$ staircase wedges, each row one qubit wider than
    the one before and offset one column to the left. The lower wedge
    is the $180^{\circ}$ rotation of the upper, and the four wedges
    pinwheel around the injected qubit.}
    \label{fig:pinwheel_d9}
\end{figure}

\section{Circuit-level noise: corner versus central}
\label{app:circuitnoise}

The Pauli web analysis of the previous sections identifies the
initialisation blind
spots, but on hardware the injected state is mostly damaged by faults
inside the extraction circuitry itself: a single failed \textsc{cnot}
can spread onto the injected qubit before any check has been
established, and whether it is caught depends on the gate ordering,
not only on the placement. The headline numbers of the injection
literature are statements at exactly this level: they are logical
error rates ($p_{\text{L}}$) of the injected state per unit physical
rate with all noise parameters set equal, $p_L \approx 46p/15$ for
Li's (unrotated surface code) corner scheme and the claimed $49p/15$
(corner) and $34p/15$ (central) of Lao and Criger (rotated surface
code). A circuit-level fault enumeration, going through every possible
single fault and classifying each as detected, harmless, or
malignant, is therefore what decides which scheme injects
better.

The circuit-level noise model of this section is inherited from
references~\cite{Li_2015, lao_criger_magic} and has four noise
parameters:
\begin{enumerate}
\item $p_2$: each \textsc{cnot} fails as one of the $15$ non-identity
two-qubit Pauli classes $P_c \otimes P_t$, each with probability
$p_2/15$, deposited on the gate's two outgoing wire segments;
\item $p_I$: each single-qubit initialisation suffers an error with
probability $p_I$;
\item $p_1$: each single-qubit gate is followed by a Pauli $X$, $Y$
or $Z$, each with probability $p_1/3$;
\item $p_M$: each ancilla readout suffers an outcome error with
probability $p_M$.
\end{enumerate}

Figure~\ref{fig:circuit3d} shows the two post-selected rounds as a
spacetime ZX-diagram at distance $3$. The flat circuit of
appendix~\ref{app:flatcircuit} redraws this as a 2D circuit
diagram, each check cell
now an explicit series of \textsc{cnot}s. Every fault class is a pair
of Pauli decorations on the two outgoing wire segments of its gate.

\begin{figure}[!h]
    \centering
    \tdplotsetmaincoords{65}{33}
    \resizebox{0.44\linewidth}{!}{
    \input{fig_circuit3d}
    }
    \caption{The circuit of figure~\ref{fig:li_noise} as a spacetime
    ZX-diagram in the style of figure~\ref{fig:web_mesh}, time running
    upwards. Every parity measurement is written out as its series of CNOTs:
    each ancilla is a vertical wire from its initialisation spider to
    its readout. The blue node at the bottom corner is the injected
    $\ket{Y}$ state. The data wires continue past the second round as
    open outputs. The two grey constant-time slices mark the ancilla
    initialisation layer of each round. The fault locations of
    figure~\ref{fig:li_noise} are the wire segments between the gates:
    the segment after each initialisation ($p_I$), the two outgoing
    segments of each CNOT ($P_c \otimes P_t$) and the segment before
    each readout spider ($p_M$). The edge connecting the two spiders of a
    CNOT carries no fault.}
    \label{fig:circuit3d}
\end{figure}

Which faults are malignant depends on what happens after the two
post-selected rounds, and references~\cite{Li_2015, lao_criger_magic}
are explicit about it: the patch is grown and ordinary decoded error
correction continues, with no further post-selection. A fault is
malignant only if no detector ever fires, comparisons against the
later rounds included, and it still flips the injected logical
$\bar{Y}$. We
adopt this convention throughout, implemented by appending further
rounds to the spacetime diagram before solving the Pauli webs.
Figure~\ref{fig:mal_webs} shows the three possible fates of a fault
at Pauli web level: caught by a detector web inside the two rounds,
malignant, or caught by a Pauli web that closes through the next round.

\begin{figure}[H]
    \centering
    \subfloat[detected]{%
    \tdplotsetmaincoords{65}{33}%
    \resizebox{0.27\linewidth}{!}{\input{fig_malweb_detected}}}%
    \hfill%
    \subfloat[malignant]{%
    \tdplotsetmaincoords{65}{33}%
    \resizebox{0.27\linewidth}{!}{\input{fig_malweb_malignant}}}%
    \hfill%
    \subfloat[caught by the next round]{%
    \tdplotsetmaincoords{65}{33}%
    \resizebox{0.27\linewidth}{!}{\input{fig_malweb_nextround}}}
    \caption{Why a segment is violet. (a) A round-$2$ $X$ fault
    (violet arrow) lands on a detector web: a detector fires. (b) A
    malignant fault: the round-$1$ $Z$ fault on the injected qubit at
    its $Z$-check tap, with the nearest X-decorated detector web
    drawn: its red decorations stop short of the violet segment, the
    fault lies on the logical $Y$ correlator
    (figure~\ref{fig:logical_cor} style, not drawn), and no detector
    web, in these rounds or any later round, reaches it. (c) A
    round-$2$ $Z$ fault that no Pauli web of the two drawn rounds sees: the
    closed web that catches it runs out of the open outputs into the
    next round (not drawn), so ordinary decoded error correction
    detects it and its segment is not violet.}
    \label{fig:mal_webs}
\end{figure}

\subsection{The error budget}
\label{sub:budget}

For the Tomita--Svore ordering that they use, Lao and Criger claim
the logical error rates of their two injection schemes to be
\begin{equation*}
p_L^{\text{corner}} = \tfrac{3}{5}\,p_2 + 2\,p_I + \tfrac{2}{3}\,p_1,
\qquad
p_L^{\text{central}} = \tfrac{3}{5}\,p_2 + p_I + \tfrac{2}{3}\,p_1
\end{equation*}
(their $p_{\text{IN}}$ is our $p_I$). Under this convention the Pauli webs
give instead, independently of the distance:
\begin{equation*}
p_L^{\text{corner}} = \tfrac{8}{15}\,p_2 + 2\,p_I + \tfrac{2}{3}\,p_1,
\qquad
p_L^{\text{central}} = \tfrac{7}{15}\,p_2 + p_I + \tfrac{2}{3}\,p_1,
\end{equation*}
with no malignant readout class in either scheme, which is why
$p_M$ does not appear and why Lao and Criger drop it. Setting all
rates equal gives $48p/15$ and $32p/15$ against their claimed
$49p/15$ and $34p/15$. The differences are not numerical
disagreements about the circuit. Three of their listed classes, one on the
corner circuit and two on the central one, are
products of two classes they also list, and such a product flips the
logical $\bar{Y}$ an even number of times (effectively no flips).
Figure~\ref{fig:mal_decoded} draws the complete unprotected set of
each scheme. In the corner scheme, the eight malignant CNOT classes
and the two malignant initialisation flips all sit on the injected
qubit and its neighbour. In the central scheme, the seven classes and
the single flip all sit around the injected qubit at the centre. In
both schemes every malignant fault lives in the first round.
\begin{figure}[H]
    \centering
    \subfloat[corner: $8$ classes, $2$ flips]{
    \tdplotsetmaincoords{65}{33}
    \resizebox{0.42\linewidth}{!}{\input{fig_mal_decoded_corner}}}
    \hfill
    \subfloat[central: $7$ classes, $1$ flip]{
    \tdplotsetmaincoords{65}{33}
    \resizebox{0.42\linewidth}{!}{\input{fig_mal_decoded_central}}}
    \caption{The complete unprotected sets under the decoded
    convention, at $d = 3$ with the Tomita--Svore ordering: violet
    segments carry the malignant CNOT classes and the malignant
    initialisation flips.}
    \label{fig:mal_decoded}
\end{figure}

Figure~\ref{fig:lc_circuit} redraws the sensitive part of Lao and
Criger's own circuit with their colour scheme, and the differences can
be read off it gate by gate: at the green \textsc{cnot}, one of
their listed classes is the product of the other two, and products of
two malignant classes flip the logical $\bar{Y}$ an even number of
times, so
no gate can carry exactly $3$ or $6$ malignant classes. Our per-gate lists
otherwise reproduce theirs exactly.

\begin{figure}[H]
    \centering
    \resizebox{0.5\linewidth}{!}{
    \begin{tikzpicture}
\begin{yquant}[operators/every barrier/.append style={red, thick}]
qubit {$X2\;\ket{+}$} x2;
qubit {$Z2\;\ket{0}$} z2;
qubit {$D2\;({\color{cyan}1})$} d2;
qubit {$D3\;({\color{cyan}2})$} d3;
qubit {$D5\;({\color{cyan}4})$} d5;
qubit {$D6\;({\color{cyan}5})$} d6;
[blue, thick, every control/.append style={fill=blue, draw=blue}, every control line/.append style={draw=blue, thick}] cnot z2 | d6;
cnot z2 | d5;
[green!60!black, thick, every control/.append style={fill=green!60!black, draw=green!60!black}, every control line/.append style={draw=green!60!black, thick}] cnot z2 | d3;
cnot d6 | x2;
[red, thick, every control/.append style={fill=red, draw=red}, every control line/.append style={draw=red, thick}] cnot d3 | x2;
measure x2;
measure z2;
\end{yquant}
\end{tikzpicture}
    }
    \caption{The sensitive part of Lao and Criger's corner circuit,
    redrawn from figure~4(b) of~\cite{lao_criger_magic} with their
    colour scheme: the blue CNOT couples $D6$ (our {\color{cyan}$5$}) to
    the $Z2$ ancilla in the first time step, the green CNOT couples the
    injected qubit $D3$ (our {\color{cyan}$2$}) to $Z2$ in the third,
    and the red CNOT couples the $X2$ ancilla to $D3$ in the fourth.
    The $X2$ ancilla is prepared in $\ket{+}$ and read out in the $X$
    basis, $Z2$ in $\ket{0}$ and the $Z$ basis. Their malignant
    classes all live on these three gates. Our enumeration agrees with Lao and
    Criger's lists on the blue ($4$) and red ($2$) gates class for
    class and correct the
    green gate from $3$ to $2$: the class $Y_C X_T$ is the product of
    $Z_C I_T$ and $X_C X_T$ and therefore flips the logical $\bar{Y}$
    an even number of times.}
    \label{fig:lc_circuit}
\end{figure}

\subsection{Orderings and placements}
\label{sub:orderings}

The two orderings are drawn in figures~\ref{fig:nz_schedule}
and~\ref{fig:sched_slices}. Valid orderings must trace an `S' or `Z'
shape across each plaquette to keep neighbouring checks commuting, a
rule due to Tomita and Svore~\cite{tomita2014low}. The N/Z interleave
is one such choice, and Tomita and Svore's own choice (standard order
for $X$-checks, a deliberate reorder for $Z$-checks so that hook
errors land perpendicular to the logical) is the ordering Lao and
Criger adopt.

\begin{figure}[H]
    \centering
    \resizebox{0.4\linewidth}{!}{
    \begin{tikzpicture}[scale=.7,every node/.style={minimum size=0.5cm},on grid]
    \node[fill=white,shape=circle,draw=black] (a1) at (0,4) {$2$};
    \node[fill=white,shape=circle,draw=black] (a2) at (4,4) {$4$};
    \node[fill=white,shape=circle,draw=black] (a3) at (0,0) {$1$};
    \node[fill=white,shape=circle,draw=black] (a4) at (4,0) {$3$};
    \begin{pgfonlayer}{background}
    \filldraw[fill=zx_green, opacity=0.65, draw=zx_green] (0.2,0.2) rectangle (3.8,3.8);
    \draw[->,black,line width=0.06cm] (0,3.3) -- (0,0.7);
    \draw[->,black,line width=0.06cm] (0.5,0.5) -- (3.5,3.5);
    \draw[->,black,line width=0.06cm] (4,3.3) -- (4,0.7);
    \end{pgfonlayer}
    \node[draw=none,fill=none] at (2,-1.4) {$X$-type: `N'};
    \begin{scope}[shift={(8,0)}]
    \node[fill=white,shape=circle,draw=black] (b1) at (0,4) {$1$};
    \node[fill=white,shape=circle,draw=black] (b2) at (4,4) {$2$};
    \node[fill=white,shape=circle,draw=black] (b3) at (0,0) {$3$};
    \node[fill=white,shape=circle,draw=black] (b4) at (4,0) {$4$};
    \begin{pgfonlayer}{background}
    \filldraw[fill=zx_red, opacity=0.65, draw=zx_red] (0.2,0.2) rectangle (3.8,3.8);
    \draw[->,black,line width=0.06cm] (0.7,4) -- (3.3,4);
    \draw[->,black,line width=0.06cm] (3.5,3.5) -- (0.5,0.5);
    \draw[->,black,line width=0.06cm] (0.7,0) -- (3.3,0);
    \end{pgfonlayer}
    \node[draw=none,fill=none] at (2,-1.4) {$Z$-type: `Z'};
    \end{scope}
    \end{tikzpicture}
    }
    \caption{The interleaved extraction schedule used for the
    circuit-level fault enumeration. Each ancilla visits its four data
    qubits in
    four time steps: $X$-type plaquettes in the `N' pattern, $Z$-type
    plaquettes in the transposed `Z' pattern, so that all checks of both
    types run simultaneously without collisions.}
    \label{fig:nz_schedule}
\end{figure}

The same enumeration, run over both orderings and both injection schemes,
gives table~\ref{tab:decoded}.

\begin{table}[H]
\centering
\begin{tabular}{lccccc}
 & \multicolumn{2}{c}{CNOT ($p_2$), $d=3$} &
 \multicolumn{2}{c}{CNOT ($p_2$), $d=5$} & init ($p_I$)\\
 & N/Z & TS & N/Z & TS & \\
\hline
corner  & $13/15$ & $8/15$ & $13/15$ & $8/15$ & $2$\\
central & $13/15$ & $7/15$ & $7/15$  & $7/15$ & $1$\\
\end{tabular}
\caption{Decoded-convention error budgets of the two injection schemes under
the two orderings.}
\label{tab:decoded}
\end{table}

\begin{figure}[H]
    \centering
    \resizebox{0.9\linewidth}{!}{\input{fig_sched_slices_d3}}
    \caption{One syndrome extraction cycle of the two orderings of
    table~\ref{tab:decoded}, layer by layer at $d = 3$, viewed from above in the
    conventions of figure~\ref{fig:slices_opt_d5}: green marks the
    control end and red the target end of every \textsc{cnot}, light
    red and light green faces are the X- and Z-checks, grey dots the
    idle data qubits. Top row: the generic N/Z interleave. Bottom row:
    Tomita and Svore's own S/Z ordering~\cite{tomita2014low}.}
    \label{fig:sched_slices}
\end{figure}

Table~\ref{tab:decoded} shows three things:
\begin{enumerate}
\item the ordering matters more than the placement: Tomita and
Svore's ordering beats the generic N/Z interleave at every entry,
because it steers hook errors perpendicular to the logical;
\item the central-qubit scheme's robust advantage is the initialisation
column, one unheralded flip against two, exactly the Pauli web level
blind spots of section~\ref{app:injection};
\item boundary effects can enter through the ordering: under N/Z the
central CNOT term improves from $13/15$ at $d = 3$ to $7/15$ at
$d = 5$, where
the centre's neighbourhood no longer touches the boundary, while the
TS entries are identical at both distances.
\end{enumerate}

\subsection{Optimised schedules and Monte-Carlo validation}
\label{sub:mc}

The Pauli webs make the fault enumeration cheap enough to search over
schedules. At $d = 3$ we score every uniform interleave, one visiting
order for the $X$-checks and one for the $Z$-checks: $76$ of the
$576$ ordered pairs measure the correct stabiliser group. Some of
these are collision-free and run in four parallel layers, like
figure~\ref{fig:nz_schedule}. In the rest, a data qubit is visited
by two checks in the same layer and the colliding gates run one
after the other, as in Li's depth-six circuits~\cite{Li_2015}. Each
schedule is scored by its leading-order logical error rate under
minimum-weight matching, since a detected fault can still be
mis-corrected by the decoder.

Within the collision-free schedules, the Tomita--Svore ordering is
already the best for both injection schemes: nothing parallel beats $8/15$
(corner) or $7/15$ (central). Serialised schedules do better:
\begin{center}
\begin{tabular}{lcccc}
 & CNOT ($p_2$) & init ($p_{I}$) & rates equal & claimed~\cite{lao_criger_magic}\\
\hline
corner, Tomita--Svore    & $8/15$ & $2$ & $48p/15$ & $49p/15$\\
corner, serialised       & $6/15$ & $2$ & $46p/15$ & \\
central, Tomita--Svore   & $7/15$ & $1$ & $32p/15$ & $34p/15$\\
central, serialised      & $3/15$ & $1$ & $\mathbf{28p/15}$ & \\
\end{tabular}
\end{center}
The corner's serialised version is exactly Li's $2p_2/5$: his
schedule transfers from the unrotated to the rotated surface code
unchanged, and both enumerations leave six malignant classes on the
same three gates. Li attributes them to the gates as $3+1+2$. The
Pauli webs give $2+2+2$. A single gate can never carry exactly three: two
malignant classes on the same gate have a product that is again a
class of that gate, fires no detector, and flips $\bar{Y}$ twice,
so a claimed third class is one such product, exactly as at Lao and
Criger's green gate in figure~\ref{fig:lc_circuit}. Li's total was
right. Only its split over the gates was not.
Figure~\ref{fig:li_split} draws the comparison. The $2+2+2$
attribution is verified by direct \texttt{stim}~\cite{gidney2021stim} sampling of the serialised schedule
(figure~\ref{fig:mc_lc}).

\begin{figure}[H]
    \centering
    \resizebox{0.78\linewidth}{!}{\begin{tabular}{c}
\begin{tikzpicture}
\begin{yquant}
qubit {$X2\;\ket{+}$} x2;
qubit {$Z2\;\ket{0}$} z2;
qubit {$D3\;({\color{cyan}2})$} d3;
qubit {$D6\;({\color{cyan}5})$} d6;
[blue, thick, every control/.append style={fill=blue, draw=blue}, every control line/.append style={draw=blue, thick}] cnot z2 | d6;
[green!60!black, thick, every control/.append style={fill=green!60!black, draw=green!60!black}, every control line/.append style={draw=green!60!black, thick}] cnot z2 | d3;
[red, thick, every control/.append style={fill=red, draw=red}, every control line/.append style={draw=red, thick}] cnot d3 | x2;
measure x2;
measure z2;
\end{yquant}
\end{tikzpicture}\\[3mm]
\begin{tabular}{lcccc}
 & {\color{blue}blue gate} & {\color{green!60!black}green gate} &
 {\color{red}red gate} & total\\
\hline
webs & $X_CX_T,\ Y_CX_T$ & $Z_CI_T,\ X_CX_T$ & $Y_CZ_T,\ Z_CZ_T$ &
 $2+2+2 = 6$\\
Li & $X_CX_T,\ Y_CX_T$ & $Z_CI_T,\ X_CX_T,\ \mathbf{Y_CX_T}$ &
 $Z_CZ_T$ & $2+3+1 = 6$\\
\end{tabular}
\end{tabular}}
    \caption{The three sensitive gates of the corner's serialised
    optimum, in the conventions of figure~\ref{fig:lc_circuit}, with
    the malignant classes each enumeration attributes to them. The
    two lists agree on the blue gate, and both give six classes in
    total. On the green gate Li's bold $Y_C X_T$ is the product of
    $Z_C I_T$ and $X_C X_T$ and is not malignant. The missing class
    is $Y_C Z_T$ on the red gate.}
    \label{fig:li_split}
\end{figure} The best schedule found for the central-qubit scheme (one example:
$X$-checks visiting NW, NE, SE, SW and $Z$-checks NE, SE, NW, SW,
with colliding gates serialised in face order) leaves three malignant
classes, all on the two CNOTs that touch the injected qubit
(figure~\ref{fig:circuit3d_opt}).
Figure~\ref{fig:slices_opt_d5} lays the cycle out slice by slice at
$d = 5$, with the two serialisation sub-layers of the colliding
layers drawn separately. The resulting $28p/15$ is
$18\%$ below the $34p/15$ of~\cite{lao_criger_magic}. %
The $576$-pair search forces all $X$-checks to share one visiting
order and all $Z$-checks another. Dropping that restriction, each of
the eight checks at $d = 3$ chooses its own order, $24^8$ schedules
in total, too many to search over exhaustively. We searched this space by
starting from a random schedule, re-ordering one check at
a time, keeping any improvement, and restarting forty times from
fresh random schedules, $2500$ schedules scored in total. Nothing beat $3/15$.
Serialisation also costs nothing in acceptance.

\begin{figure}[tbp]
    \centering
    \tdplotsetmaincoords{65}{33}
    \resizebox{0.40\linewidth}{!}{
    \input{fig_circuit3d_opt}
    }
    \caption{The optimised central schedule as a spacetime ZX-diagram
    in the style of figure~\ref{fig:circuit3d}, with the
    complete unprotected set of the scheme in violet: the
    initialisation segment of the injected qubit ($1$ flip class) and
    the outgoing segments of the two CNOTs that touch it in the first
    round ($2+1$ malignant classes). Every other wire segment of the
    whole spacetime feeds a detector or is harmless. Compare
    figure~\ref{fig:mal_decoded}: optimising the placement and the
    schedule has pulled the entire residual fault set onto the injected
    qubit itself~\cite{injection_webs_code}.}
    \label{fig:circuit3d_opt}
\end{figure}

\begin{figure}[tbp]
    \centering
    \resizebox{0.60\linewidth}{!}{\input{fig_slices_opt_d5}}
    \caption{One syndrome extraction cycle of the optimised central
    schedule at $d = 5$, slice by slice, viewed from above as a 2D
    ZX diagram: for every \textsc{cnot} of the slice, green marks the
    control end and red the target end. Light red and light green
    faces are the X- and Z-checks, grey dots the idle data qubits,
    and the blue circle the injected qubit. The schedule is
    serialised, not parallel: layers $1$ and $2$ each contain eight
    colliding pairs (a Z-check and an X-check tapping the same data
    qubit), executed serially, $a$ before $b$, in the face-scan order
    that \texttt{stim} and the spacetime diagrams use. Both sub-layers of gates
    sit inside the same \texttt{stim} tick, the serialisation being the gate
    order within it. The injected qubit's own collision sits in layer
    $2$. Layers $3$ and $4$ are collision-free and run fully
    parallel~\cite{injection_webs_code}.}
    \label{fig:slices_opt_d5}
\end{figure}

The malignant classes themselves do not depend on the distance:
the same classes sit on the same gates at every odd distance from $3$ to $11$, because every
undetected class lives on the CNOTs touching the injected qubit (for
the corner under their schedule, also its neighbour), and the
neighbourhood of those gates is the same on every patch with
$d \geq 3$.
Increasing the distance, going to a larger surface code lattice,
adds fault locations, but every added location feeds a detector. Three things do depend on the distance:
\begin{enumerate}
\item the decoding graph: whether a detected class is mis-corrected is
a property of the whole matching graph, and at $d = 3$ a hook-type
residual aliases with a single-qubit error while at $d \geq 5$ it
does not, so the minimum-weight-matching part of the objective was
recomputed at every distance (it vanishes for all the quoted
schedules at all of them);
\item the generic N/Z pattern: its central CNOT term is $13/15$ at $d = 3$
but $7/15$ at $d \geq 5$, a boundary effect with no analogue for the
optimised schedules;
\item everything beyond leading order: every added check
contributes fault locations whose detected faults are rejected, so
the acceptance falls with the patch size, and the quadratic
coefficient of section~\ref{sub:tinj} is a property of the full
error graph, hence of the distance.
\end{enumerate}

\begin{figure}[H]
    \centering
    \includegraphics[width=0.95\linewidth]{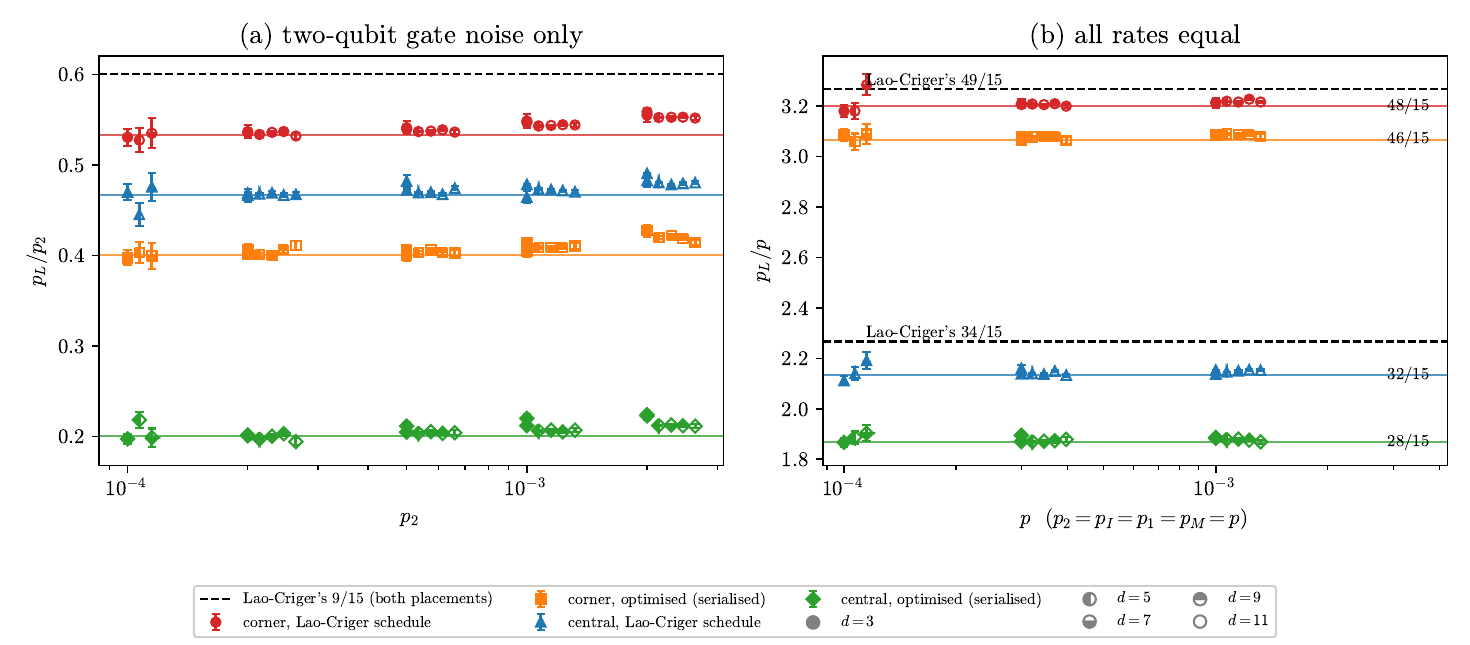}
    \caption{Monte-Carlo simulations at $d = 3$ to $11$ (marker fill
    solid to open with increasing distance). (a) CNOT noise only: sampled $p_L/p_2$
    for all four schemes against the Pauli web-derived slopes $8/15$,
    $6/15$, $7/15$ and $3/15$ (solid) and the claimed $9/15$
    (dashed). (b) All rates equal: the totals $48p/15$, $46p/15$,
    $32p/15$ and $28p/15$ against the claimed $49p/15$ and $34p/15$
    (dashed). Error bars are one standard deviation. The drift at
    larger $p$ is the $O(p^2)$ term.}
    \label{fig:mc_lc}
\end{figure}

Pauli webs see only undetected faults. A detected fault can still
become a logical error when the matching mis-corrects it. The search
therefore scores each candidate with the Pauli webs and the decoder
together, and the optimised schedule choice is certified by
sampling. The logical error claims above are verified using direct
stabiliser sampling (figure~\ref{fig:mc_lc}): \texttt{stim}
circuits of the full protocol at every odd distance from $3$ to
$11$, up to $1.8$ billion shots per point, post-selection exactly as
in the protocol, and minimum-weight matching on the accepted shots.
The sampled $p_L/p_2$ converges at small $p_2$ to the enumerated
coefficients at every distance.

\section{Parametric ZX-diagrams}
\label{sec:paramzx}

In section~\ref{app:circuitnoise} the Pauli webs enumerated the
malignant faults at leading order, and \texttt{stim} sampling
confirmed the resultant logical error rates. Sampled rates carry statistical uncertainty. The Pauli web enumeration is
exact but stops at leading order. Parameterised
diagrams~\cite{tsim2026, sutcliffe2025param} remove both
restrictions: each noise channel enters as binary phase variables,
the rewrites carry them along, and one reduction covers all fault
configurations at once, giving closed forms for the logical error
rate and the acceptance probability at all orders and all noise
strengths. Sampling then serves only as a cross-check. This section
develops the machinery on the injection protocol.

Figure~\ref{fig:tsim_warmup} introduces the parametric ZX
notation from \texttt{tsim} on a distance-$2$ repetition code
block, modified from the worked
example of~\cite{tsim2026}: two data qubits prepared in $\ket{0}$,
their $ZZ$ parity measured twice ($r_1$, $r_2$), an $X$ error
$e_1$ between the two rounds, another, $e_2$, after the second
round, and readouts $m_a$, $m_b$. A detector is an XOR of
measurement outcomes that is deterministic without noise, here
$D = r_1 \oplus r_2$, and the logical readout is
$L = m_a$. The outcomes $r_k$, $m_a$, $m_b$ and the errors $e_i$ are all
binary variables carried along symbolically, so the reduction is
simple parity algebra.
Everything that follows is this example at scale.
Figure~\ref{fig:paramzx_overview} sketches the pipeline on the
injection protocol itself.

\begin{figure}[H]
    \centering
    \resizebox{0.92\linewidth}{!}{\begin{tikzpicture}[scale=1.0,
  zpd/.style={minimum size=4.6mm, font={\footnotesize\boldmath},
    shape=rectangle, rounded corners=2mm, inner sep=1.1mm, draw=black,
    fill={rgb,255: red,216; green,248; blue,216}},
  xpd/.style={minimum size=4.6mm, font={\footnotesize\boldmath},
    shape=rectangle, rounded corners=2mm, inner sep=1.1mm, draw=black,
    fill={rgb,255: red,248; green,216; blue,216}},
  xorc/.style={shape=circle, draw=black, inner sep=0.25mm,
    font={\scriptsize},
    fill={rgb,255: red,248; green,216; blue,216}},
  wire/.style={-, line width=1.4pt, draw=black},
  cwire/.style={-, line width=0.4pt, draw=black},
  ann/.style={draw=none, fill=none, scale=0.68},
  vio/.style={draw=none, fill=none, scale=0.68, text=violet}]
\node[xpd] (a0) at (0, 1.5) {};
\node[zpd] (a1) at (1.3, 1.5) {};
\node[xpd] (ae1) at (2.6, 1.5) {\textcolor{violet}{$\pi e_1$}};
\node[zpd] (a2) at (3.9, 1.5) {};
\node[xpd] (ae2) at (5.2, 1.5) {\textcolor{violet}{$\pi e_2$}};
\node[xpd] (am) at (6.4, 1.5) {$m_a\pi$};
\draw[wire] (a0) -- (a1); \draw[wire] (a1) -- (ae1);
\draw[wire] (ae1) -- (a2); \draw[wire] (a2) -- (ae2);
\draw[wire] (ae2) -- (am);
\node[xpd] (b0) at (0, 0) {};
\node[zpd] (b1) at (1.3, 0) {};
\node[zpd] (b2) at (3.9, 0) {};
\node[xpd] (bm) at (6.4, 0) {$m_b\pi$};
\draw[wire] (b0) -- (b1); \draw[wire] (b1) -- (b2);
\draw[wire] (b2) -- (bm);
\node[xpd] (r1) at (1.3, 0.75) {$r_1\pi$};
\node[xpd] (r2) at (3.9, 0.75) {$r_2\pi$};
\draw[wire] (a1) -- (r1); \draw[wire] (r1) -- (b1);
\draw[wire] (a2) -- (r2); \draw[wire] (r2) -- (b2);
\node[xorc] (Dx) at (2.6, -1.0) {$\oplus$};
\draw[cwire] (r1) -- (Dx);
\draw[cwire] (r2) -- (Dx);
\draw[cwire] (Dx) -- (7.05, -1.0);
\node[ann, anchor=west] at (7.05, -1.0) {$D = r_1 \oplus r_2$};
\draw[cwire] (am) -- (7.05, 1.5);
\node[ann, anchor=west] at (7.05, 1.5) {$L = m_a$};
\draw[cwire] (bm) -- (7.05, 0);
\node[draw=none, fill=none, scale=0.9] at (9.35, 0.75)
  {$\xrightarrow{\ \texttt{full\_reduce}\ }$};
\node[zpd] (D0) at (11.0, 1.5) {$\pi e_1$};
\node[ann, anchor=west] at (11.5, 1.5) {$D = 0$ (accept)};
\node[zpd] (Lb) at (11.0, 0) {$\pi[e_1{\oplus}e_2]$};
\node[ann, anchor=west] at (11.85, 0) {$L$ (flip)};
\end{tikzpicture}}
    \caption{The \texttt{tsim} notation on a distance-$2$
    repetition code block, modified from the worked example
    of~\cite{tsim2026}. Left: the noisy circuit, read left to
    right, with the two $ZZ$ parity measurements $r_1$, $r_2$, the
    parameterised $X$ errors $e_1$, $e_2$ (violet) and the final
    readouts $m_a$, $m_b$. Thick wires are quantum, thin wires
    classical: each measurement outcome leaves on a thin wire, and
    the declared detector $D = r_1 \oplus r_2$ and logical readout
    $L = m_a$ are XOR combinations of them (the unused $m_b$ is
    summed over). Right: after the parameter-safe reduction, $D$ and
    $L$ are single spiders carrying XOR forms of the error
    bits.}
    \label{fig:tsim_warmup}
\end{figure}

\begin{figure}[H]
    \centering
    \pgfdeclarelayer{edgelayer}
\pgfdeclarelayer{nodelayer}
\pgfsetlayers{edgelayer,nodelayer,main}
\tikzset{
  zd/.style={shape=circle, minimum size=1.6mm, inner sep=0pt,
    draw=black, fill={rgb,255: red,216; green,248; blue,216}},
  xd/.style={zd, fill={rgb,255: red,248; green,216; blue,216}},
  zpd/.style={minimum size=3.4mm, font={\scriptsize\boldmath},
    shape=rectangle, rounded corners=1.4mm, inner sep=0.7mm,
    draw=black, fill={rgb,255: red,216; green,248; blue,216}},
  xpd/.style={zpd, fill={rgb,255: red,248; green,216; blue,216}},
  thad/.style={shape=rectangle, minimum size=1.9mm, inner sep=0pt,
    draw=black, fill={rgb,255: red,255; green,255; blue,128}},
  tnone/.style={draw=none, fill=none, inner sep=0.4mm,
    font={\scriptsize}},
  tedge/.style={-, thick},
  thedge/.style={-, dashed, dash pattern=on 2pt off 0.5pt, thick,
    draw={rgb,255: red,68; green,136; blue,255}}}
\begin{tabular}{ccc}
\resizebox{0.40\linewidth}{!}{\input{fig_tsim_stage_pre_Y}} &
\hspace*{-9mm}$\xrightarrow{\;\texttt{full\_reduce}\;}$ &
\resizebox{0.46\linewidth}{!}{\begin{tikzpicture}
  \begin{pgfonlayer}{nodelayer}
    \node [style=zd] (0) at (0.00, -2.60) {};
    \node [style=zpd] (10) at (0.00, 0.00) {$\pi[e_{0}{\oplus}e_{3}{\oplus}e_{4}{\oplus}e_{5}{\oplus}e_{6}{\oplus}e_{7}{\oplus}e_{9}{\oplus}e_{10}{\oplus}e_{11}{\oplus}e_{15}{\oplus}e_{16}]$};
    \node [style=zpd] (13) at (4.31, 0.00) {$\pi e_{7}$};
    \node [style=zd] (15) at (0.00, -10.40) {};
    \node [style=zd] (29) at (0.00, -5.20) {};
    \node [style=zd] (35) at (0.00, -7.80) {};
    \node [style=zd] (48) at (1.50, -10.40) {};
    \node [style=zpd] (49) at (4.89, -2.60) {$\pi[e_{1}{\oplus}e_{3}{\oplus}e_{5}{\oplus}e_{7}{\oplus}e_{9}{\oplus}e_{11}{\oplus}e_{15}{\oplus}r_{0}{\oplus}r_{10}]$};
    \node [style=zpd] (60) at (1.19, -2.60) {$\pi e_{2}$};
    \node [style=zpd] (62) at (1.19, -5.20) {$\pi r_{4}$};
    \node [style=zpd] (88) at (9.23, 0.00) {$\tfrac{\pi}{2}{+}\pi[e_{3}{\oplus}e_{5}{\oplus}e_{6}{\oplus}e_{7}{\oplus}e_{8}{\oplus}e_{9}{\oplus}e_{10}{\oplus}e_{11}{\oplus}e_{15}{\oplus}e_{16}{\oplus}r_{3}]$};
    \node [style=zpd] (99) at (1.19, -7.80) {$\pi r_{7}$};
    \node [style=zd] (100) at (8.51, -2.60) {};
    \node [style=zd] (101) at (3.00, -10.40) {};
    \node [style=zd] (102) at (4.50, -10.40) {};
    \node [style=zd] (103) at (14.06, 0.00) {};
    \node [style=zd] (104) at (2.38, -5.20) {};
    \node [style=zd] (106) at (6.00, -10.40) {};
    \node [style=zd] (107) at (2.38, -7.80) {};
    \node [style=zd] (108) at (7.50, -10.40) {};
    \node [style=zd] (109) at (9.00, -10.40) {};
    \node [style=zd] (113) at (10.50, -10.40) {};
    \node [style=zd] (114) at (12.00, -10.40) {};
    \node [style=zd] (115) at (13.50, -10.40) {};
    \node [style=zpd] (120) at (9.70, -2.60) {$\pi e_{2}$};
    \node [style=zd] (122) at (15.00, -10.40) {};
    \node [style=zpd] (124) at (17.75, 0.00) {$\tfrac{\pi}{2}{+}\pi[e_{3}{\oplus}e_{5}{\oplus}e_{7}{\oplus}e_{9}{\oplus}e_{11}{\oplus}e_{15}{\oplus}r_{16}]$};
    \node [style=zd] (125) at (21.43, 0.00) {};
    \node [style=tnone] (131) at (4.89, -1.60) {$D_{0}$};
    \node [style=tnone] (132) at (9.23, 1.00) {$D_{1}$};
    \node [style=tnone] (133) at (1.19, -4.20) {$D_{2}$};
    \node [style=tnone] (134) at (1.19, -6.80) {$D_{3}$};
    \node [style=tnone] (135) at (17.75, 1.00) {$L$};
    \node [style=tnone] (136) at (5.79, -3.60) {$r_{0}$};
    \node [style=tnone] (137) at (3.00, -9.40) {$r_{1}$};
    \node [style=tnone] (138) at (4.50, -9.40) {$r_{2}$};
    \node [style=tnone] (139) at (10.13, -1.00) {$r_{3}$};
    \node [style=tnone] (140) at (2.09, -6.20) {$r_{4}$};
    \node [style=tnone] (141) at (12.00, -9.40) {$r_{5}$};
    \node [style=tnone] (142) at (6.00, -9.40) {$r_{6}$};
    \node [style=tnone] (143) at (2.09, -8.80) {$r_{7}$};
    \node [style=tnone] (144) at (7.50, -9.40) {$r_{8}$};
    \node [style=tnone] (145) at (9.00, -9.40) {$r_{9}$};
    \node [style=tnone] (146) at (1.19, -3.60) {$r_{10}$};
    \node [style=tnone] (147) at (15.00, -9.40) {$r_{11}$};
    \node [style=tnone] (148) at (1.50, -9.40) {$r_{12}$};
    \node [style=tnone] (149) at (10.50, -9.40) {$r_{13}$};
    \node [style=tnone] (150) at (0.00, -9.40) {$r_{14}$};
    \node [style=tnone] (151) at (13.50, -9.40) {$r_{15}$};
    \node [style=tnone] (152) at (18.65, -1.00) {$r_{16}$};
  \end{pgfonlayer}
  \begin{pgfonlayer}{edgelayer}
    \draw [style=thedge] (0) to (49);
    \draw [style=thedge] (10) to (124);
    \draw [style=thedge] (10) to (88);
    \draw [style=thedge] (13) to (88);
    \draw [style=thedge] (15) to (150);
    \draw [style=thedge] (29) to (62);
    \draw [style=thedge] (35) to (99);
    \draw [style=thedge] (48) to (148);
    \draw [style=thedge] (49) to (100);
    \draw [style=tedge] (49) to (136);
    \draw [style=tedge] (49) to (131);
    \draw [style=thedge] (49) to (120);
    \draw [style=thedge] (49) to (60);
    \draw [style=thedge] (60) to (146);
    \draw [style=thedge] (62) to (104);
    \draw [style=tedge] (62) to (140);
    \draw [style=tedge] (62) to (133);
    \draw [style=thedge] (88) to (103);
    \draw [style=tedge] (88) to (139);
    \draw [style=tedge] (88) to (132);
    \draw [style=thedge] (88) to (124);
    \draw [style=thedge] (99) to (107);
    \draw [style=tedge] (99) to (143);
    \draw [style=tedge] (99) to (134);
    \draw [style=thedge] (101) to (137);
    \draw [style=thedge] (102) to (138);
    \draw [style=thedge] (106) to (142);
    \draw [style=thedge] (108) to (144);
    \draw [style=thedge] (109) to (145);
    \draw [style=thedge] (113) to (149);
    \draw [style=thedge] (114) to (141);
    \draw [style=thedge] (115) to (151);
    \draw [style=thedge] (122) to (147);
    \draw [style=thedge] (124) to (125);
    \draw [style=tedge] (124) to (152);
    \draw [style=tedge] (124) to (135);
  \end{pgfonlayer}
\end{tikzpicture}}\\
{\footnotesize the noisy protocol as parsed} & &
{\footnotesize detectors, coins and observable, as XOR forms}
\end{tabular}

    \caption{The parametric pipeline for the $\ket{Y}$ injection
    campaign of the optimised central scheme, drawn from the graphs
    of \texttt{tsim}~\cite{tsim2026} itself at $d = 3$. Left: the noisy protocol as
    \texttt{tsim} parses it, with the error parameters following the
    spiders. The detectors $D_j$, the measurement outcomes $r_k$ and the logical
    observable $L$ are open legs. Right: after the
    parameter-safe reduction, each
    detector carries an XOR form of the error variables, and $L$ is
    attached to a spider carrying an XOR form of its own. Post-selecting the detectors on the trivial value (binary
    $0$), summing the measurement outcomes that no detector
    constrains, and weighting the error variables by their
    probabilities (each two-qubit Pauli class with $p_2/15$, and so
    on) turns the post-selected probabilities into exact
    closed forms.}
    \label{fig:paramzx_overview}
\end{figure}

\texttt{tsim} models Pauli channels as parameterised vertices
carrying binary variables~\cite{tsim2026,sutcliffe2025param}, and its
rewrite engine reduces diagrams without evaluating the variables (we
use the same machinery through the framework extensions
of~\cite{wan2026cultivation}). Injecting a fault as such a variable
$s$, leaving the injected qubit's wire open and reducing, the entire
protocol, over a thousand spiders at $d = 9$, contracts to a single spider behind a Hadamard on that wire, a $Z$-phase $3\pi/2 + \pi s$: the
residual logical error in closed form. The gadget is identical at
every odd distance we have reduced, up to $d = 29$, where eleven and
a half thousand spiders contract to the same two vertices. We will see why this happens at every odd distance in
figure~\ref{fig:cone_identical}. This one small object also gives
us everything we need in the next section: substituting values for
the fault variable reproduces the frame corrections and the damage
of each fault class.

The acceptance probability and the post-selected logical error rate
follow exactly, as expectations over the reduced object: fix the
detector variables to zero, sum the outcome variables that no
detector constrains, and weight the error variables by their channel
law. For a single two-qubit depolarising channel on the CNOT
coupling the injected qubit to its $X$-check, the acceptance
probability, the joint probability of accepting and flipping the
logical $\bar{Y}$, and the resulting logical error rate are
\begin{equation}\label{eq:onechannel}
P_{\mathrm{acc}} = 1 - \tfrac{8}{15}p_2, \qquad
P(\text{flip logical and accepted}) = \tfrac{4}{15}p_2, \qquad
p_L = \frac{4p_2}{15 - 8p_2},
\end{equation}
closed forms at all orders in which the enumeration is visible
directly:
of the fifteen fault classes, eight fire a detector, four pass and
flip, three pass and are harmless.

The multi-channel closed form is itself documented at this level.
With an initialisation-flip channel $q$ on the injected qubit (a
single site's preparation flip, varied independently of the model's
$p_I$) and
two-qubit depolarising channels $p_2$ on its four check CNOTs, the
contraction of the whole protocol gives the logical error rate of
the injected $\ket{Y}$ state, written $p_L^{\ket{Y}}$, exactly:
\begin{multline}\label{eq:ymulti}
225\,(8p_2-15)^2 \; p^{\ket{Y}}_L(p_2, q) = 65536p_2^4 q - 32768p_2^4
- 245760p_2^3 q + 122880p_2^3\\
+ 345600p_2^2 q - 165600p_2^2 - 216000p_2 q + 81000p_2 + 50625 q,
\end{multline}
with leading term $13p_2/5$ at $q = p_2$. The reduction terminates
in a bare scalar, since the protocol is Clifford. The formula is
verified numerically, by an exhaustive branch sum through an
independent contraction path and by direct sampling, and setting
$q = 0$ and all but the $X$-check CNOT channel to zero recovers the
$p_L$ of equation~\eqref{eq:onechannel}.
With the preparation channel off ($q = 0$),
\begin{equation}\label{eq:q0}
p_L\big|_{q=0} = \frac{8p_2\,(10125 - 20700\,p_2 + 15360\,p_2^2
- 4096\,p_2^3)}{225\,(15-8p_2)^2},
\end{equation}
with leading term $8p_2/5 = 24\,p_2/15$, the summed flip classes of
the injected qubit's four check \textsc{cnot}s. Equation~\eqref{eq:ymulti} does
not depend on the distance, and the contraction returns the identical
rational function at every odd distance from $3$ to $29$.
Figure~\ref{fig:cone_identical} draws a translucent cone
around the neighbourhood of the injected qubit that holds every
noisy channel, at $d = 5$ and at $d = 9$: the sub-circuit inside
the cone is the same in the two panels, and faults never propagate
beyond it. Only the patch around the cone grows with the distance.

\begin{figure}[H]
    \centering
    \subfloat[$d=5$]{
    \tdplotsetmaincoords{65}{33}
    \resizebox{0.31\linewidth}{!}{\input{fig_cone_d5}}}
    \hfill
    \subfloat[$d=9$]{
    \tdplotsetmaincoords{65}{33}
    \resizebox{0.41\linewidth}{!}{\input{fig_cone_d9}}}
    \caption{The same fault cone in a growing patch: the noisy
    protocol at $d = 5$ and $d = 9$ with the neighbourhood of the
    injected qubit that holds every noisy channel enclosed in a
    translucent cone. The sub-circuit inside the cone is identical
    in the two panels.}
    \label{fig:cone_identical}
\end{figure}

\begin{figure}[H]
    \centering
    \includegraphics[width=0.62\linewidth]{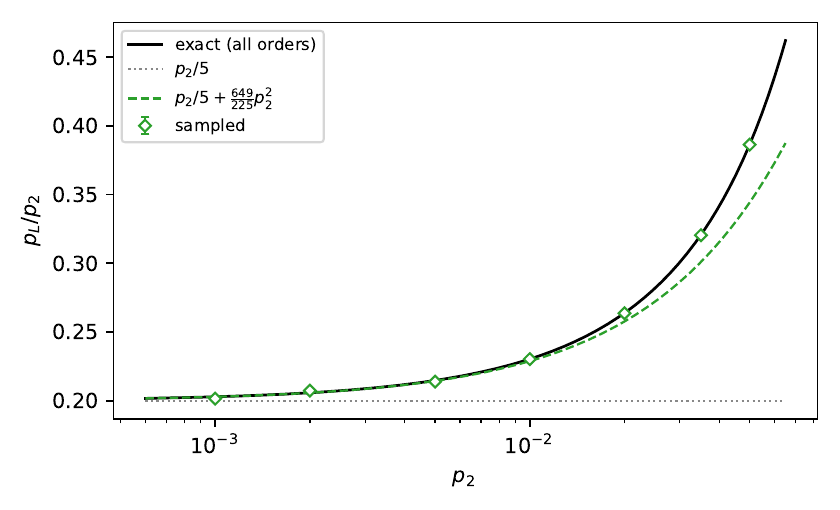}
    \caption{The exact all-orders logical error rate of the
    optimised central scheme for injected $\ket{Y}$ at $d = 3$
    (solid), from the parity expansion over the \texttt{tsim}-derived
    detector forms, against direct \texttt{stim} sampling of the
    full protocol (markers, up to $1.5\times10^8$ shots per point)
    and the first- and second-order truncations (dotted, dashed).}
    \label{fig:exact_curve}
\end{figure}

The reduction is one route to the exact rates. The linearity of
fault propagation is another. An exact enumeration of single faults and fault pairs in
the sampled circuit (each class inserted deterministically, its
detector and readout effect read off, pairs composed by the linearity
of Pauli frames) gives
\begin{equation}\label{eq:ypair}
p_L^{\ket{Y}} = \tfrac{1}{5}\,p_2 + \tfrac{649}{225}\,p_2^2 +
O(p_2^3)
\end{equation}
for CNOT noise at $d = 3$, the leading term reproducing the
$3p_2/15$ of section~\ref{sub:mc}. The quadratic term is what bends the
sampled points of figure~\ref{fig:t_inj} away from their dotted
leading-order lines.
The pair enumeration extends to all orders. Every fault effect is
linear over $\mathbb{F}_2$, so each detector and the observable is
an XOR of the error bits. For a single parity bit $b$ the indicator
of $b = 0$ is $\tfrac{1}{2}(1 + (-1)^b)$. Multiplying one indicator
per detector, one for the observable, and expanding the product,
the post-selected probability becomes a finite sum of expectations
of $(-1)^{e_{i_1} \oplus \cdots \oplus e_{i_k}}$. Each expectation
factorises over the independent channels, a channel contributes
$1 - 16p_2/15$ whenever any of its bits enter the parity and $1$
otherwise. The sum therefore evaluates to a closed
rational function of $p_2$, with series
\begin{equation}\label{eq:yseries}
p_L^{\ket{Y}} = \tfrac{1}{5}\,p_2 + \tfrac{649}{225}\,p_2^2
+ \tfrac{48761}{3375}\,p_2^3 + O(p_2^4),
\end{equation}
the first two coefficients reproducing the pair enumeration of
equation~\eqref{eq:ypair}.
Figure~\ref{fig:exact_curve} checks the exact function against
direct sampling up to $p_2 = 0.05$.

The same machinery recreates the full error budgets. The remaining
noise locations of the model enter the same way, one or two bits
each (figure~\ref{fig:full_budget}). Classifying all of them returns
\begin{equation}\label{eq:budget}
p_L^{\text{central}} = \tfrac{3}{15}\,p_2 + p_I + \tfrac{2}{3}\,p_1
+ 0\,p_M
\end{equation}
for the optimised central scheme and
$\tfrac{7}{15}\,p_2 + p_I + \tfrac{2}{3}\,p_1 + 0\,p_M$ for the
Tomita--Svore ordering: the equations of section~\ref{sub:budget}
again, now from fault insertion (the corner budgets follow
identically, $8/15$ and $6/15$ with $2\,p_I$). The same expansion
goes through unchanged, each location contributing $1$ or its own
$1 - \tfrac{16p_2}{15}$, $1 - 2p_I$, $1 - \tfrac{4p_1}{3}$ or
$1 - 2p_M$, and gives the exact $p_L(p_2, p_I, p_1, p_M)$. With
all rates equal,
\begin{equation}\label{eq:equalrates}
p_L = \tfrac{28}{15}\,p + \tfrac{1759}{225}\,p^2
+ \tfrac{15467}{1125}\,p^3 + O(p^4),
\end{equation}
the headline of section~\ref{sub:mc} with its exact higher orders.
At
\begin{equation*}
(p_2, p_I, p_1, p_M) = (0.02, 0.01, 0.015, 0.01)
\end{equation*}
the exact function gives $0.02590$ against sampling:
$0.025862(36)$.

The four routes to these numbers, the Pauli web enumeration at
leading order, the fault insertion, the parametric contraction and
direct sampling, agree wherever they overlap.

\begin{figure}[H]
    \centering
    \resizebox{0.9\linewidth}{!}{\begin{tikzpicture}[scale=1.0,
  zpd/.style={minimum size=4.6mm, font={\footnotesize\boldmath},
    shape=rectangle, rounded corners=2mm, inner sep=1.1mm, draw=black,
    fill={rgb,255: red,216; green,248; blue,216}},
  xpd/.style={minimum size=4.6mm, font={\footnotesize\boldmath},
    shape=rectangle, rounded corners=2mm, inner sep=1.1mm, draw=black,
    fill={rgb,255: red,248; green,216; blue,216}},
  hedge/.style={-, dashed, dash pattern=on 2pt off 0.5pt, thick,
    draw={rgb,255: red,68; green,136; blue,255}},
  wire/.style={-, thick, draw=black},
  ann/.style={draw=none, fill=none, scale=0.68},
  vio/.style={draw=none, fill=none, scale=0.68, text=violet}]
\node[xpd] (ai) at (0, 1.5) {};
\node[zpd] (ac) at (1.6, 1.5) {};
\node[xpd] (am) at (3.2, 1.5) {};
\draw[wire] (ai) -- (ac); \draw[wire] (ac) -- (am);
\node[vio] at (0, 0.95) {$+\,\pi e_{\mathrm{I}}$};
\node[vio, anchor=west] at (3.5, 1.5) {$+\,\pi e_{\mathrm{M}}$};
\node[ann] at (0, 1.98) {init};
\node[ann] at (3.2, 1.98) {readout};
\node[zpd] (di) at (0, 0) {$\frac{\pi}{2}$};
\node[xpd] (dt) at (1.6, 0) {};
\node[draw=none, fill=none] (dout) at (3.2, 0) {};
\draw[wire] (di) -- (dt); \draw[wire] (dt) -- (dout.center);
\node[ann] at (3.45, 0) {$\cdots$};
\node[vio] at (0, -0.62) {$+\,\pi e^{x}_{1}, \pi e^{z}_{1}$};
\node[ann] at (0, 0.6) {rotation};
\draw[wire] (ac) -- (dt);
\node[vio] at (2.35, 1.05)
  {$+\,\pi e^{x}_{a}, \pi e^{z}_{a}$};
\node[vio] at (2.42, -0.42)
  {$+\,\pi e^{x}_{d}, \pi e^{z}_{d}$};
\node[draw=none, fill=none, scale=0.9] at (4.55, 0.75)
  {$\xrightarrow{\ \texttt{full\_reduce}\ }$};
\node[zpd] (D0) at (6.3, 1.5) {$\pi[{\textstyle\bigoplus} e]$};
\node[ann, anchor=west] at (6.85, 1.5) {$D_i = 0$ (accept)};
\node[zpd] (Lb) at (6.3, 0) {$\pi[{\textstyle\bigoplus} e]$};
\node[ann, anchor=west] at (6.85, 0) {$L$ (flip)};
\draw[hedge] (D0) -- (Lb);
\node[ann, text width=3.1cm, align=left] at (6.9, -0.95)
  {every location a factor $1$ or\\
   $1{-}\tfrac{16p_2}{15}$, $1{-}2p_I$, $1{-}\tfrac{4p_1}{3}$,
   $1{-}2p_M$};
\end{tikzpicture}}
    \caption{The full-budget computation, schematically: every noise
    location of the model enters the campaign diagram as one or two
    binary parameters (the four CNOT class bits, one initialisation
    bit, two rotation bits, one readout bit), the reduction leaves
    each detector and the observable carrying an XOR form of them,
    and each expectation of the parity expansion picks up one factor
    per location: $1$, or the location's own $1 - \tfrac{16p_2}{15}$,
    $1 - 2p_I$, $1 - \tfrac{4p_1}{3}$ or $1 - 2p_M$.}
    \label{fig:full_budget}
\end{figure}

\section{Injecting $\ket{T}$ states}
\label{sub:tinj}

The optimised central scheme injects arbitrary states, not only
$\ket{Y}$.
For a general injected state $\ket{\psi}$ the leading CNOT
infidelity is $\sum_P n_P \left(1 - |\langle\psi|P|\psi\rangle|^2
\right) p_2/15$, where $n_P$ is the number of undetected fault
classes acting as the logical $\bar{P}$, $P \in \{X, Y, Z\}$:
each such class deposits $\bar{P}$ on the injected state, and the
projective readout registers an error with the Born probability
$1 - |\langle\psi|P|\psi\rangle|^2$. The
$n_P$ depend on the circuit, not on the injected state: for the
optimised central scheme at $d = 3$, $(n_X, n_Y, n_Z) = (2, 1, 1)$. For $\ket{T}$ the weights are $1$ for $\bar{Z}$ and
$\tfrac{1}{2}$ for $\bar{X}$ and $\bar{Y}$~\cite{Litinski_2019}, so the optimised central
scheme predicts $2.5\,p_2/15 = \tfrac{1}{6}\,p_2$ against
$3\,p_2/15$ for $\ket{Y}$: the
magic state itself is slightly better protected than the Clifford
test state, because half of the residual fault classes only
partially damage it. The $p_1$ channel on the single-qubit preparation rotation
contributes the same way to both states, $2p_1/3$ each.

\begin{figure}[H]
    \centering
    \includegraphics[width=0.62\linewidth]{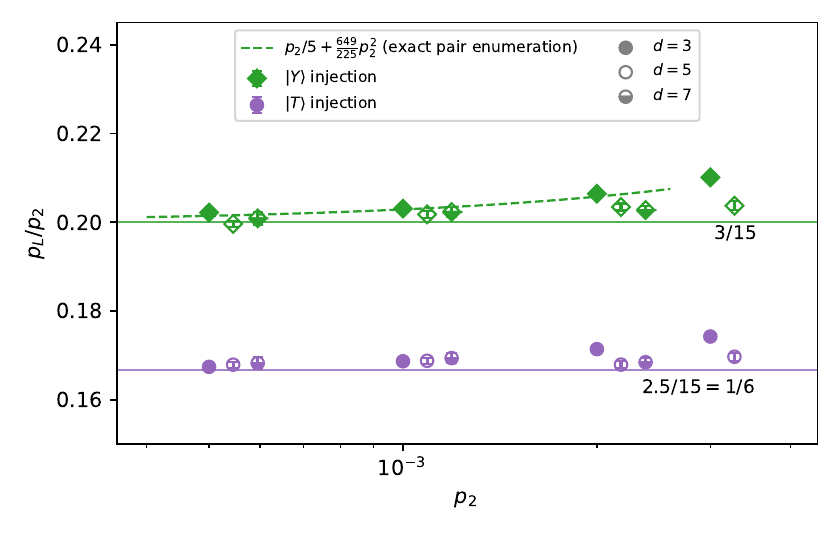}
    \caption{Sampled logical error rate of the optimised central scheme
    for injected $\ket{T}$ against injected $\ket{Y}$, from the
    stabiliser-rank simulation with the projective readout of
    section~\ref{sub:tinj}, at $d=3$ (filled), $d=5$ (open) and
    $d=7$ (half-filled),
    depolarising CNOT noise only, $p_2$ from $5\times10^{-4}$ to
    $3\times10^{-3}$ with up to $10^9$ shots per point. Solid lines are the type-resolved
    predictions, $2.5\,p_2/15$ for $\ket{T}$ and $3\,p_2/15$ for
    $\ket{Y}$~\cite{injection_webs_code}.}
    \label{fig:t_inj}
\end{figure}

Since the $\ket{T}$ prediction cannot be checked with a
stabiliser sampler, a ZX-based stabiliser-rank simulator
(\texttt{tsim}~\cite{tsim2026}) was used: a single $T$ gate costs
stabiliser rank two, and with the readout's $T^{\dagger}$ included
the rank stays two~\cite{wan2026cultivation}, so the full injection circuit samples at
near-Clifford cost. The circuit ends with a noiseless single-qubit readout: the patch
is measured down to the single injected qubit, and that
qubit is projected onto $\{\ket{T}, \ket{T^{\perp}}\}$, the
noiseless version of the faulty $T$ measurement
of~\cite{Litinski_2019}. At $d = 3$, $5$ and $7$ the sampled ratios match
the predicted $3/15$ for $\ket{Y}$ and $2.5/15 = 1/6$ for $\ket{T}$
(figure~\ref{fig:t_inj}). Both states drift above their leading
lines, following the quadratic terms computed in
section~\ref{sec:paramzx}.

\begin{figure}[H]
    \centering
    \includegraphics[width=0.62\linewidth]{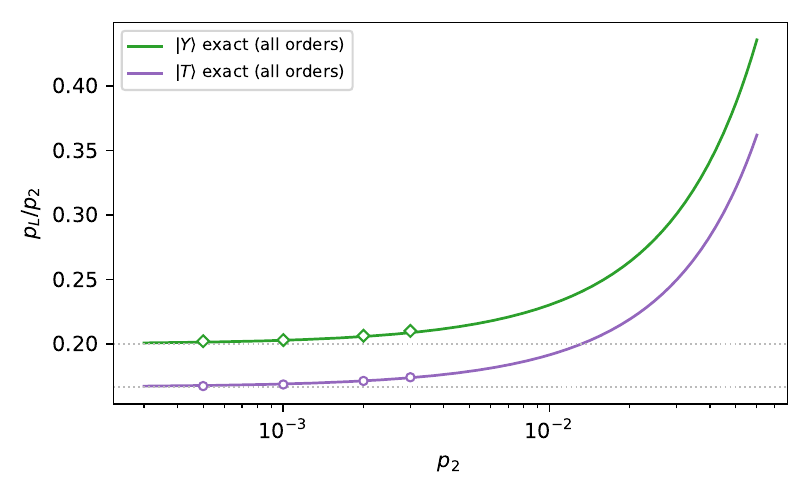}
    \caption{The exact all-orders $p_L/p_2$ for injected $\ket{Y}$
    and $\ket{T}$ (optimised central scheme, $d = 3$, CNOT noise),
    with the sampled campaign points of figure~\ref{fig:t_inj}.
    Dotted lines are the leading $1/5$ and $1/6$. All eight sampled
    points agree with the exact curves within $1.5$ standard
    deviations.}
    \label{fig:exactT}
\end{figure}

The exact curve extends to $\ket{T}$. The reduction also tracks
which residual logical error, $\bar{X}$, $\bar{Y}$ or $\bar{Z}$,
each accepted fault configuration leaves behind, and weighting
these by the damage weights of the opening paragraph gives
\begin{equation}\label{eq:tseries}
p_L^{\ket{T}} = \tfrac{1}{6}\,p_2 + \tfrac{1067}{450}\,p_2^2
+ \tfrac{40486}{3375}\,p_2^3 + O(p_2^4)
\end{equation}
exactly, with leading term the predicted $2.5\,p_2/15$. The same
construction returns the $\ket{Y}$ series of
section~\ref{sec:paramzx} unchanged. Figure~\ref{fig:exactT}
draws both exact functions through the sampled points.

The multi-channel closed form of equation~\eqref{eq:ymulti}
extends to $\ket{T}$ by the same contraction, with one difference:
the protocol is no longer Clifford, so the reduction stops at a
small remnant instead of a bare scalar, the three-spider chain of
figure~\ref{fig:tsim_remnant}. Its phases carry the surviving XOR
forms of the error bits, and averaging over the error bits with
their probabilities gives
\begin{multline}\label{eq:tmulti}
450\,(8p_2-15)^2 \; p_L(p_2, q) = 65536p_2^4 q - 65536p_2^4
- 245760p_2^3 q + 245760p_2^3 \\
+ 345600p_2^2 q - 331200p_2^2 - 216000p_2 q + 162000p_2 + 50625 q,
\end{multline}
with leading term $21p_2/10$ at $q = p_2$. The formula is
verified numerically, by the branch sum and by direct sampling,
like its $\ket{Y}$ counterpart, and the contraction returns the
identical rational function at every odd distance from $3$ to
$29$. Against $13p_2/5$ for $\ket{Y}$ the magic state is better
protected by exactly $q/2$: a preparation flip fully decoheres
$\ket{Y}$ but only half-decoheres $\ket{T}$. At $q = 0$ the two
functions coincide, equation~\eqref{eq:q0}.

\begin{figure}[H]
    \centering
    \resizebox{0.72\linewidth}{!}{
    \begin{tikzpicture}[scale=1.0,
  zpd/.style={minimum size=5mm, font={\footnotesize\boldmath},
    shape=rectangle, rounded corners=2mm, inner sep=1.2mm, draw=black,
    fill={rgb,255: red,216; green,248; blue,216}},
  hedge/.style={-, dashed, dash pattern=on 2pt off 0.5pt, thick,
    draw={rgb,255: red,68; green,136; blue,255}}]
\node[zpd] (a) at (0,0) {$\frac{\pi}{4}$};
\node[zpd] (b) at (3.1,0) {$0$};
\node[zpd] (c) at (6.6,0) {$\frac{7\pi}{4}+b\pi$};
\draw[hedge] (a) -- (b);
\draw[hedge] (b) -- (c);
\node[draw=none,fill=none,scale=0.72] at (0,-0.75)
  {$+\,\pi e_1$};
\node[draw=none,fill=none,scale=0.72] at (3.1,-0.75)
  {$+\,\pi(e_0{\oplus}e_4{\oplus}e_6{\oplus}e_{10}{\oplus}e_{16})$};
\node[draw=none,fill=none,scale=0.72] at (6.6,-0.75)
  {$+\,\pi(e_1{\oplus}e_3{\oplus}e_5{\oplus}e_7{\oplus}e_9{\oplus}e_{11}{\oplus}e_{15})$};
\end{tikzpicture}
    }
    \caption{The reduced parameterised diagram behind the
    multi-channel closed form: three $Z$-spiders joined by Hadamard
    edges,
    each phase carrying an XOR form of the channel error bits $e_i$.
    The readout bit $b$ enters only the final phase. The same
    three-spider object underlies both symbolic derivations of the
    formula~\cite{injection_webs_code}.}
    \label{fig:tsim_remnant}
\end{figure}

Figure~\ref{fig:tsim_stages} follows the $d = 3$ reduction stage
by stage: every stage of the pipeline removes deterministic
structure only, the error parameters on the nine violet segments
are carried along by every rewrite, and everything that reaches the
deflation detectors cancels against them. Only the three XOR forms
supported on the injected qubit's wire survive.
Figure~\ref{fig:mini3d} shows the same reduction at $d = 3$, $5$,
$7$ and $9$: the spacetime diagram it starts from grows
quadratically with the distance, but each diagram collapses to the
same three-spider chain of figure~\ref{fig:tsim_remnant}.

\begin{figure}[H]
    \centering
    \resizebox{0.95\linewidth}{!}{\pgfdeclarelayer{edgelayer}
\pgfdeclarelayer{nodelayer}
\pgfsetlayers{edgelayer,nodelayer,main}
\tikzset{
  zd/.style={shape=circle, minimum size=1.6mm, inner sep=0pt,
    draw=black, fill={rgb,255: red,216; green,248; blue,216}},
  xd/.style={zd, fill={rgb,255: red,248; green,216; blue,216}},
  zpd/.style={minimum size=3.4mm, font={\scriptsize\boldmath},
    shape=rectangle, rounded corners=1.4mm, inner sep=0.7mm,
    draw=black, fill={rgb,255: red,216; green,248; blue,216}},
  xpd/.style={zpd, fill={rgb,255: red,248; green,216; blue,216}},
  thad/.style={shape=rectangle, minimum size=1.9mm, inner sep=0pt,
    draw=black, fill={rgb,255: red,255; green,255; blue,128}},
  tnone/.style={draw=none, fill=none, inner sep=0.4mm,
    font={\scriptsize}},
  tedge/.style={-, thick},
  thedge/.style={-, dashed, dash pattern=on 2pt off 0.5pt, thick,
    draw={rgb,255: red,68; green,136; blue,255}}}
\begin{tabular}{ccc}
\resizebox{0.44\linewidth}{!}{\input{fig_tsim_stage_pre}} &
$\xrightarrow{\;\text{\footnotesize fuse}\;}$ &
\resizebox{0.44\linewidth}{!}{\input{fig_tsim_stage_fuse}}\\
{\footnotesize (a) $153$ vertices} & &
{\footnotesize (b) $115$ vertices}\\[2.5mm]
\resizebox{0.44\linewidth}{!}{\input{fig_tsim_stage_id}} &
$\xrightarrow{\;\text{\footnotesize reduce}\;}$ &
\resizebox{0.44\linewidth}{!}{\begin{tikzpicture}
  \begin{pgfonlayer}{nodelayer}
    \node [style=zd] (0) at (0.00, -2.60) {};
    \node [style=zpd] (9) at (0.00, 0.00) {$\tfrac{\pi}{4}$};
    \node [style=zpd] (10) at (2.76, 0.00) {$\pi[e_{0}{\oplus}e_{4}{\oplus}e_{6}{\oplus}e_{10}{\oplus}e_{16}]$};
    \node [style=zpd] (13) at (5.32, 0.00) {$\pi e_{7}$};
    \node [style=zd] (15) at (0.00, -10.40) {};
    \node [style=zd] (29) at (0.00, -5.20) {};
    \node [style=zd] (35) at (0.00, -7.80) {};
    \node [style=zd] (48) at (1.50, -10.40) {};
    \node [style=zpd] (49) at (4.89, -2.60) {$\pi[e_{1}{\oplus}e_{3}{\oplus}e_{5}{\oplus}e_{7}{\oplus}e_{9}{\oplus}e_{11}{\oplus}e_{15}{\oplus}r_{0}{\oplus}r_{10}]$};
    \node [style=zpd] (60) at (1.19, -2.60) {$\pi e_{2}$};
    \node [style=zpd] (62) at (1.19, -5.20) {$\pi r_{4}$};
    \node [style=zpd] (88) at (7.88, 0.00) {$\pi[e_{6}{\oplus}e_{8}{\oplus}e_{10}{\oplus}e_{16}{\oplus}r_{3}]$};
    \node [style=zpd] (99) at (1.19, -7.80) {$\pi r_{7}$};
    \node [style=zd] (100) at (8.51, -2.60) {};
    \node [style=zd] (101) at (3.00, -10.40) {};
    \node [style=zd] (102) at (4.50, -10.40) {};
    \node [style=zd] (103) at (10.35, 0.00) {};
    \node [style=zd] (104) at (2.38, -5.20) {};
    \node [style=zd] (106) at (6.00, -10.40) {};
    \node [style=zd] (107) at (2.38, -7.80) {};
    \node [style=zd] (108) at (7.50, -10.40) {};
    \node [style=zd] (109) at (9.00, -10.40) {};
    \node [style=zd] (113) at (10.50, -10.40) {};
    \node [style=zd] (114) at (12.00, -10.40) {};
    \node [style=zd] (115) at (13.50, -10.40) {};
    \node [style=zpd] (120) at (9.70, -2.60) {$\pi e_{2}$};
    \node [style=zd] (122) at (15.00, -10.40) {};
    \node [style=zpd] (123) at (13.76, 0.00) {$\tfrac{7\pi}{4}{+}\pi[e_{3}{\oplus}e_{5}{\oplus}e_{7}{\oplus}e_{9}{\oplus}e_{11}{\oplus}e_{15}]$};
    \node [style=zpd] (124) at (17.33, 0.00) {$\pi r_{16}$};
    \node [style=zd] (125) at (18.59, 0.00) {};
    \node [style=tnone] (131) at (4.89, -1.60) {$D_{0}$};
    \node [style=tnone] (132) at (7.88, 1.00) {$D_{1}$};
    \node [style=tnone] (133) at (1.19, -4.20) {$D_{2}$};
    \node [style=tnone] (134) at (1.19, -6.80) {$D_{3}$};
    \node [style=tnone] (135) at (17.33, 1.00) {$L$};
    \node [style=tnone] (136) at (5.79, -3.60) {$r_{0}$};
    \node [style=tnone] (137) at (3.00, -9.40) {$r_{1}$};
    \node [style=tnone] (138) at (4.50, -9.40) {$r_{2}$};
    \node [style=tnone] (139) at (8.78, -1.00) {$r_{3}$};
    \node [style=tnone] (140) at (2.09, -6.20) {$r_{4}$};
    \node [style=tnone] (141) at (12.00, -9.40) {$r_{5}$};
    \node [style=tnone] (142) at (6.00, -9.40) {$r_{6}$};
    \node [style=tnone] (143) at (2.09, -8.80) {$r_{7}$};
    \node [style=tnone] (144) at (7.50, -9.40) {$r_{8}$};
    \node [style=tnone] (145) at (9.00, -9.40) {$r_{9}$};
    \node [style=tnone] (146) at (1.19, -3.60) {$r_{10}$};
    \node [style=tnone] (147) at (15.00, -9.40) {$r_{11}$};
    \node [style=tnone] (148) at (1.50, -9.40) {$r_{12}$};
    \node [style=tnone] (149) at (10.50, -9.40) {$r_{13}$};
    \node [style=tnone] (150) at (0.00, -9.40) {$r_{14}$};
    \node [style=tnone] (151) at (13.50, -9.40) {$r_{15}$};
    \node [style=tnone] (152) at (18.23, -1.00) {$r_{16}$};
  \end{pgfonlayer}
  \begin{pgfonlayer}{edgelayer}
    \draw [style=thedge] (0) to (49);
    \draw [style=thedge] (9) to (10);
    \draw [style=thedge] (10) to (123);
    \draw [style=thedge] (13) to (88);
    \draw [style=thedge] (15) to (150);
    \draw [style=thedge] (29) to (62);
    \draw [style=thedge] (35) to (99);
    \draw [style=thedge] (48) to (148);
    \draw [style=thedge] (49) to (100);
    \draw [style=tedge] (49) to (136);
    \draw [style=tedge] (49) to (131);
    \draw [style=thedge] (49) to (120);
    \draw [style=thedge] (49) to (60);
    \draw [style=thedge] (60) to (146);
    \draw [style=thedge] (62) to (104);
    \draw [style=tedge] (62) to (140);
    \draw [style=tedge] (62) to (133);
    \draw [style=thedge] (88) to (103);
    \draw [style=tedge] (88) to (139);
    \draw [style=tedge] (88) to (132);
    \draw [style=thedge] (88) to (123);
    \draw [style=thedge] (99) to (107);
    \draw [style=tedge] (99) to (143);
    \draw [style=tedge] (99) to (134);
    \draw [style=thedge] (101) to (137);
    \draw [style=thedge] (102) to (138);
    \draw [style=thedge] (106) to (142);
    \draw [style=thedge] (108) to (144);
    \draw [style=thedge] (109) to (145);
    \draw [style=thedge] (113) to (149);
    \draw [style=thedge] (114) to (141);
    \draw [style=thedge] (115) to (151);
    \draw [style=thedge] (122) to (147);
    \draw [style=thedge] (123) to (124);
    \draw [style=thedge] (124) to (125);
    \draw [style=tedge] (124) to (152);
    \draw [style=tedge] (124) to (135);
  \end{pgfonlayer}
\end{tikzpicture}}\\
{\footnotesize (c) $109$ vertices} & &
{\footnotesize (d) $52$ vertices}
\end{tabular}
}
    \caption{The four stages of the reduction at $d=3$:
    \texttt{tsim}'s graph of the noisy protocol with the errors as
    parameters, the four postselection detectors $D_0,\dots,D_3$, the
    measurement outcomes $r_0,\dots,r_{16}$ and the logical
    observable $L$ pinned as open legs. (a) The circuit as parsed:
    green and red dots are Z- and X-spiders, boxed spiders carry a
    phase or a $\pi[\,\cdot\,]$ error parameter. (b) Fusing adjacent
    same-colour spiders and (c) removing identities strip
    deterministic structure while every rewrite carries the error
    variables along. (d) After the parameter-safe
    \texttt{full\_reduce}: the unconstrained outcomes are uniformly
    random bits on disconnected legs, $D_2$ and $D_3$ read single
    outcomes directly,
    $D_0$ and $D_1$ have absorbed the syndrome XORs
    $e_1{\oplus}e_3{\oplus}e_5{\oplus}e_7{\oplus}e_9{\oplus}e_{11}{\oplus}e_{15}$
    and $e_6{\oplus}e_8{\oplus}e_{10}{\oplus}e_{16}$ that make them
    fire, and the observable leg $L$ is attached to the three-spider chain
    of figure~\ref{fig:tsim_remnant}, whose middle spider carries
    $e_0{\oplus}e_4{\oplus}e_6{\oplus}e_{10}{\oplus}e_{16}$. Setting
    the detector legs to zero and summing the unconstrained outcomes
    is the
    postselected acceptance and error probability
    computation~\cite{injection_webs_code}.}
    \label{fig:tsim_stages}
\end{figure}

\begin{figure}[H]
    \centering
    \subfloat[$d=3$: $82$ spiders]{
    \tdplotsetmaincoords{65}{33}
    \resizebox{0.40\linewidth}{!}{\input{fig_mini3d_d3}}}
    \hfill
    \subfloat[$d=5$: $258$ spiders]{
    \tdplotsetmaincoords{65}{33}
    \resizebox{0.42\linewidth}{!}{\input{fig_mini3d_d5}}}\\
    \subfloat[$d=7$: $530$ spiders]{
    \tdplotsetmaincoords{65}{33}
    \resizebox{0.44\linewidth}{!}{\input{fig_mini3d_d7}}}
    \hfill
    \subfloat[$d=9$: $898$ spiders]{
    \tdplotsetmaincoords{65}{33}
    \resizebox{0.46\linewidth}{!}{\input{fig_mini3d_d9}}}
    \caption{Before the reduction, at every distance: the spacetime
    ZX-diagram of the noisy protocol behind the multi-channel closed
    form (initialisations, one optimised round, deflation readout,
    the injected qubit's wire open at the top), in the style of
    figure~\ref{fig:circuit3d}. The nine violet segments are the noise
    channel locations: the initialisation edge of the injected qubit
    and the outgoing edges of its four check CNOTs, the same nine
    segments on every patch. The parameter-safe reduction contracts
    each of these diagrams to the identical three-spider chain of
    figure~\ref{fig:tsim_remnant} (in about ten milliseconds), which
    is why the closed form cannot depend on the
    distance~\cite{injection_webs_code}.}
    \label{fig:mini3d}
\end{figure}

\section{Summary}
\label{sec:summary}

We analysed state injection on the rotated
surface code with Pauli webs. The Pauli webs identify the post-selection
checks and the malicious initialisation errors of the corner and
central placements, and under circuit-level noise serialised
schedules bring the central scheme to
$\tfrac{3}{15}\,p_2$ and the corner scheme to $\tfrac{6}{15}\,p_2$
at leading order. Since one round of 15-to-1 distillation outputs at
$35p^3$~\cite{bravyi2005universal}, the $34p/15 \rightarrow 28p/15$
improvement roughly halves the output distilled state's error.
Parameterised ZX-diagrams then upgrade the
leading-order enumeration to exact closed forms at all orders and
all noise strengths, verified against direct sampling, and the
contraction returns the same rational function at every odd
distance from $3$ to $29$. The same machinery covers arbitrary
injected states, with $\ket{T}$ slightly better protected than
$\ket{Y}$, $\tfrac{1}{6}\,p_2$ against $\tfrac{1}{5}\,p_2$ at
leading order.

\clearpage
\section{Acknowledgements}
 Kwok Ho Wan wishes to thank Zhenghao Zhong for a decade of friendship and also for inviting him to Oxford as a visitor whilst he was in between jobs. Zhenghao Zhong wants to thank Kwok Ho Wan for motivating him to write up these old ideas from their Imperial College London days. Zhenghao Zhong was supported by the ERC Consolidator Grant \# 864828 ``Algebraic Foundations of Supersymmetric Quantum Field Theory'' (SCFTAlg). Zhenghao Zhong acknowledges the TikZ code in figure \ref{fig:surface_code_Y_inj_post_selection} stems from Kwok Ho Wan's personal lecture notes at Imperial College London and has explicit consent to use it. The wording in parts of this manuscript had been refined using LLMs.

\bibliography{main}

\appendix

\section{The flat circuit-level diagram of the corner scheme}
\label{app:flatcircuit}

Figure~\ref{fig:li_noise} is the circuit-level diagram of the
corner scheme at $d = 3$ drawn as a 2D circuit.

\begin{figure}[H]
    \centering
    \resizebox{0.74\linewidth}{!}{
    \input{fig_circuit_full}
    }
    \caption{The circuit-level diagram of the corner scheme at
    $d = 3$ as a 2D circuit: nine data wires initialised as in
    figure~\ref{fig:inj_1} (blue node the $\ket{Y}$ state), one wire
    per ancilla, and two rounds of extraction in the interleaved
    scheduling of figure~\ref{fig:nz_schedule}, the second
    post-selected. Faults enter at the violet locations:
    initialisation flips ($p_I$, violet arrows, the two malignant
    ones on {\color{cyan}$2$} and {\color{cyan}$5$}), the fifteen
    two-qubit Pauli classes of each CNOT ($p_2/15$ each, on its two
    outgoing segments) and readout flips ($p_M$). A violet number
    next to a gate counts its malignant classes under the decoded
    convention of section~\ref{app:circuitnoise}: they sum to $13$,
    all in round $1$ and all around the injected corner, the
    $(13/15)\,p_2 + 2\,p_I$ of
    table~\ref{tab:decoded}~\cite{Li_2015,injection_webs_code}.}
    \label{fig:li_noise}
\end{figure}

\section{ZZ injection}
\label{app:zz}

ZZ injection~\cite{singh2022injection} places no bare physical state
in the initial time slice: the patch is first prepared
fault-tolerantly in a logical Clifford state, and the magic is rotated
in afterwards with a single two-qubit gate. At distance $2$:
\begin{enumerate}
\item initialise all four data qubits in $\ket{+}$ and measure one
round of parities: this prepares $\ket{\bar{+}}$ fault-tolerantly,
with the green plaquette parity post-selectable and the two $Z$-type
boundary outcomes absorbed into the Pauli frame;
\item apply $ZZ(\theta) = e^{-i\theta Z \otimes Z/2}$ to the row pair
{\color{cyan}$0$}, {\color{cyan}$2$} of
figure~\ref{fig:zz_injection}: the product
$Z_{{\color{cyan}0}} Z_{{\color{cyan}2}}$ is a representative of
$\bar{Z}$ (a column pair is a $Z$-type check and acts trivially), so
the gate is the logical $e^{-i\theta\bar{Z}/2}$ and $\theta = \pi/2$
yields $\ket{\bar{Y}}$;
\item keep measuring parities at distance $2$, restarting whenever a
post-selectable parity disagrees, then grow to the target distance.
\end{enumerate}

\begin{figure}[!h]
    \centering
    \resizebox{0.35\linewidth}{!}{
    \begin{tikzpicture}[scale=.7,every node/.style={minimum size=0.35cm},on grid]
    \node[fill=white,shape=circle,draw=black] (n1) at (2,6) {${\ket{+}}$};
    \node[fill=white,shape=circle,draw=black] (n0) at (2,2) {${\ket{+}}$};
    \node[fill=white,shape=circle,draw=black] (n3) at (6,6) {${\ket{+}}$};
    \node[fill=white,shape=circle,draw=black] (n2) at (6,2) {${\ket{+}}$};
    \node[draw=none,fill=none,text=cyan] at (1.1,1.1) {$0$};
    \node[draw=none,fill=none,text=cyan] at (1.1,6.9) {$1$};
    \node[draw=none,fill=none,text=cyan] at (6.9,1.1) {$2$};
    \node[draw=none,fill=none,text=cyan] at (6.9,6.9) {$3$};
    \begin{pgfonlayer}{background}
    \filldraw[fill=zx_green, opacity=0.65, draw=zx_green] (2.2,2.2) rectangle (5.8,5.8);
    \draw[zx_red,fill=zx_red,opacity=0.65](1.8,2.2) to[curve through={(0,4)}](1.8,5.8);
    \draw[zx_red,fill=zx_red,opacity=0.65](6.2,2.2) to[curve through={(8,4)}](6.2,5.8);
    \path [-,line width=0.1cm,black,opacity=1] (n0) edge node {} (n1);
    \path [-,line width=0.1cm,black,opacity=1] (n2) edge node {} (n3);
    \path [-,line width=0.1cm,black,opacity=1] (n0) edge node {} (n2);
    \path [-,line width=0.1cm,black,opacity=1] (n1) edge node {} (n3);
    \node[fill=black,circle,draw=black,scale = 1.5, text=white] at (4,4) {$+1$};
    \node[fill=white,circle,draw=black,scale = 0.25] at (0,4) {};
    \node[fill=white,circle,draw=black,scale = 0.25] at (8,4) {};
    \end{pgfonlayer}
    \draw[violet, line width=0.12cm, dashed] (2.6,2) -- (5.4,2);
    \node[draw=none,fill=none,text=violet] at (7.4,0.6) {$ZZ(\theta)$};
    \draw[->,violet] (6.4,0.8) to[bend left=15] (4.6,1.8);
    \end{tikzpicture}
    }
    \caption{ZZ injection~\cite{singh2022injection} on a distance $2$
    rotated surface code, drawn in the conventions of
    figure~\ref{fig:surface_code_Y_inj_post_selection}. All four data
    qubits start in $\ket{+}$, so the green $X$-type plaquette parity is
    post-selectable (`$+1$') and the patch encodes $\ket{\bar{+}}$
    fault-tolerantly. The two-qubit rotation $ZZ(\theta)$ (dashed purple)
    acts on the row pair {\color{cyan}$0$}, {\color{cyan}$2$}, whose
    product $Z_{{\color{cyan}0}}Z_{{\color{cyan}2}}$ is a representative
    of $\bar{Z}$, so $\theta = \pi/2$ yields $\ket{\bar{Y}}$.}
    \label{fig:zz_injection}
\end{figure}

In ZX-calculus, the gate is the phase gadget of
figure~\ref{fig:zz_gadget}, and
at $\theta = \pi/2$ its leaf is the blue $\pi/2$ node, so the diagram
stays Clifford and the Pauli webs apply unchanged. The unprotected
locations are confined to the gadget instead of a qubit world-line.
The price is a native data-to-data two-qubit gate at the injection
site~\cite{singh2022injection,gidney2023cleaner}.

\begin{figure}[!h]
    \centering
    \resizebox{0.9\linewidth}{!}{
    \begin{tikzpicture}[scale=.7,every node/.style={minimum size=0.55cm},on grid]
    \node[draw=none,fill=none] at (-3.4,0) {$e^{-i\theta Z/2}\;=$};
    \draw[black,line width=0.05cm] (-1.6,0) -- (1.6,0);
    \node[fill=zx_green,shape=circle,draw=black] at (0,0) {$\theta$};

    \begin{scope}[shift={(8.6,0)}]
    \node[draw=none,fill=none] at (-3.4,0) {$e^{-i\theta Z\otimes Z/2}\;=$};
    \draw[black,line width=0.05cm] (-1.6,1) -- (1.6,1);
    \draw[black,line width=0.05cm] (-1.6,-1) -- (1.6,-1);
    \node[fill=zx_green,shape=circle,draw=black] (t1) at (0,1) {};
    \node[fill=zx_green,shape=circle,draw=black] (t2) at (0,-1) {};
    \node[fill=zx_red,shape=circle,draw=black] (c) at (1.4,0) {};
    \node[fill=zx_green,shape=circle,draw=black] (l) at (2.8,0) {$\theta$};
    \draw[black,line width=0.05cm] (t1) -- (c);
    \draw[black,line width=0.05cm] (t2) -- (c);
    \draw[black,line width=0.05cm] (c) -- (l);
    \end{scope}

    \begin{scope}[shift={(17.4,0)}]
    \node[draw=none,fill=none] at (-3.0,0) {$ZZ(\tfrac{\pi}{2})\;=$};
    \draw[black,line width=0.05cm] (-1.6,1) -- (1.6,1);
    \draw[black,line width=0.05cm] (-1.6,-1) -- (1.6,-1);
    \node[fill=zx_green,shape=circle,draw=black] (u1) at (0,1) {};
    \node[fill=zx_green,shape=circle,draw=black] (u2) at (0,-1) {};
    \node[fill=zx_red,shape=circle,draw=black] (c2) at (1.4,0) {};
    \node[fill=zx_blue,shape=circle,draw=black] (l2) at (2.8,0) {};
    \draw[black,line width=0.05cm] (u1) -- (c2);
    \draw[black,line width=0.05cm] (u2) -- (c2);
    \draw[black,line width=0.05cm] (c2) -- (l2);
    \end{scope}
\end{tikzpicture}
    }
    \caption{Building $ZZ(\theta) = e^{-i\theta Z \otimes Z/2}$
    in the ZX-notation, left to right: the
    single-qubit rotation $e^{-i\theta Z/2}$ is a green spider with
    phase $\theta$ (half-angle convention: the spider phase equals
    the rotation angle; in the convention $e^{i\theta Z}$ it would be
    $-2\theta$), the phase-free red spider collects the $Z$-parity of
    the two wires and feeds it to the phase leaf, and at
    $\theta = \pi/2$ the leaf is a blue node following
    figure~\ref{fig:blue_notation}.}
    \label{fig:zz_gadget}
\end{figure}

Grown to distance $5$ (the seed round, the gadget, the $21$ growth
initialisations, two full rounds of parities), the Pauli webs verify the
scheme: the solver certifies that the diagram prepares
$\ket{\bar{Y}}$ of the distance $5$ code, $36$ parity outcomes are
post-selectable, and exactly four fault locations are malicious, all
on the gadget (an $X$ on either tapped wire before the next round, an
$X$ or $Z$ on the collector line). Figure~\ref{fig:zz_d5_web} draws
the logical $Y$ correlator web.

\begin{figure}[!h]
    \centering
    \tdplotsetmaincoords{70}{23}
    \resizebox{0.6\linewidth}{!}{
    \input{fig_zz_d5_web}
    }
    \caption{ZZ injection grown from distance $2$ to distance $5$, as a
    spacetime ZX-diagram in the style of figure~\ref{fig:web_mesh}: the
    four seed qubits and their round at the bottom, the $ZZ(\pi/2)$ phase
    gadget, the $21$ growth initialisations, and two full rounds at
    distance $5$. The Pauli web drawn is the logical $Y$ correlator: it enters
    at the phase gadget and exits on the $\bar{Y}$ representative at the
    final time slice. Green edges carry a $Z$-type decoration and red
    edges an $X$-type one. The blue node is the gadget's phase leaf, a
    green spider with phase $\pi/2$~\cite{injection_webs_code}.}
    \label{fig:zz_d5_web}
\end{figure}

\section{Hook injection}
\label{app:hook}

Hook injection~\cite{gidney2023cleaner} moves the rotation of
appendix~\ref{app:zz} onto a measurement ancilla: a hook error, a
single ancilla fault halfway through an extraction circuit that
propagates onto two data qubits, is used on purpose, with an
intentional rotation in place of the fault. At distance $2$:
\begin{enumerate}
\item initialise all four data qubits in $\ket{0}$ and measure one
round of parities: this prepares $\ket{\bar{0}}$ fault-tolerantly,
with the two red $Z$-type boundary checks post-selectable;
\item during the very first extraction of the green plaquette, order
the CNOTs so that a column pair, {\color{cyan}$0$} and
{\color{cyan}$1$}, comes last, and insert the rotation
$e^{-i\theta X/2}$ on the ancilla after the second CNOT
(figure~\ref{fig:hook_ladder}): pushed through the two remaining
CNOTs it becomes a rotation about
$X_{{\color{cyan}0}} X_{{\color{cyan}1}}$, a representative of
$\bar{X}$, and $\theta = \pi/2$ produces $\ket{-\bar{Y}}$, a
Pauli-frame relabel of $\ket{\bar{Y}}$;
\item verification rounds at distance $2$ and growth to the target
distance follow as in the other schemes.
\end{enumerate}

\begin{figure}[!h]
    \centering
    \resizebox{0.62\linewidth}{!}{
    \begin{tikzpicture}[scale=.7,every node/.style={minimum size=0.4cm},on grid]
    \foreach \y/\i in {8/3, 6/2, 4/1, 2/0} {
        \node[fill=zx_red,shape=circle,draw=black] (i\i) at (0,\y) {};
        \node[draw=none,fill=none,text=cyan] at (-1,\y) {$\i$};
        \draw[black,line width=0.05cm] (i\i) -- (14,\y);
    }
    \node[fill=zx_green,shape=circle,draw=black] (anc) at (1,0) {};
    \node[draw=none,fill=none] at (-1.2,0) {ancilla};
    \draw[black,line width=0.05cm] (anc) -- (5,0);
    \draw[black,line width=0.05cm] (9,0) -- (13,0);
    \draw[-, dashed, dash pattern=on 2pt off 1.2pt, line width=0.06cm,
          draw=zx_hadamard] (5,0) -- (9,0);
    \node[fill=zx_green,shape=circle,draw=black] (cap) at (13,0) {};
    \node[fill=zx_green,shape=circle,draw=black] (c1) at (3,0) {};
    \node[fill=zx_red,shape=circle,draw=black] (t1) at (3,6) {};
    \draw[black,line width=0.05cm] (c1) -- (t1);
    \node[fill=zx_green,shape=circle,draw=black] (c2) at (5,0) {};
    \node[fill=zx_red,shape=circle,draw=black] (t2) at (5,8) {};
    \draw[black,line width=0.05cm] (c2) -- (t2);
    \node[fill=zx_blue,shape=circle,draw=black] (rot) at (7,0) {};
    \node[fill=zx_green,shape=circle,draw=black] (c3) at (9,0) {};
    \node[fill=zx_red,shape=circle,draw=black] (t3) at (9,4) {};
    \draw[black,line width=0.05cm] (c3) -- (t3);
    \node[fill=zx_green,shape=circle,draw=black] (c4) at (11,0) {};
    \node[fill=zx_red,shape=circle,draw=black] (t4) at (11,2) {};
    \draw[black,line width=0.05cm] (c4) -- (t4);
    \end{tikzpicture}
    }
    \caption{The hooked extraction of the green $X$-type plaquette at
    distance $2$, as a ZX-diagram read left to right. The data qubits
    {\color{cyan}$0$}--{\color{cyan}$3$} start in $\ket{0}$ (red spiders).
    The ancilla starts in $\ket{+}$ (green spider), interacts with the
    data qubits by CNOTs (green control on the ancilla wire, red target
    on the data wire), and is finally measured in the $X$ basis (green
    spider, post-selected). The rotation $e^{-i\theta X/2}$ at
    $\theta = \pi/2$, an $X$-type $\pi/2$ spider drawn as a blue node
    between two Hadamard edges (figure~\ref{fig:blue_notation}, and the
    spider phase equals the rotation angle in the half-angle
    convention), is inserted after the second CNOT.}
    \label{fig:hook_ladder}
\end{figure}

Grown to distance $5$ as in appendix~\ref{app:zz}, the Pauli webs verify
the scheme at scale: the solver certifies that the grown diagram
prepares the $\bar{Y}$ eigenstate of the distance $5$ code, $37$
parity outcomes are post-selectable, and exactly four fault locations
are malicious, all on the ancilla segment carrying the rotation
(three $Z$-type, one $X$-type). Figure~\ref{fig:hook_d5_web} draws
the logical $Y$ correlator web.

\begin{figure}[!h]
    \centering
    \tdplotsetmaincoords{70}{23}
    \resizebox{0.6\linewidth}{!}{
    \input{fig_hook_d5_web}
    }
    \caption{Hook injection grown from distance $2$ to distance $5$, as a
    spacetime ZX-diagram in the style of figure~\ref{fig:web_mesh}: the
    all-$\ket{0}$ seed round with the hook gadget, the growth
    initialisations, and two full rounds at distance $5$. The Pauli web drawn
    is the logical $Y$ correlator, entering at the hook gadget and
    exiting on the $\bar{Y}$ representative at the final time slice.
    The gadget's rotation leaf, an $X$-type $\pi/2$ spider, is drawn as
    a blue node on a Hadamard edge, following
    figure~\ref{fig:blue_notation}~\cite{injection_webs_code}.}
    \label{fig:hook_d5_web}
\end{figure}

\section{Unitary-encoder injection}
\label{app:unitary}

A unitary encoder injects with no measurements at all: the input
qubit holds the state $\ket{\psi}$ to inject, the other $d^2 - 1$
data qubits start in $\ket{+}$ or $\ket{0}$, and layers of CNOTs
spread the input across the patch until the logical state is prepared
exactly (figure~\ref{fig:unitary_cone}). Such circuits exist at
nearest-neighbour depth $2d$~\cite{higgott2021optimal}, at
$d + O(1)$~\cite{claes2025lowerdepth}, and at $O(\log d)$ with
non-local gates~\cite{tsai2025unitary}.

\begin{figure}[!h]
    \centering
    \resizebox{0.5\linewidth}{!}{
    \begin{tikzpicture}[scale=.7,every node/.style={minimum size=0.4cm},on grid]
    \node[fill=zx_green,shape=circle,draw=black] (i0) at (0,8) {};
    \node[draw=none,fill=none] at (-1.6,8) {$\ket{+}$};
    \node[fill=zx_red,shape=circle,draw=black] (i1) at (0,6) {};
    \node[draw=none,fill=none] at (-1.6,6) {$\ket{0}$};
    \node[draw=none,fill=none] at (-1.6,4) {$\ket{\psi}$};
    \node[fill=zx_red,shape=circle,draw=black] (i3) at (0,2) {};
    \node[draw=none,fill=none] at (-1.6,2) {$\ket{0}$};
    \node[fill=zx_green,shape=circle,draw=black] (i4) at (0,0) {};
    \node[draw=none,fill=none] at (-1.6,0) {$\ket{+}$};
    \draw[black,line width=0.05cm] (i0) -- (10,8);
    \draw[black,line width=0.05cm] (i1) -- (10,6);
    \draw[black,line width=0.05cm] (-0.6,4) -- (10,4);
    \draw[black,line width=0.05cm] (i3) -- (10,2);
    \draw[black,line width=0.05cm] (i4) -- (10,0);
    \node[fill=zx_green,shape=circle,draw=black] (c1) at (3,4) {};
    \node[fill=zx_red,shape=circle,draw=black] (t1) at (3,6) {};
    \draw[black,line width=0.05cm] (c1) -- (t1);
    \node[fill=zx_green,shape=circle,draw=black] (c2) at (4.2,4) {};
    \node[fill=zx_red,shape=circle,draw=black] (t2) at (4.2,2) {};
    \draw[black,line width=0.05cm] (c2) -- (t2);
    \node[fill=zx_green,shape=circle,draw=black] (c3) at (7,6) {};
    \node[fill=zx_red,shape=circle,draw=black] (t3) at (7,8) {};
    \draw[black,line width=0.05cm] (c3) -- (t3);
    \node[fill=zx_green,shape=circle,draw=black] (c4) at (7,2) {};
    \node[fill=zx_red,shape=circle,draw=black] (t4) at (7,0) {};
    \draw[black,line width=0.05cm] (c4) -- (t4);
    \end{tikzpicture}
    }
    \caption{The light cone of a unitary encoder, as a ZX-diagram read
    left to right (schematic: the real circuits act on the full two
    dimensional patch~\cite{higgott2021optimal,claes2025lowerdepth}). The
    input wire carries the state $\ket{\psi}$ to inject. Each CNOT layer
    (green control spider, red target spider) spreads it one step
    further, while the $\ket{+}$ and $\ket{0}$ ancilla wires supply the
    stabilisers. There are no measurements anywhere, so no Pauli webs
    terminate and nothing can be post-selected.}
    \label{fig:unitary_cone}
\end{figure}

The Pauli webs make the trade-off immediate. With no measurement spiders
there is nothing to post-select on, so the acceptance probability is
one. But the logical $Y$ correlator web runs from the input wire
through the whole cone, and any single fault on that path is
undetectable: the preparation has fault distance one at every code
distance~\cite{higgott2021optimal,tsai2025unitary}. Unitary encoders
are therefore used as the growth stage of a scheme seeded by the
measurement-based methods of the previous appendices.

At distance $5$ (input qubit at the corner, CNOT chains down the
first column and along the bottom row, one round of parity
measurements) the Pauli webs verify this picture: the solver certifies the
encoding, and fifteen fault locations are malicious, against two for
the central-qubit scheme of appendix~\ref{app:injection} and four for
the gadget schemes of appendices~\ref{app:zz} and~\ref{app:hook}.
Figure~\ref{fig:unitary_d5_web} draws the logical $Y$ correlator
web.

\begin{figure}[!h]
    \centering
    \tdplotsetmaincoords{70}{23}
    \resizebox{0.6\linewidth}{!}{
    \input{fig_unitary_d5_web}
    }
    \caption{A unitary encoder at distance $5$, as a spacetime
    ZX-diagram in the style of figure~\ref{fig:web_mesh}: the input wire
    at the corner, CNOT chains spreading it across the patch, one round
    of parity measurements. The Pauli web drawn is the logical $Y$ correlator,
    from the input wire to the $\bar{Y}$ representative at the final
    time slice. Every fault location along this Pauli web that precedes the
    parity measurements is malicious~\cite{injection_webs_code}.}
    \label{fig:unitary_d5_web}
\end{figure}

\end{document}